\newif\ifsubmode
\submodefalse

\ifsubmode
  \documentclass[12pt,preprint]{aastex}
  \received{}
  \revised{}
  \accepted{}
\else
  \documentclass{emulateapj}

  \usepackage{apjfonts}
\fi
\usepackage{epsf}

\newcommand{\hdf}{HDF--N}
\newcommand{\hdfs}{HDF--S}

\newcommand{\hst}{\textit{HST}}

\newcommand{\wfu}{\hbox{$U_{300}$}}
\newcommand{\wfb}{\hbox{$B_{450}$}}
\newcommand{\wfv}{\hbox{$V_{606}$}}
\newcommand{\wfi}{\hbox{$I_{814}$}}

\newcommand{\nicj}{\hbox{$J_{110}$}}
\newcommand{\nich}{\hbox{$H_{160}$}}
\newcommand{\mstar}{\hbox{$\mathcal{M}^\ast$}}

\newcommand{\lya}{Lyman~$\alpha$}

\newcommand{\lsim}{\lesssim}
\newcommand{\gsim}{\gtrsim}
\newcommand{\lstar}{\hbox{$L^\ast$}}

\newcommand{\royalsociety}{Phil.\ Trans.\ R.\ Soc.\ Lond.\ A}
\newcommand{\etal}{et al.}
\newcommand{\eg}{e.g.}
\newcommand{\ie}{i.e.}

\newcommand{\msol}{\hbox{$\mathcal{M}_\odot$}}

\newcommand{\kms}{\hbox{km~s$^{-1}$}}

\newcommand{\infinity}{\hbox{$\infty$}}

\newcommand{\uit}{\textit{UIT}}

\newcommand{\tablethree}{
\tabletypesize{\small}
\tablewidth{0pt}
\tablecaption{Galaxy Component Models\label{table:models}}
\tablehead{ \colhead{} & \colhead{} & \colhead{ } &
 \colhead{$r_{1/2}$} & \colhead{} &
 \colhead{$\mathcal{M}/L_B$} & \colhead{  $\wfv - \nicj$ } & \colhead
 {$\wfi - \nich$} \\
 \colhead{Component} & \colhead{$t$} & \colhead{ $t/\tau$ } &
 \colhead{[kpc]} & \colhead{Profile} & \colhead{ [$\mathcal{M}_\odot /
 L_{B,\odot}$]} & \colhead{$z = 1$ } & \colhead{$z=2.3$ } \\
 \colhead{(1)} & \colhead{(2)} & \colhead{ (3) } &
 \colhead{(4)} & \colhead{(5)} & \colhead{(6)} & \colhead{(7)} & \colhead{(8)}}
\startdata
disk model 1 & 0.3 Gyr & 0.6 & 7.5 & Exponential & 0.11\phn & \phs0.41 & \phs0.36 \\
\phm{disk} model 2 & 0.1 Gyr & 1.0 & 7.5 & Exponential & 0.065 & \phs0.15 &
\phs0.08 \\
\phm{disk} model 3 & 0.1 Gyr & 4.0 & 7.5 & Exponential & 0.26\phn & \phs1.45 &
\phs1.67 \\
spheroid  & 2.8 Gyr& \infinity & 2.3 & de~Vaucouleurs & 2.7\phn\phn & \phs3.06 & \phs4.83\\
\ion{H}{2} region & 50 Myr & $\approx 0$ & \nodata  & Point source &
0.032 & $-0.16$ & $-0.27$
\enddata
\tablecomments{(1) Galaxy component; (2) Age of stellar population; (3)
Ratio of age to the star--formation rate e--folding timescale; (4)
Half--light radius; (5) Profile; (6) Stellar mass to blue light ratio
in solar units; observed frame colors at (7) $z=1$, and (8) $z=2.3$.
}
}

\newcommand{\figonecap}{Rest--frame UV and optical images of 28 galaxies
  in the $0.7 \leq z \leq 1.4$ sample with the highest luminosities in
  the rest--frame $B$--band from the HDF-N. The left panel for each
  galaxy shows the WFPC2 F606W image, which corresponds to the
  rest--frame mid--UV wavelengths.  The right panel shows the
  NICMOS F110W image, which depicts each galaxy approximately in the
  rest--frame $B$--band. Galaxies span luminosities $-22.7 \leq M(B)
  \leq -20.4$, and have redshifts as indicated in each
  image. Redshifts shown within parentheses indicate cases where only
  photometrically derived redshifts are available.  Note the diversity
  in the galaxies appearances as a function of rest--frame wavelength.
  Each panel is $5\times 5$~arcsec$^{2}$, which corresponds to $\simeq
  40\times40$~kpc$^2$ in the rest--frame. \label{fig:montage_lowz1}}

\newcommand{\figtwocap}{Same as Figure~\ref{fig:montage_lowz1}, but here
the figure shows the 28 galaxies from the $1.9 \le z \le 3.3$ galaxy
sample with the highest luminosities in rest--frame $B$--band.  The
left panel for each galaxy shows the WFPC2 F606W image, which
corresponds to rest--frame far--UV wavelengths. The right panel shows
the NICMOS F160W image, which corresponds approximately to the
rest--frame $B$--band (i.e., the same rest--frame band as F110W for
galaxies at $z\sim 1$).  Note the lack of morphological transformation
between wavelengths, even though these images span a longer wavelength
baseline compared to Figure~\ref{fig:montage_lowz1}.  The galaxies
span luminosities $-22.5 \leq M(B) \leq -20.7$ (cf.\ to the similar
range of galaxy $B$--band luminosities shown in
Figure~\ref{fig:montage_lowz1}; see also Dickinson \etal\ 2003). The
size of each panel is $5\times5$~arcsec$^2$, which corresponds to
$\simeq 40\times 40$~kpc$^2$ in the
rest--frame.\label{fig:montage_hiz1}}

\newcommand{\figthreecap}{Distribution of galaxy half--light radii
versus absolute magnitude in rest--frame light at 4400~\AA.  The
absolute magnitudes are presented in AB units for the default
cosmology.  The plot shows the radius--luminosity distribution for the
\hdf\ $z\sim 2.3$ (open stars) and $z\sim 1$ (filled squares) galaxy
samples, along with reference data collected from the RC3 catalog (de
Vaucouleurs \etal\ 1991).  The solid line indicates the relation for
local E/S0 galaxies \citep{bin84}, and the dashed and dotted lines
indicate the relation for local Sab/Sbc and Scd/Sdm/Irr galaxies,
respectively \citep{fre70,roc96}. Note that the NICMOS F160W PSF
corresponds to a half--light radius of $\simeq 1.2-1.4$~kpc\ for
$z\sim 0.7-3.0$. \label{fig:radlum}}

\newcommand{\figfourcap}{Distribution of galaxy sizes for the $z\sim
2.3$ and $z\sim 1$ \hdf\ samples for objects with $M(B) \leq -20.0$.
In both panels, the shaded histograms show the measured distributions.
In the \textit{right} panel, the thick, solid lines illustrate the
observed distribution of the $z\sim 1$ galaxies as measured from their
appearances when simulated at $z=2.7$  (see text), and normalized to
the total number of objects in the shaded histogram of each
panel. Note that the redshift range, $1.9 \leq z \leq 3.0$, contains
approximately $2.3\times$ the co-moving volume as $0.7 \leq z \leq
1.4$, and so the deficit of large galaxies at $z\sim 2.3$ has even
greater significance than what is illustrated in the figure.  In each
panel, the dotted line indicates the half--light radius of the NICMOS
PSF.\label{fig:radhists}}

\newcommand{\figfivecap}{The internal color dispersion measured between the
(rest--frame) UV--optical colors for the \hdf\ galaxy
samples as a function of redshift. The plot shows the internal
dispersion: $\xi(\wfv,\nicj)$ and
$\xi(\wfi,\nich)$ for the $z\sim 1$ and $z\sim 2.3$ galaxies,
respectively.  Error bars indicate the approximate $1\sigma$
uncertainties (see text).  The dashed lines in each panel
show the derived internal color dispersion for two local galaxies (M81
and M74) after convolving with the NICMOS F160W PSF, and resampling to
the same pixel scale as the high--redshift \hdf\ galaxies.
\label{fig:hdfxi}}

\newcommand{\figsixcap}{The internal color differences of the \hdf\
galaxies with the highest $\xi(\wfv,\nicj)$ values in the $z\sim 1$
sample.  For each galaxy, the  panels display the \wfv\ and \nicj\
images (as labeled); the right--most panel shows the difference
between the images after scaling to remove the mean total color, \ie,
these residuals show $I_1 - \alpha I_2 - \beta$, where $I_1$, and
$I_2$ refer to the two images given in the panel inset as
$\xi(I_1,I_2)$, and $\alpha$ and $\beta$ are the fitted constants (see
text for full description).  The corresponding internal color
dispersion value is shown in each color--residual image. Note that the
color residuals are not the same as the definition of internal color
dispersion (the latter is the squared sum of the former; divided by
the squared sum of the image flux).  Light/Dark regions correspond to
features with internal colors that are redder/bluer than the mean
total color.  Each panel has dimensions of $5\times 5$~arcsec$^2$,
which corresponds to $\simeq 40\times 40$~kpc$^2$ in the rest--frame
for these galaxies.
\label{fig:montage_xi_lowz}}

\newcommand{\figsevencap}{The internal color differences of the \hdf\
  galaxies with the highest $\xi(\wfi,\nich)$ values in the $z\sim
  2.3$ sample.  For each galaxy, the  panels display the \wfi\ and
  \nich\ images (as labeled); the right--most panel shows the
  difference between the images after scaling to remove the mean total
  color (see caption of fig~\ref{fig:montage_xi_lowz}). Each panel has
  dimensions of $5\times 5$~arcsec$^2$, which corresponds to $\simeq
  40\times 40$~kpc$^2$ in the rest--frame for these
  galaxies.\label{fig:montage_xi_hiz} }

\newcommand{\figeightcap}{ The effects of surface--brightness dimming
on the internal color dispersion.  The figure compares the internal
color dispersion values of galaxies in the $z\sim 1$ sample with $M(B)
\leq -20$, and the internal color dispersion of these galaxies
simulated at $z = 2.3$.  For each simulated galaxy, the \wfi\ and
\nich\ images were constructed by resampling their \wfv\ and \nicj\
images at a pixel scale corresponding to $z=2.3$, and appropriately
dimming their surface brightness.  Note that the simulated
$\xi(\wfi,\nich)$ may be \textit{underestimated} because this color at
this redshift is slightly more sensitive to heterogeneity in a
galaxy's stellar populations compared with $\xi(\wfv,\nicj)$ at $z\sim
1$, see \S~\ref{section:discussion}.\label{fig:xivxisim}}

\newcommand{\figninecap}{The galaxy size--stellar-mass distribution
  for the $z\sim 1$ and 2.3 samples from the \hdf.  Stellar mass
  estimates correspond to the derived values from Papovich et al.\
  (2001) and Dickinson et al.\ (2003).  Error bars on the stellar
  masses are shown for one--third of the objects to illustrate
  characteristic uncertainties.  The dashed horizontal line
  shows the stellar mass of a present--day \mstar-galaxy
  \citep[see,][with adjustments to the default cosmology]{col01}. The
  triple--dot--dashed and dot-dashed lines show the limiting
  stellar--mass for a maximally old object observed at the
  ($10\sigma$) flux limit of the data at $z=2.3$ and 1, respectively.
  The data are sensitive to all sources with masses greater than these
  lines.   Objects below these lines have lower mass-to-light ratios
  than the maximal model and are readily detected, but the
  completeness below these lines is less secure.  
\label{fig:r50vmass} }

\newcommand{\figtencap}{\footnotesize Distribution of galaxy internal color
dispersion and total colors of galaxies at $z\sim 1$ (\textit{top
panels}) and $z\sim 2.3$ (\textit{bottom panels}) in the \hdf.  The
\textit{left} panels show the derived internal color dispersion
between the UV--optical bands as a function of the total UV--optical
color.  The blue, dashed horizontal lines indicate a fiducial value
$\xi = 0.05$.  The \textit{right} panels show the (total) color--color
plots for each galaxy sample.  The open stars in the $z\sim 1$ plot
correspond to objects with insecure redshifts (see text).  The open
boxes in the $z\sim 2.3$ plot correspond to objects with $z\sim 2.9$,
where the \wfb--band flux is attenuated by intergalactic \lya, which
increases the observed $\wfb - \wfi$ color.  In these plots, the three
curves illustrate the colors in the observed frame for an evolving
stellar population with three monotonic star--formation histories in
the observed frame for $z=1$ (\wfu\wfv\nicj) and $z=2.3$
(\wfb\wfi\nich) respectively.  The blue curve with star symbols
corresponds to a stellar population with a constant star formation
rate.  The green curve with triangle symbols corresponds to a stellar
population with an exponentially decaying star--formation rate and
e--folding time constant, $\tau = 1$~Gyr. The red curve with circles
corresponds to a stellar population formed in an instantaneous
burst. The  symbols along each curve indicate the colors of the
stellar population at an age of $\log ( t / \mathrm{yr}) = 7, 8, 8.5,
8.7, 9, 9.5, 9.7, 10$.  The arrows illustrate the reddening vector for
a color excess, $E(B-V) = 0.2$, and assuming a starburst--like
extinction law (Calzetti \etal\ 2000).
\label{fig:colorcolor}}

\newcommand{\figelevencap}{Distribution of galaxies internal color
  dispersion as a function of total UV--optical color.  Data and
  symbols definitions are the same as in Figure~\ref{fig:colorcolor}.
  The curves  in the  correspond to the simple models discussed in the
  text.  The black curves show the expected colors of galaxies with an
  old central bulge and star-forming exponential disk with different
  star--formation histories (model 1, dashed lines; model 2, solid
  lines; see text).  The red curves illustrate the expected colors for
  objects with additional star-forming, \ion{H}{2} regions within
  older, passively evolving stellar populations in a spheroid and disk
  (model 3).  \label{fig:colorvxi_models} }

\newcommand{\figtwelvecap}{The measured internal UV--optical color
  dispersion from simulated galaxies as a function of the fraction of
stellar mass in the young stellar populations of the disk, \ie,
$\mathcal{M}_\mathrm{disk} / (\mathcal{M}_\mathrm{disk} +
\mathcal{M}_\mathrm{bulge})$.  The three curves correspond to galaxies
with the spheroidal components and disk components listed in
Table~\ref{table:models} (model 1, \textit{dashed} line;  model 2,
\textit{thick} line; model 3, \textit{thin} line) viewed at
$z=1$.\label{fig:sim_massratio}} 

\newcommand{\figthirteencap}{Distribution of concentration and asymmetry
  parameters of the $z\sim 1$ (\textit{Left} panel) and $z\sim 2.3$
  (\textit{Right} panel) galaxy samples.  Filled stars denote objects
  with $\xi(\wfv,\nicj) \ge 0.05$ and $\xi(\wfi,\nich) \ge 0.05$ in
  the $z\sim 1$ and $z\sim 2.3$ panels, respectively.  Objects with
  internal color dispersion values below this value are indicated by
  filled circles.  The dashed lines denote rough fiducial regions in
  the $C$--$A$ diagram that correspond to early (ellipticals,
  early--type spirals), middle (mid--type spirals), and late
  (irregulars and late--type spirals)  Hubble--type galaxies (see
  Conselice \etal\ 2000).
\label{fig:ca}}

\shorttitle{DIVERSITY IN THE STELLAR--POPULATIONS OF GALAXIES}
\shortauthors{PAPOVICH ET AL.}

\begin{document}

\slugcomment{Accepted for Publication in the Astrophysical Journal}
\ifsubmode
\title{ THE ASSEMBLY OF DIVERSITY IN THE MORPHOLOGIES AND
  STELLAR POPULATIONS OF HIGH--REDSHIFT GALAXIES\altaffilmark{1}} 
\else
\title{ THE ASSEMBLY OF DIVERSITY IN THE MORPHOLOGIES AND
  STELLAR POPULATIONS \\ OF HIGH--REDSHIFT GALAXIES\altaffilmark{1}} 
\fi

\author{\sc Casey Papovich}
\affil{Steward Observatory, University of Arizona, 933 North Cherry
  Avenue, Tucson, AZ 85721; papovich@as.arizona.edu}
\ifsubmode
\author{\sc Mark Dickinson\altaffilmark{2,3}}
\affil{National Optical Astronomy Observatory, 950 North Cherry
Avenue, Tucson, AZ 85719; med@noao.edu} 
\author{\sc Mauro Giavalisco }
\affil{Space Telescope Science Institute, 3700 San Martin Drive,
Baltimore, MD 21218; mauro@stsci.edu} 
\author{\sc Christopher
J. Conselice\altaffilmark{4}}\affil{Palomar Observatory, California
Institute of Technology,  MS 105--24, Pasadena, CA 91125;
cc@astro.caltech.edu}
\else
\author{\vspace{-12pt}
\sc Mark Dickinson\altaffilmark{2,3}} 
\affil{National Optical Astronomy Observatory, 950 North Cherry
Avenue, Tucson, AZ 85719; med@noao.edu} 
\author{\vspace{-12pt} \sc Mauro Giavalisco }
\affil{Space Telescope Science Institute, 3700 San Martin Drive,
Baltimore, MD 21218; mauro@stsci.edu} 
\author{\vspace{-12pt} \sc Christopher
J. Conselice\altaffilmark{4}}\affil{Palomar Observatory, California
Institute of Technology,  MS 105--24, Pasadena, CA 91125;
cc@astro.caltech.edu}
\fi
\and
\author{\sc Henry C. Ferguson\altaffilmark{3}}
\affil{Space Telescope Science Institute, 3700 San Martin Drive,
  Baltimore, MD 21218; ferguson@stsci.edu}
\altaffiltext{1}{Based on observations taken with the NASA/ESA Hubble
Space Telescope, which is operated by the Association of Universities
for Research in Astronomy, Inc.\ (AURA) under NASA contract
NAS5--26555.  These observations are associated with programs
GO/DD--6337 and GO--7817.}    
\altaffiltext{2}{also Space Telescope Science Institute}
\altaffiltext{3}{also Department of Physics and Astronomy; The Johns
  Hopkins University, 3400 N. Charles St., Baltimore, MD 21218}
\altaffiltext{4}{NSF Fellow}

%%%%%%%%%%%%%%%%%%%%%%%%%%%%%%%%%%%%%%%%%%%%%%%%%%%%%%%%%%%%%%%%%%%%%%

\begin{abstract}

We have used deep images from the \textit{Hubble Space Telescope} to
study the evolution in the morphologies, sizes, stellar--masses,
colors, and internal color dispersion of galaxies in the Hubble Deep
Field North. We selected two galaxy samples at $0.7 \leq z \leq 1.4$
and $1.9 \leq z \leq 3.0$ from a near--infrared, flux--limited catalog
with complete photometric and spectroscopic redshift information. At
$z\sim 1$ the majority of galaxies with $M(B) \leq -20.5$ have
rest--frame optical morphologies of early--to--mid-type
Hubble--sequence galaxies, and many galaxies show strong
transformations between their rest--frame UV and optical
morphologies. In stark contrast, galaxies at $z\sim 2.3$ all have
compact and irregular rest--frame optical morphologies with little
difference between their rest--frame UV--optical morphologies, and with no
clearly evident Hubble--sequence candidates.   The mean galaxy size
increases from $z\sim 2.3$ to 1 by roughly 40\%, and the number
density of galaxies larger than 3~kpc increases by a factor of
$\approx 7$.  The size--luminosity distribution of $z\sim 1$ galaxies
is broadly consistent with that in the local universe, allowing for
passive evolution.   However, we argue that galaxies at $z\sim 2.3$
are not the fully formed progenitors of present--day galaxies, and
they must continue to grow in both their sizes and stellar masses.  We
have quantified the differences in morphology by measuring the
galaxies' internal UV--optical color dispersion, which constrains the
amount of current star--formation relative to older stellar
populations.   The mean and scatter in the galaxies' UV--optical total
colors and internal color dispersion increase from $z\sim 2.3$ to 1.
At $z\sim 1$ many galaxies with large internal color dispersion are
spirals, with a few irregular and interacting systems.   Few $z\sim
2.3$ galaxies have high internal color dispersion, and we infer that
those that do are also actively merging.  Using simple models, we
interpret the change in the total color and internal color dispersion
as evidence for the presence of older and more diverse stellar
populations at $z\sim 1$ that are not generally present at $z\gsim 2$.
The $z\sim 2.3$ galaxies do not increase their stellar diversity  as
rapidly as they could given basic timescale arguments and simple
models.  We conclude that the star--formation histories of galaxies at
$z\gsim 2$ are dominated by discrete, recurrent starbursts, which
quickly homogenize the galaxies' stellar content and are possibly
associated with mergers.   The increase in the diversification of
stellar populations by $z\sim 1$ implies that merger--induced
starbursts occur less frequently than at higher redshifts, and more
quiescent modes of star--formation become the dominant mechanism. This
transition in the mode of star formation coincides with the emergence
of Hubble--sequence galaxies, which seems to occur around $z\sim 1.4$.

\end{abstract}
 
 
%%%%%%%%%%%%%%%%%%%%%%%%%%%%%%%%%%%%%%%%%%%%%%%%%%%%%%%%%%%%%%%%%%%%%%

\section{Introduction}

Galaxy morphologies stem from complex formation and evolutionary
records.  The fact that present--day galaxies span such a great range
of morphological type and that galaxy morphology correlates with
optical luminosity, color, size, and mass \citep[\eg,][]{rob94}
indicates that galaxy appearances depend strongly on their assembly
histories.    The quantitative study of the morphological properties
of galaxies at high redshifts has advanced rapidly due primarily to
the high--angular resolution  of the \textit{Hubble Space Telescope}
(\hst).  The \hst\ instruments provide imaging with
angular resolution of $\sim 0\farcs 1$, which corresponds to physical
scales of less than 1~kpc for cosmologically distant galaxies.
% (see, \eg, Papovich \etal\ 2003).

Many surveys using \hst\ imaging have broadly shown that while
the local Hubble Sequence seems to exist for $z\lsim 1-1.5$
\citep[\eg,][]{vdb96,vdb01,gri96,bri98,lil98,sim99,dok00,sta04,rav04},
the morphological mix of galaxies at these redshifts shows an increase
in the abundance of irregular and peculiar galaxies, particularly at
fainter magnitudes \citep[\eg][]{dri95a,ell97,bri98,bri00}.  However,
the formation and evolution of such systems is unclear.  Several
studies have concluded that a large fraction of giant spiral and other
early--type galaxies were already in place by moderate redshifts
\citep{vdb96,abr96,bri98,lil98,im99,sim99,im02,mou04,rav04,sta04} and
that the  characteristic galaxy size has changed little since $z\sim 1$
\citep{lil98}.  Moreover, little discernible evolution is evident in
such systems even when uncertainties in the morphological
classification and other selection effects are included
\citep{mar98,sim99,rav04}, which reinforces the conclusion that some
fraction of the Hubble--sequence galaxy population had assembled by
$z\sim 1$.

How these large, symmetric Hubble--sequence galaxies form and assemble
their diverse stellar populations remains enigmatic.  Logically, they
descend from higher--redshift progenitors, but it is difficult to link
the populations.   The Hubble sequence seems to no longer apply at
$z\gsim 1.5-2$ (van den Bergh \etal\ 2001; van den Bergh 2002).  The
observed morphologies of galaxies at $z\gsim 2$ are generally highly
irregular or centrally compact (Giavalisco, Steidel, \& Macchetto
1996; Lowenthal \etal\ 1997; Dickinson 2000; Daddi \etal\ 2004), but some large
disk--galaxy candidates exist
\citep[\eg,][]{lab03,con04}.  The sizes, spatial densities, and
stellar masses of galaxies at these redshifts are comparable to those
of present--day bulges and lower--redshift galactic spheroidal
components \citep{gia96,gia98,low97,gia01,pap01,sha01}, and thus it is
tempting to equate them with a one--to--one correspondence.  However,
the space density of galaxies at $z\sim 2-3$ may be inconsistent with
the fraction of bulge--dominated galaxies observed at moderate
redshift ($z\sim 1$) in the \hdf\ \citep{mar98}. Moreover, although
the stellar--mass density at $z\sim 1$ is dominated by red, evolved
galaxies \citep{bel04}, the majority of the galaxy stellar mass at
$z\gsim 2$ exists in blue, irregular galaxies
\citep[\eg,][]{dic03,con04b} with some contribution from
dust--enshrouded, massive starbursts \citep{for04,van04}. 

The physical processes that govern how
galaxies observed at $z\gsim 2$ evolve into galaxies at more modest
redshifts ($z\sim 1$) are poorly understood.  This
is partly due to difficulties in comparing the properties of
galaxies observed at different rest--frame wavelengths and with
different telescope and detector combinations.  It is entirely
expected that all observable galaxy properties (\eg, luminosities,
colors, masses, morphologies, densities) evolve substantially over
galaxies' lifetimes, and the subsequent observational biases further
frustrate studies of galaxy evolution when examining galaxy
populations based on some selection criteria.

One complication in the studies of galaxy morphology results from the
fact that a fixed bandpass samples lower rest--frame wavelengths with
increasing redshift.  For example, observations at optical wavelengths
probe only the rest--frame UV portions of the spectral energy
distributions (SEDs) of galaxies at $z\gsim 1$, and there is little
relation between rest-frame UV and optical properties of galaxies.
Local galaxies have been shown to exhibit strong transformations
between their morphologies at rest--frame UV and optical wavelengths
\citep[see, \eg][]{kuc00,kuc01,mar01,win02,pap03}.  The
UV emission from galaxies is generally dominated by the light from
young, massive stars, and as a result a galaxy's morphology at UV
wavelengths can appear as a later--type system compared to that
derived at optical wavelengths.   The composition (ratio of the number
of young to old stars) and configuration (heterogeneity in the
distribution of young and old stars) of the stellar populations within
elliptical, spiral, and irregular galaxies varies widely.  Galaxy
morphology can vary strongly from optical to UV wavelengths, and comparing
the morphological properties of galaxies at UV wavelengths
likely provides an incomplete evolutionary picture. 

To compare the appearances between high--redshift and local galaxies,
one requires observations at near--infrared (NIR) wavelengths, which
probe rest--frame optical wavelengths for $z\gsim 1$ galaxies. Several
recent studies have focused on the morphological properties of
high--redshift galaxies in their rest--frame optical using \hst\ high
angular resolution NICMOS data \citep[\eg][]{tep98, dic00b, cor01,
mot02, con03, sta04} and ground--based data \citep[\eg][]{lab03,tru04}, and
this was a strong motivation behind our NICMOS imaging survey of the
\hdf.  This dataset combined with the \hst/WFPC2 data (Williams et
al.\ 1996) provides high angular resolution images ($\sim 0\farcs1$)
in six bandpasses spanning $0.3-1.6$~\micron, which permits
comparisons between galaxy morphologies at rest--frame UV--optical
wavelengths as a function of redshift for $0 < z \lsim 3$.

In this paper, we investigate the distribution of the morphologies,
sizes, colors, and internal colors of high--redshift galaxies ($z\sim
0.7-3.0$) in the \hdf\ using data available from the \hst\ datasets.
In a previous paper \citep[hereafter P03]{pap03}, we presented a
diagnostic to measure a galaxy's internal color dispersion, which
quantifies the differences in galaxy morphology as a function of
observed wavelength (i.e., it quantifies the morphological
$K$--correction). Here, we apply this statistic to the rest--frame
UV--optical colors of a sample of galaxies selected from the deep
\hst/NICMOS observations of the \hdf.   In \S~2, we present a summary
of the data and reduction procedures for the \hdf\ data and define
galaxy samples used here.  In \S~3, we describe the galaxy
morphologies, and quantify their sizes, colors, and internal color
dispersion.  In \S~4, we discuss differences in these quantities
versus redshift, and compare these with simple scenarios of galaxy
evolution.  In \S~5, we summarize our results.  Throughout this work,
we assume a cosmology with $\Omega_M = 0.3$, $\Lambda = 0.7$, and a
Hubble parameter of $h=0.7$ (where the present--day Hubble constant is
$H_0 = 100\,h$~\kms\ Mpc$^{-1}$).  All magnitudes have units in the AB
system \citep{oke96}, $m_\mathrm{AB} = 31.4 - 2.5\log( \langle f_\nu
\rangle /\mathrm{1\, nJy})$.  We also refer to the four WFPC2 (F300W,
F450W, F606W, F814W)  and the two NICMOS (F110W, F160W) bandpasses
used on the \hdf\ as \wfu, \wfb, \wfv, \wfi, \nicj, \nich,
respectively.

%%%%%%%%%%%%%%%%%%%%%%%%%%%%%%%%%%%%%%%%%%%%%%%%%%%%%%%%%%%%%%%%%%%%%%

\section{The Data and Sample Definition}\label{section:hdfsample}

The complete \hdf\ was observed with NICMOS for GO proposal 7817
during the UT 1998 June NICMOS refocus campaign.\footnote{During the
1998 refocus campaign (\hst\ Cycle 7N) the \hst\ secondary was moved
to the optimal focus for the NICMOS camera 3.  This has not been done
in cycles following the installation of the NICMOS Cooling System
(NCS) during the \hst\ servicing mission 3B.  As a result, the image
quality of the HDF--N mosaic is significantly improved over what is
currently available with NICMOS post--NCS.  For example, the PSF FWHM
of the HDF-N mosaic, $\approx 0\farcs22$ in \nich, is approximately
60\% smaller than that of similar observations of the Hubble
Ultra--Deep Field (R.~I.\ Thompson, private communication; see also
Roye \etal\ 2003). }  The details of the observations, data reduction,
and object cataloging will be presented elsewhere (see also Dickinson
\etal\ 2000; Papovich \etal\ 2001; Dickinson \etal\ 2003; Stanford
\etal\ 2004).  Briefly, we observed the complete \hst/WFPC2 field of
the \hdf\ with eight  sub--fields with the NICMOS Camera 3.  Each
sub--field was imaged three times through both the F110W and F160W
filters, and each image was dithered into three separate exposures
(resulting in nine independent, dithered positions in each
sub--field), with a net exposure time of 12600s per filter.  The data
were processed using standard STScI pipeline routines and custom
software, and combined into a single mosaic that was registered to the
\hdf\ WFPC2 images (Williams \etal\ 1996) using the ``drizzling''
method of Fructher \& Hook (2002).  The WFPC2 and \nicj\ images were
then convolved to match the \nich\ point--spread function
($\mathrm{FWHM} \approx 0\farcs22$) in order to ensure that photometry
was measured through similar apertures in each image.  Our tests on
point sources in the PSF--matched images indicate that photometry
between bands is accurate to within 5\% for aperture radii greater
than $0\farcs1$.  Object catalogs were constructed using SExtractor
\citep{ber96} by detecting objects in a combined $\nicj + \nich$ image
and measuring fluxes through matched apertures in all individual
bands.  As in Stanford \etal\ (2004), we will refer to the complete
galaxy catalog as the HDF NICMOS Mosaic, or HNM, catalog.  Relative
photometry from the \hst\ WFPC2 and NICMOS data was measured through
matched elliptical apertures defined using the radial moments of each
galaxies' F110W + F160W light profile (\ie, the SExtractor MAG\_AUTO
flux measurements). The difference between the flux measured in the
total (MAG\_AUTO) apertures and the isophotal (MAG\_ISO) apertures has
a median value of $\approx 6$\% for all of the galaxies with $\nich
\leq 25.5$.

In this work, we define two galaxy samples from the \hdf\ data 
selected on the basis of redshift and limiting magnitude:  a
moderate--redshift sample, $0.7 \leq z \leq 1.4$ (with a mean redshift
$z\simeq 1$); and a high--redshift sample, $1.9 \leq z \leq 3.0$ (with
a mean redshift $z\simeq 2.3$).  We have used the most complete
collection of HDF-N spectroscopic redshifts known (see Dickinson
\etal\ 2003; and references therein). For the remaining galaxies, we
have used the photometric--redshift catalog of \citet{bud00} for the
\hdf\ based on the combined WFPC2 and NICMOS data.  

Because we analyze the two-dimensional galaxy appearances, we have
selected galaxies on the basis of signal--to--noise ratio (S/N) down
to a limiting isophotal level, which is more appropriate than a pure,
flux--limited sample as galaxies of larger sizes have lower S/N for a
fixed flux density.  We also select the objects in the bandpass that
approximately corresponds to the rest-frame $B$--band emission.  We
select all galaxies at $z\sim 1$ with $\mathrm{S/N}(\nicj) \geq 20$
within a limiting isophote above $\mu(\nicj) \leq
26.5$~mag~arcsec$^{-2}$.  Similarly, we select galaxies at $z\sim 2.3$
with $\mathrm{S/N}(\nich) \geq 20$ within a limiting isophote above
$\mu(\nich) \leq 26.5$~mag~arcsec$^{-2}$.  Our tests have shown that
these limits on object S/N within an isophotal aperture are reasonable
for measuring galaxy morphological and structural parameters (see P03
and discussion below).  These limits in S/N correspond to approximate
magnitude limits measured within the elliptical apertures (\ie, the
SExtractor MAG\_AUTO) of $\nicj \sim \nich \sim 25.5$.  The final
galaxy samples have 113 objects at $z\sim 1$ and 53 objects at $z\sim
2.3$ that satisfy these selection criteria in S/N and redshift. Of the
samples, 69\% of the $z\sim 1$ galaxies and 62\% of the $z\sim 2.3$
galaxies have spectroscopic redshifts.   In the following analysis, we
also compare the properties of the galaxy samples down to a fixed
absolute magnitude, $M(B) \leq -20.0$, where both samples are mostly
complete (see Dickinson et al.\ 2003).  This luminous sub--sample
contains 41 objects at $z\sim 1$ and 46 objects at $z\sim 2.3$.

The redshift ranges used here were chosen to take advantage of two
attributes of the data.  The apparent galaxy size remains roughly
constant for $0.7 \lsim z \lsim 3.0$ because the angular--diameter
distance --- the ratio of an object's physical size to its angular
size --- changes little ($< 18$\%) for the default cosmology: from
7.1~kpc arcsec$^{-1}$ at $z =  0.7$ to 8.5~kpc arcsec$^{-1}$ at $z\sim
1.6$ (this is generally true for any plausible cosmology).   Thus, the
\hst\ images provide roughly constant angular resolution of $\sim
1.5$~kpc for all the galaxies in this dataset nearly independent of
redshift.  Note, however, that the co-moving volume contained in the
redshift slice $1.9 \leq z \leq 3.0$ is 2.3 times than contained by
$0.7 \leq z \leq 1.4$.   

The other advantage comes from the fact that at these redshifts
combinations of the WFPC2 and NICMOS bandpasses probe approximately
the same wavelengths in the rest--frame, which allows us to minimize
morphological $K$--corrections between the two samples.   We broadly
define two rest--frame wavelength regions: the mid--UV ($\sim 2400 -
3800$~\AA), and optical ($\sim 4000-6000$~\AA).  For the $0.7 \le z
\le 1.4$ galaxy sample, the mid-UV--, and optical--wavelength regions
correspond to the \wfv\ and \nicj\ bandpasses, whereas at $1.9 \le z
\le 3.0$ these rest--frame wavelength regions correspond to the \wfi\
and \nich\ bandpasses.  We note that using a single bandpass to track
a specific rest--frame wavelength for a wide redshift interval does
induce overlap in wavelength and redshift coverage due to the width
and central wavelengths of the of the \hst\ filter curves.  For
example, due to its large width, the centroid of the \nicj\ filter
samples $\lambda_0 \sim 4600-6500$~\AA\ for $z\sim 0.7-1.4$, and
$\lambda_0 \sim 2750-3800$~\AA\ at $z\sim 1.9-3.0$.  One could devise
a method to interpolate the observed bandpasses to acquire the flux at
a common wavelength.  However, this would in principle hinder the
analysis here, as one of the goals of this work is to contrast and
compare the individual color features of galaxies for common
rest--frame wavelengths.  Interpolation between bands would only
suppress any intrinsic signal.  Therefore, we have opted to use the
fixed bandpasses as approximate surrogates for the rest--frame colors.
Furthermore, our tests on the UV--optical colors of local galaxies
(P03) has shown that the internal color dispersion is most sensitive
to changes due to the the heterogeneity of the stars that contribute
to the flux redward and blueward of the Balmer and 4000\AA\ breaks.
For the galaxies considered here, these breaks are always confined
between the chosen bandpasses.

%%%%%%%%%%%%%%%%%%%%%%%%%%%%%%%%%%%%%%%%%%%%%%%%%%%%%%%%%%%%%%%%%%%%%%

\pagebreak

\section{Galaxy Morphological and Structural Properties}

\ifsubmode
\begin{figure}
\plotone{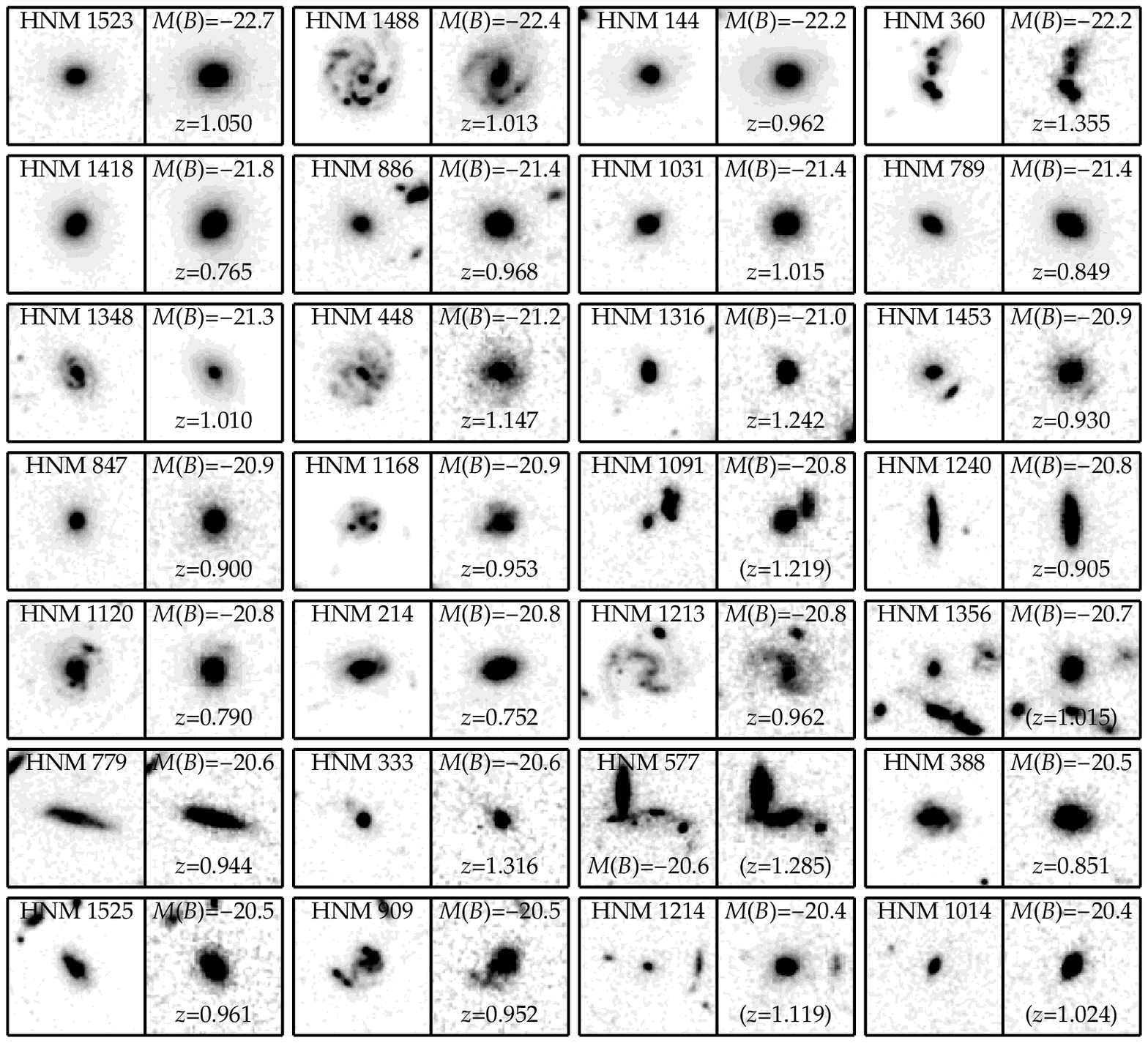}
\caption\figonecap
\end{figure}
\begin{figure}
\plotone{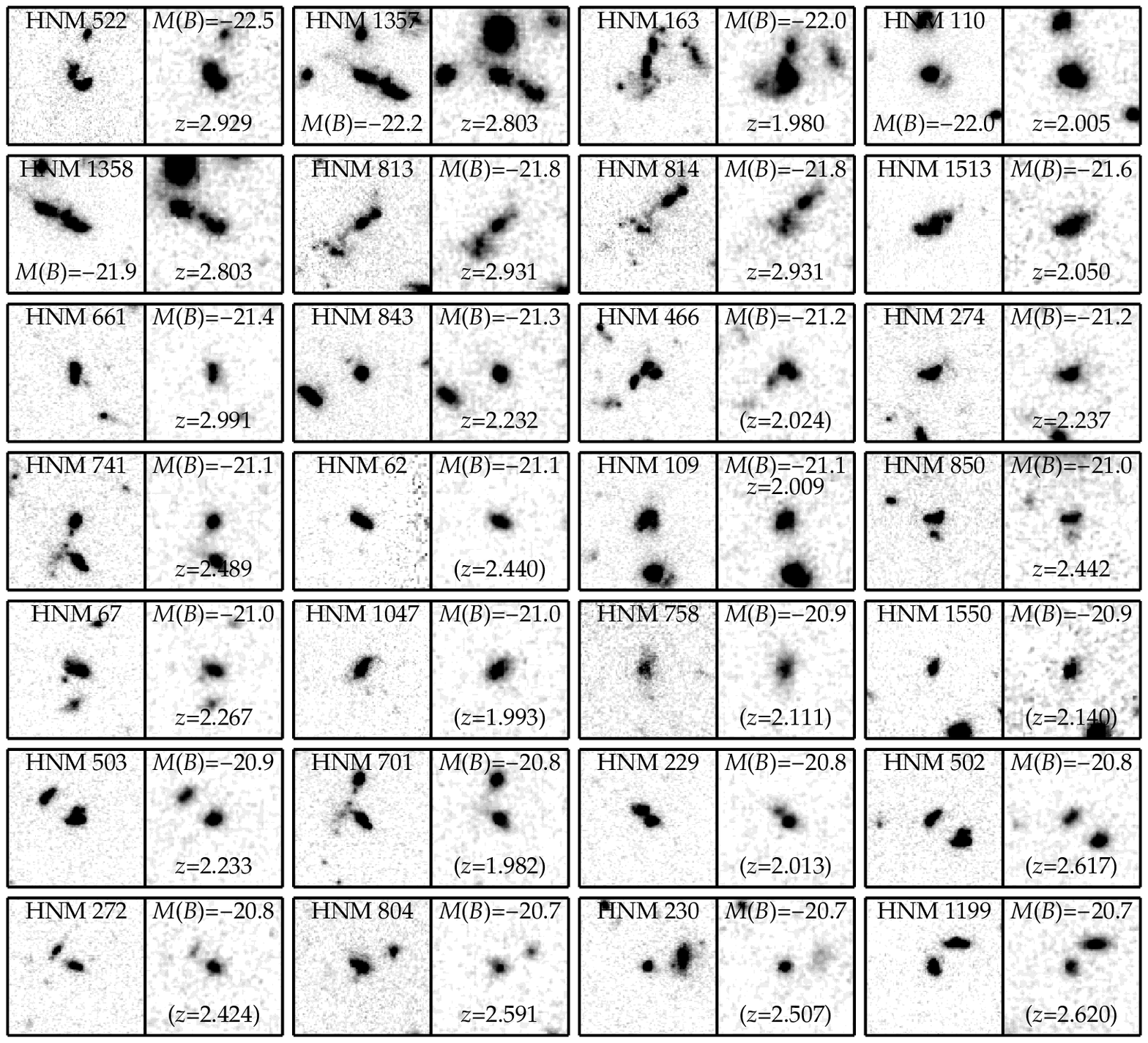}
\caption\figtwocap
\end{figure}
\fi

Figures~\ref{fig:montage_lowz1} and \ref{fig:montage_hiz1} show images
of the 28 most luminous galaxies (in the rest--frame $B$--band) in
each of the $z\sim 1$ and $z\sim 2.3$ samples.  These figures
illustrate the morphologies of the brightest galaxy members of each
sample.  Although the galaxies in each sample span comparable
rest--frame luminosities, $-20.5 \gsim M(B) \gsim -22$, their
morphologies are distinctly different.  Many of
the galaxies at $z\sim 1$ have regular and symmetric morphologies,
particularly in the rest-frame optical images, and are similar to
``normal'' Hubble--sequence galaxies.  Our by--eye classification of
the $\nicj$ morphologies finds that of the brightest $z\sim 1$
galaxies $\simeq 43$\% (12/28) have elliptical, lenticular, or
spheroidal morphologies (HNM 144, 789, 847, 886, 1014, 1091, 1031, 1214,
1356, 1418, 1453, 1523), $\simeq 43$\% (12/28) have spiral or
disk morphologies (HNM 214, 388, 448, 577, 779, 1120, 1168, 1213,
1240, 1348, 1488, 1525), and only $\simeq 14$\% (4/28) have irregular or
compact morphologies (HNM 333, 360, 909, 1316).  Many of the objects
show strong differences between their morphologies at UV and
optical wavelengths (i.e., strong morphological $K$--corrections),
which is particularly apparent in those galaxies with spiral and/or
disk morphologies.   

\ifsubmode
\else
\begin{figure*}[th]
\plotone{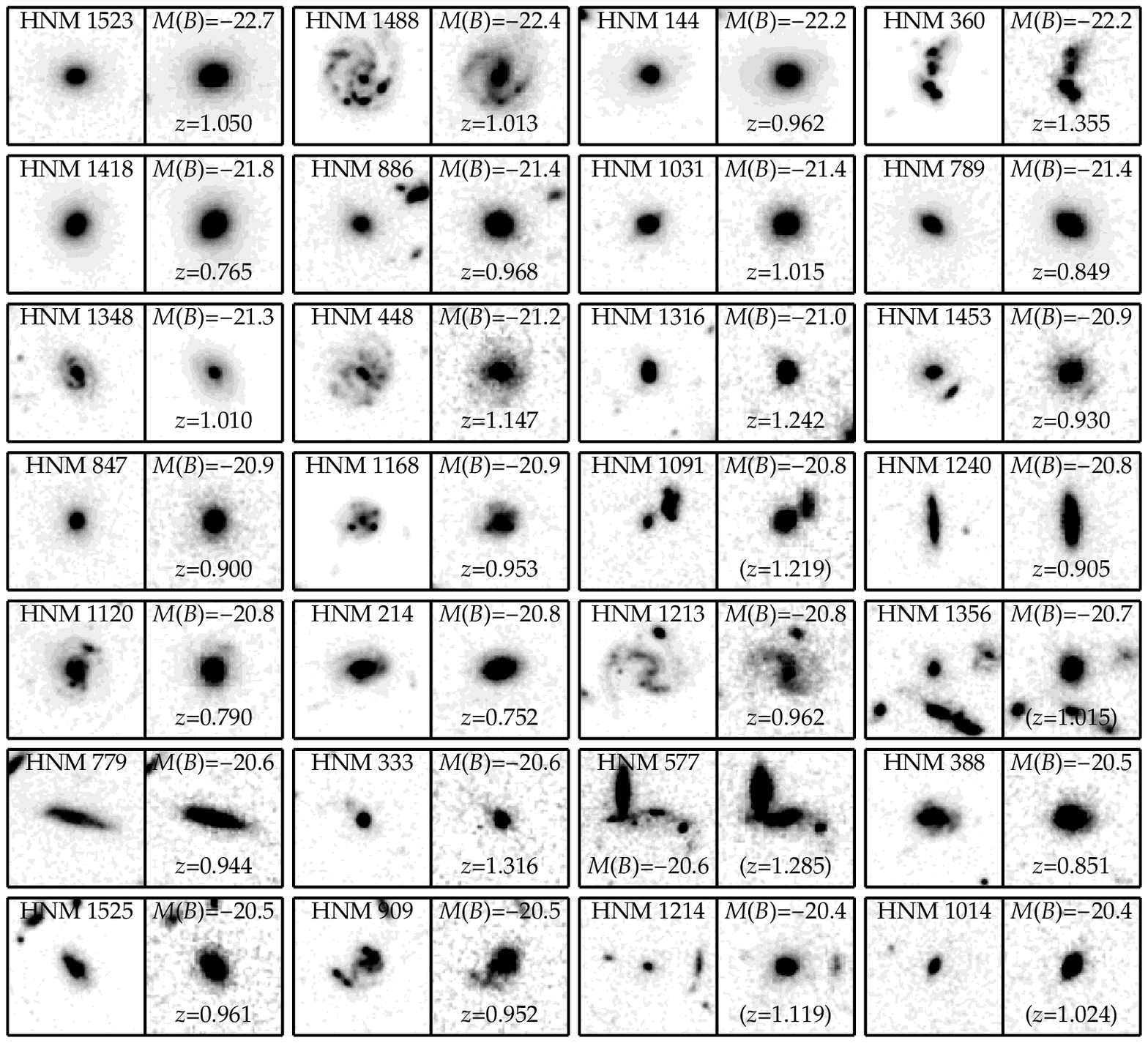}
\caption\figonecap
\end{figure*}
\begin{figure*}[th]
\plotone{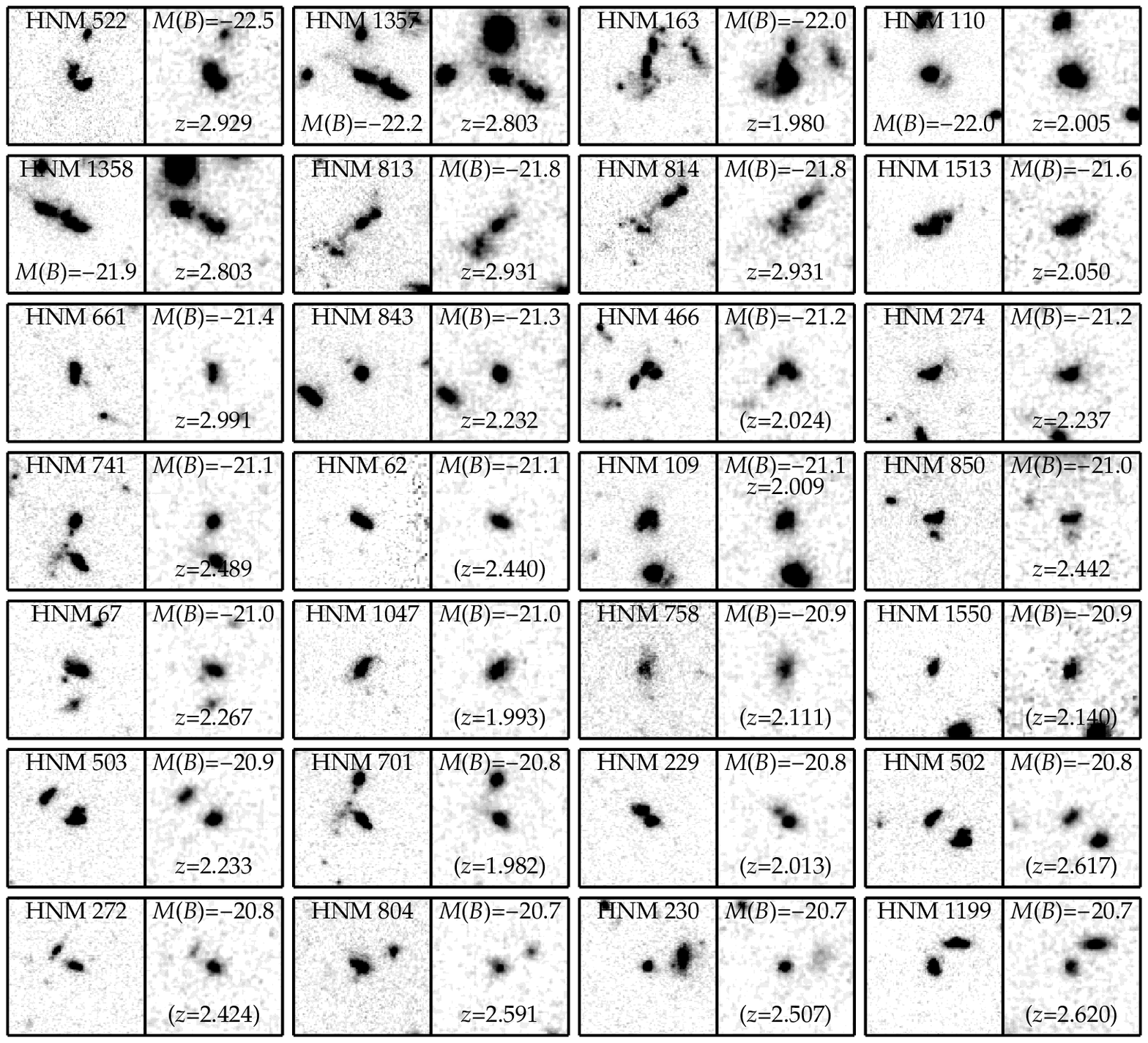}
\caption\figtwocap
\end{figure*}
\fi

In contrast, the brightest galaxies at $z\sim 2.3$ have either
irregular (HNM 552, 1357, 163, 1358, 813, 814, 1513, 109, 850, 758,
701, 502, 804) or compact morphologies (HNM 110 661, 843, 466, 274,
741, 62, 67, 1047, 1550, 503, 229, 272, 230, 1199). Unlike the
galaxies at $z\sim 1$, there are no galaxies with disk--like or
spiral--galaxy morphologies at $z\sim 2.3$: no galaxies show evidence
for spiral arms or for being edge--on disk systems.  This is broadly
similar to the morphological descriptions of $z\sim 3$ galaxies based
solely on rest--frame UV observations \citep{gia96,low97}.  The
majority of the optically bright galaxies at $z\sim 2.3$ have similar
morphologies in the \wfv-- (rest--frame 1500~\AA) and \nich--
(rest--frame 4400~\AA), which is even more significant because this
wavelength baseline is longer than that shown for the $z\sim 1$ sample
in Figure~\ref{fig:montage_lowz1}.  The lack of significant
morphological $K$--correction combined with the fact that these
galaxies have blue UV--optical colors (Dickinson
\etal\ 2003) implies that recent star formation dominates the
UV--optical morphologies, which we discuss in more detail below (see
also the discussion in P03). 

\subsection{The Galaxy Size--Luminosity Relationship
\label{section:radlum}}

The panels in figures~\ref{fig:montage_lowz1} and
\ref{fig:montage_hiz1} show the images of the $z\sim 1$ and 2.3
galaxies at the same physical scale (i.e., each panel is approximately 
40~kpc $\times$~40~kpc).  Qualitatively, the typical galaxy size
appears to increase from $z\sim 2.3$ to $z\sim 1$, with the most
dramatic evolution in the rest--frame $B$--band images.  In this
section, we quantify this observed evolution in terms of the galaxy
size--luminosity distribution.

We derived approximate absolute $B$--band magnitudes by  interpolating
the \wfv \wfi \nicj \nich\ photometry to measure the apparent magnitude
at rest--frame 4400~\AA\ and applying the distance moduli with the
default cosmology for the observed redshift.  We quantify galaxy sizes
with the half--light radius $r_{1/2}$, which is defined as the radius
of the aperture that encompasses 50\% of the object's total flux as
derived by the SExtractor software.  Our tests using simulated data
with noise properties similar to that of the \hdf\ indicates that the
SExtractor measurements of the half--light radius are accurate ($<
10$\% rms) for $\mathrm{S/N} > 20$, which is satisfied for the galaxy
samples used here.

Figure~\ref{fig:radlum} shows the distribution of the derived
half--light radii for the \hdf\ galaxies as a function of their
absolute magnitudes at rest-frame 4400~\AA.   The upper envelope of
rest--frame $B$--band luminosities is roughly the same at $z\sim
1$ and $z\sim 2.3$ \citep[see also the discussion above;][]{dic03}.
These luminosities are comparable to those of the largest present--day
early--type spirals and ellipticals
\citep{dev91,ben92}. Many galaxies in both samples have luminosities
well in excess of the present--day characteristic
``$L^\ast$'',\footnote{Norberg et al.\ 2002 find a present--day
characteristic luminosity of $M(B)^\ast = -20.5$ [AB], after slight
adjustments in cosmology and bandpass.} although the majority of
galaxies extend to fainter luminosities.

\ifsubmode
\begin{figure}[thp]
\epsscale{0.8}
\plotone{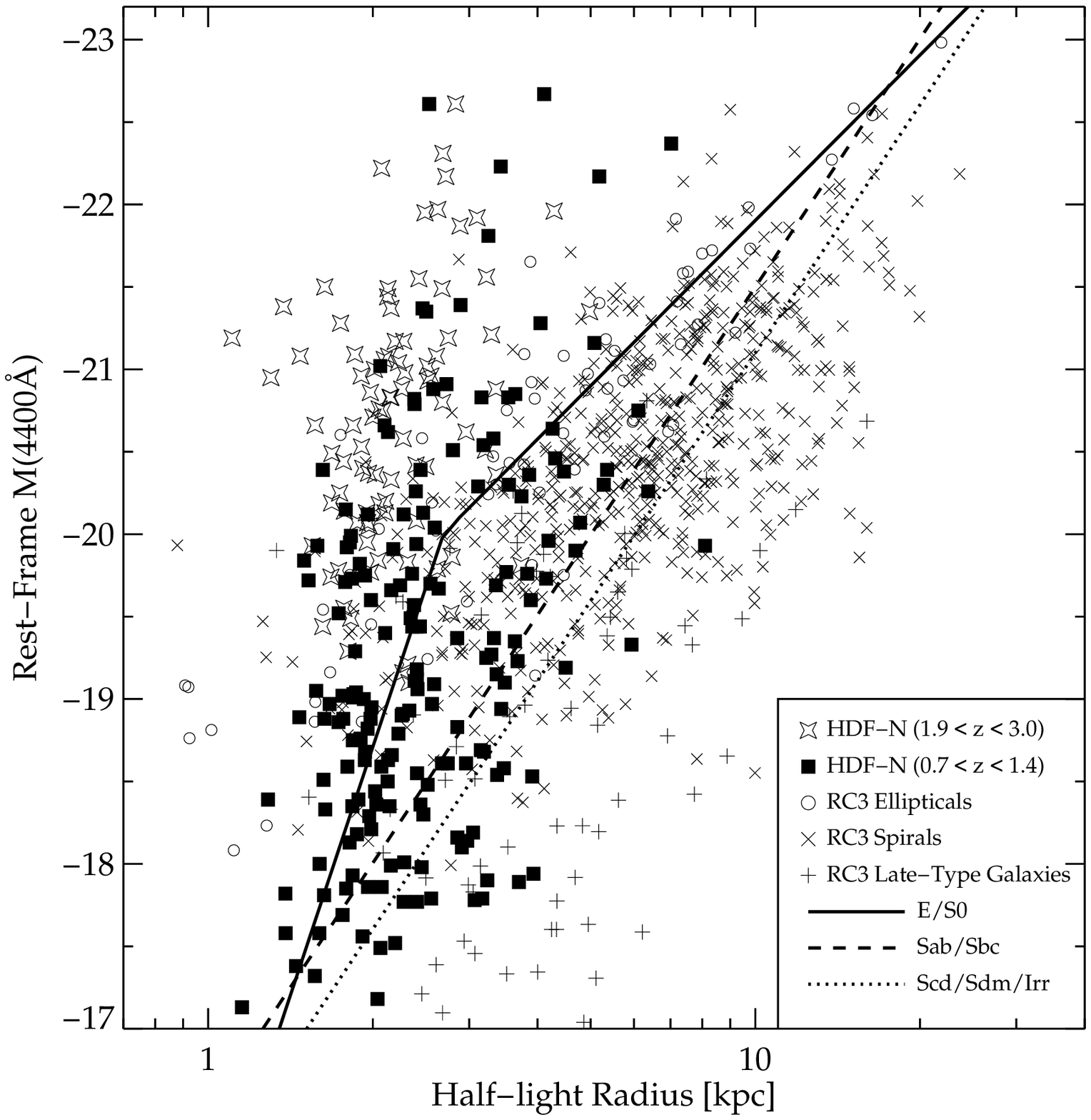}
\caption{\figthreecap}
\epsscale{1.0}
\end{figure}
\else
\begin{figure*}[th]
\epsscale{1.2}
\plotone{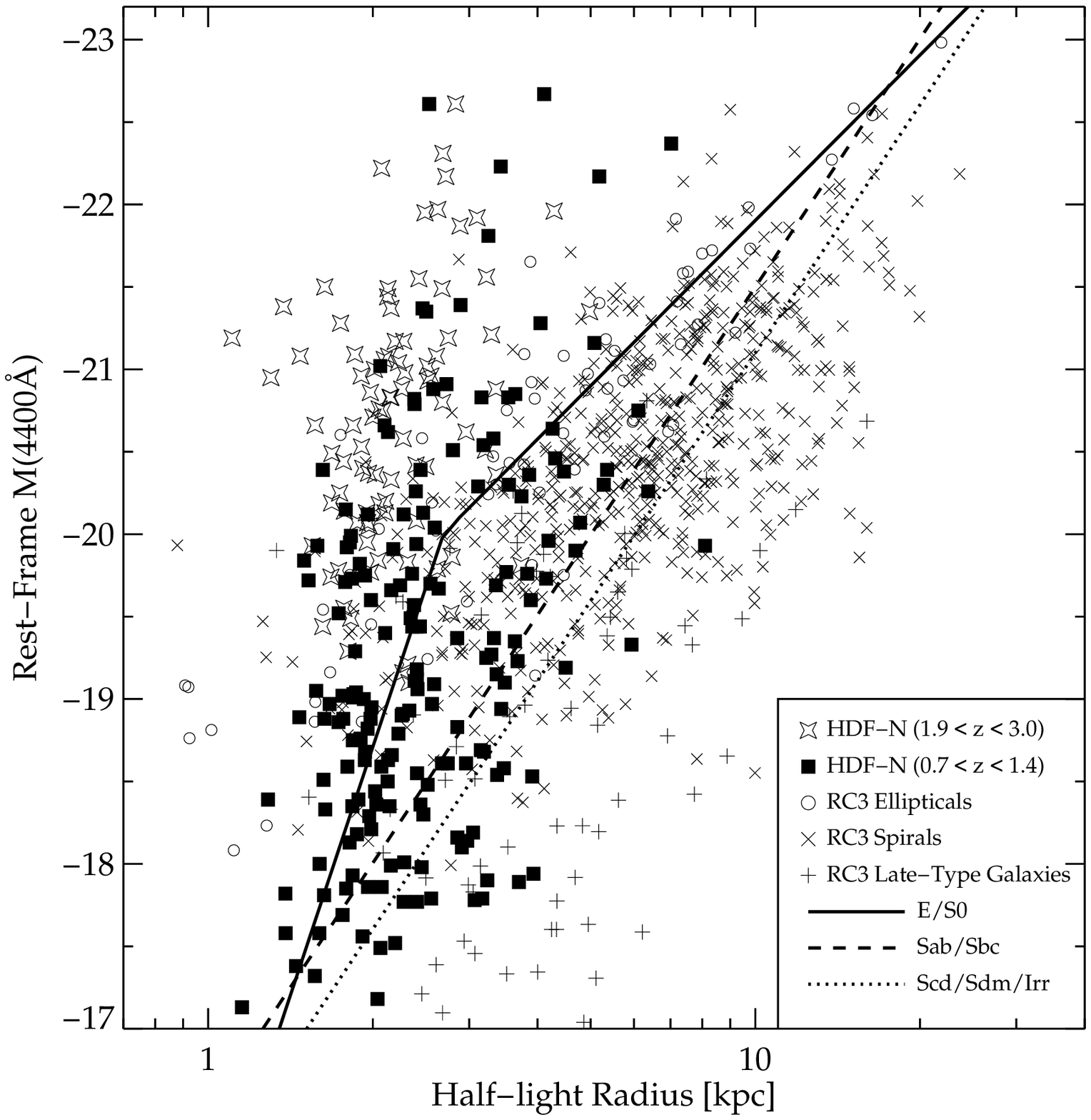}
\caption{\figthreecap}
\epsscale{1.0}
\end{figure*}
\fi

The average galaxy half--light radius grows with time from $z\sim 2.3$
to 1.  The $z\sim 2.3$ galaxies have relatively small intrinsic
sizes, $1 \lsim r_{1/2} \lsim 4$~kpc, with a mean value $\langle
r_{1/2}\rangle \simeq 2.3\pm 0.3$~kpc for $M(B) \leq -20.0$, where the
error represents the uncertainty on the mean.  In contrast, the measured
radii of the $z\sim 1$ galaxies are larger for comparable absolute
magnitudes.  The mean half--light radius for galaxies with $M(B)
\leq -20.0$ is $\langle r_{1/2}\rangle \simeq 3.2\pm 0.3$~kpc, which
is roughly 40\% higher than the average size of the $z\sim 2.3$
galaxies.  We find that based on the two--dimensional
Kolmogoro--Smirnov (KS) test \citep{pea83,fas87} the $z\sim 2.3$ and
$z\sim 1$ galaxy samples, for $M(B)\leq M^\ast(B)$, trace different
distributions at the $\simeq 99.96$\% confidence level ($\simeq
3.6\sigma$).  The luminosity--size distribution is qualitatively
similar to results reported by \citet{tru04} for the \hdfs, but a
quantitative comparison is beyond the scope of this work given the
differences in rest--frame bandpass and sample redshifts.

In addition, there is a dramatic increase in the number density of
large galaxies from $z\sim 2.3$ to 1.  Considering only galaxies with
$r_{1/2} > 3$~kpc and $M(B) \leq -20.0$,  we find 24 galaxies at
$z\sim 1$ and 8 at $z\sim 2.3$.  Because the co-moving volume of the
$z\sim 2.3$ redshift range is larger by a factor of 2.3, there is an
increase in the number of large galaxies by a factor of $\approx 7$
between these redshifts.  Increasing the size limit to $> 4$~kpc makes
the evolution even more dramatic: there are 13 such galaxies at $z\sim
1$ compared with only 2 at $z\sim 2.3$, but this comparison suffers
from obvious small--number statistics.   Note that our magnitude
limits are also conservative.  If we include fainter sources (e.g., to
$M(B) \leq -19$), then the observed change is even greater: the number
density of sources with $r_{1/2} > 3$~kpc increases by a factor of
$\approx 11$ from $z\sim 2.3$ to $z\sim 1$, but at the fainter
magnitudes the $z\sim 2.3$ sample probably suffers from some
incompleteness.

Note that we made no attempt to remove the contribution of the image
PSF from the measured half--light radii.  The NIC3 F160W PSF has a
small intrinsic half--light radius, $r_{1/2,\mathrm{PSF}}\simeq
0.15\arcsec$, which for these redshifts and default cosmology
corresponds to $r_{1/2}\simeq 1.2-1.4$~kpc.  The majority of objects
have $r_{1/2} \gsim 2$~kpc, to which the PSF contributes $< 20$\%.
However, the PSF dominates the extrinsic sizes for objects with
$r_{1/2} \lsim r_{1/2,\mathrm{PSF}}$.   Because the PSF can only
increase the measured half--light radii, its effects will tend to
\textit{decrease} the amount of intrinsic size evolution between
$z\sim 2.3$ and $z\sim 1$.  Therefore, if anything our result
underestimates the intrinsic evolution. 

There is a deficit of \hdf\ galaxies at $0.7 \le z \le 1.4$ in the
\hdf\ with $r_{1/2} \gsim 8$~kpc.  Although this result could be
interesting for the evolution and growth of galaxies for $z\sim 0-1$,
it likely stems from the fact that the \hdf\ data probes a small
volume.   The present--day bivariate (radius--, luminosity--)
distribution function \citep{dej00} predicts that there should be
$\simeq 1.6\pm 1.3$ objects with $r_{1/2} \geq 8$~kpc in the
co--moving volume of the \hdf\ for $0.7 \leq z \leq 1.4$.   Therefore,
the lack of galaxies with large radial sizes in the \hdf\ at these
redshifts is not strongly inconsistent with the present--day galaxy
population: the volume is simply too small to expect many giant
galaxies.  This conclusion is reinforced by the presence of large
galaxies in the disk--galaxy samples of \citet{lil98},
\citet{sim99}, and \citet{rav04} to $z\sim 0.7-1$, which are drawn
from redshift surveys that cover larger solid angles.  However, if the
local galaxy size--luminosity distribution were applicable at higher
redshifts,  then one would expect to find $\simeq 5\pm 2$ galaxies with $1.9
\leq z \leq 3.0$ and $r_{1/2} \geq 8$~kpc, and none are observed.
Note that although \citet{lab03} report several galaxies with $r_{1/2}
\simeq 4-6$~kpc at $z\sim 2-3$ in the HDF--S, there are no candidates
at these redshifts with larger radii.

The (rest--frame) $B$--band half--light radii of the
$z\sim 2.3$ galaxies are consistent with those reported from $z\sim 3$
galaxies from observations solely in the rest--frame UV
\citep{gia96,low97}.  This implies that UV--selected galaxies at these
redshifts have high surface brightness at both rest--frame UV and
optical wavelengths, and are not generally UV--bright star--forming
regions within otherwise normal, larger host galaxies.  

\subsection{Effects of Surface Brightness on Galaxy
Sizes}\label{section:sb_sizes} 

We have considered the effects of surface--brightness dimming, which
rapidly increase with redshift (the bolometric surface brightness
scales as $(1+z)^{-4}$).   We simulated images of the luminous ($M(B)
\leq -20.0$) galaxies from the $z\sim 1$ sample as they would appear
at $z= 2.7$ (i.e., somewhat higher than the mean redshift $z  = 2.3$
in order to achieve more conservative limits).  The general details of
this process are presented in P03, but briefly, we resampled the
NICMOS \nicj\ images of the $z\sim 1$ galaxies to the physical pixel
scale for $z=2.7$.  We then appropriately reduced the pixel intensity
to account for surface--brightness dimming and inserted them into
blank regions in the NICMOS \nich\ images.  Because the relative
resolution does not change substantially from $z = 0.7$ to 2.7, we
have made the approximation that the image PSF does not vary in the
simulated images.  To compare these simulated $z=2.7$ galaxies with
the $z\sim 2.3$ \hdf\ galaxy sample, we repeated the source detection
and photometry processes using the SExtractor software to each
simulated ``redshifted'' galaxy.  The galaxy absolute magnitudes and
characteristic radii were calculated on the simulated images in
exactly the manner described previously.

\ifsubmode
\begin{figure}
\epsscale{1.1}
\plottwo{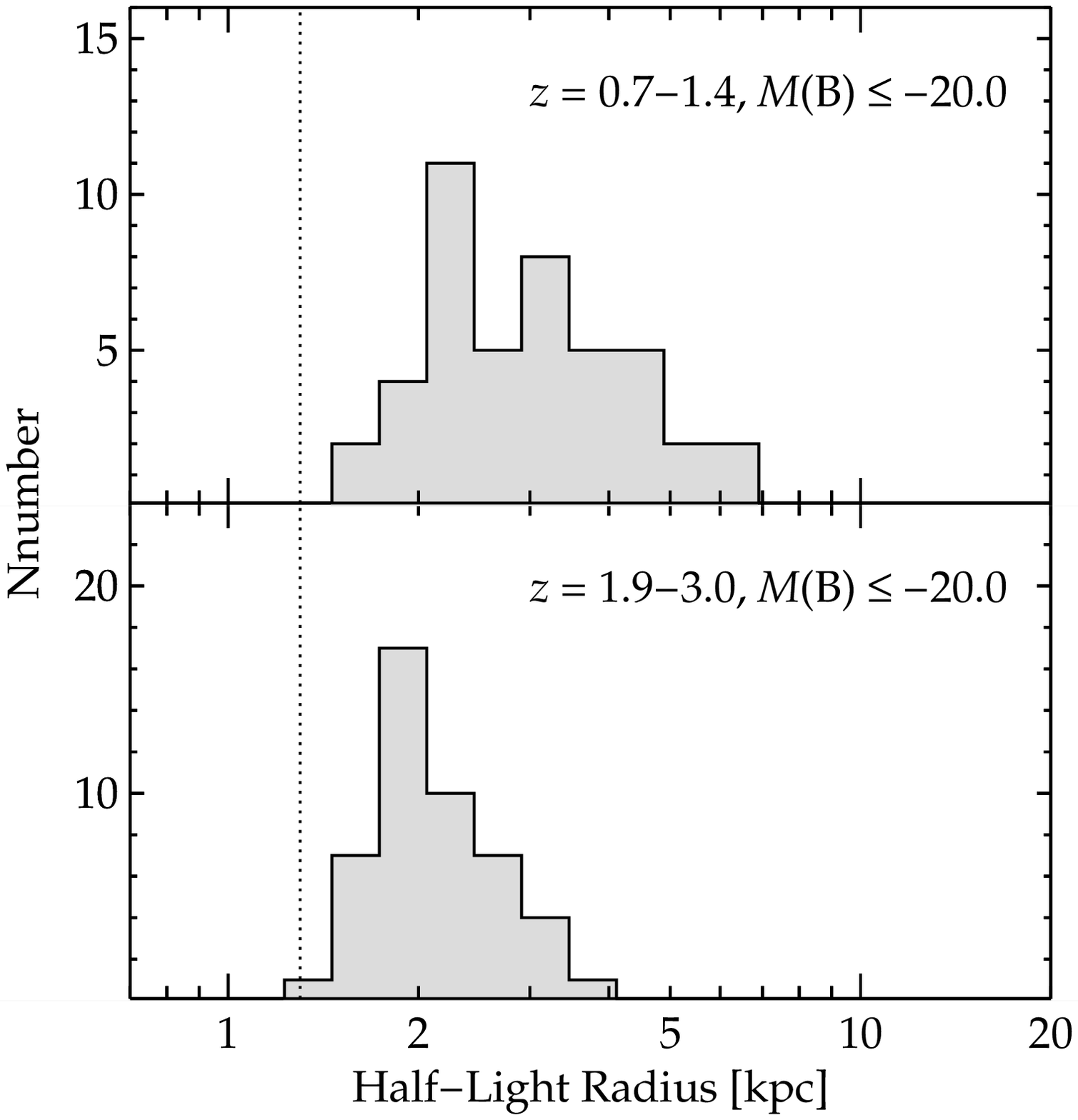}{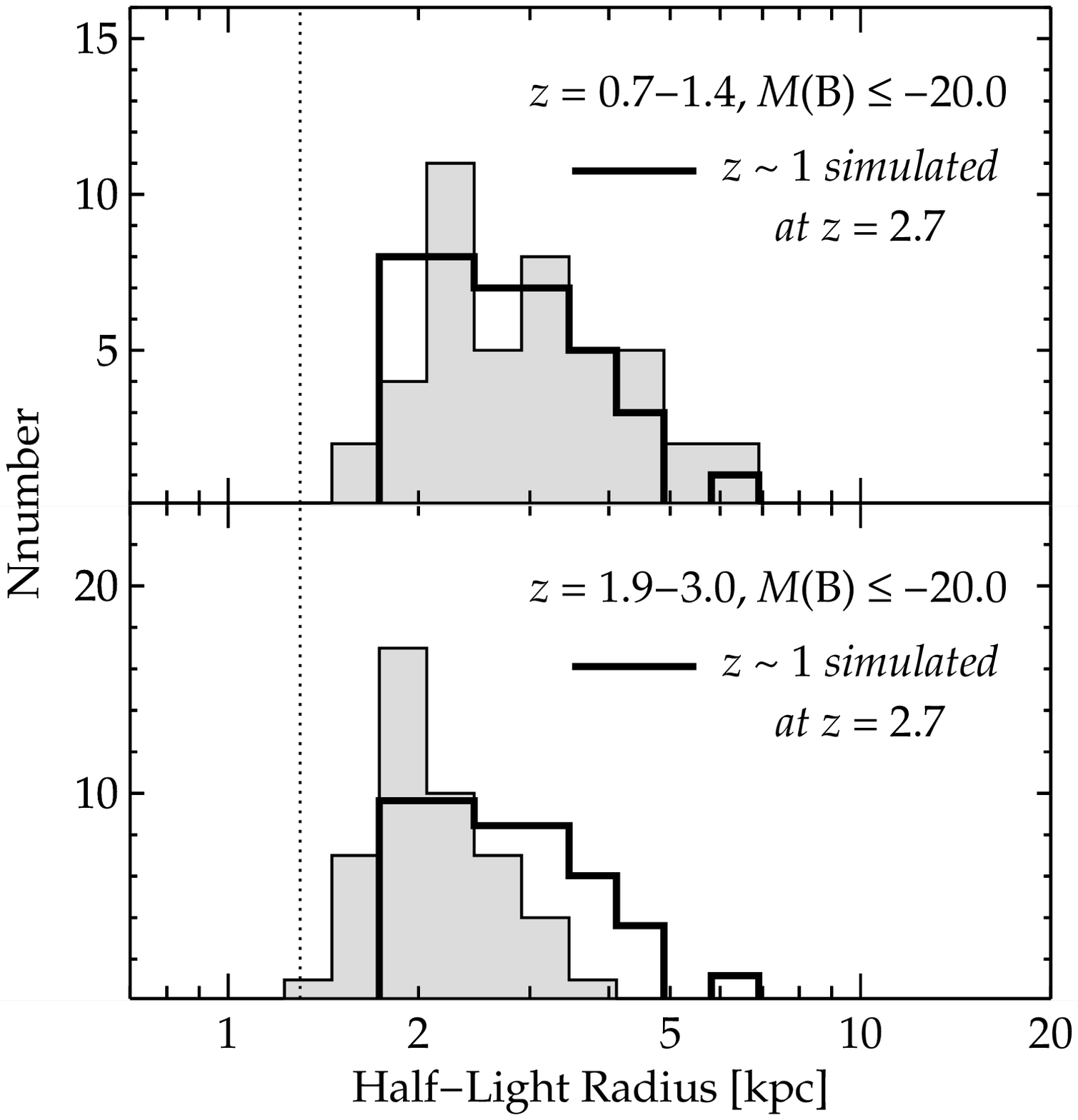}
\epsscale{1.0}
\caption{\figfourcap}
\end{figure}
\else
\begin{figure*}[th]
\epsscale{1.1}
\plottwo{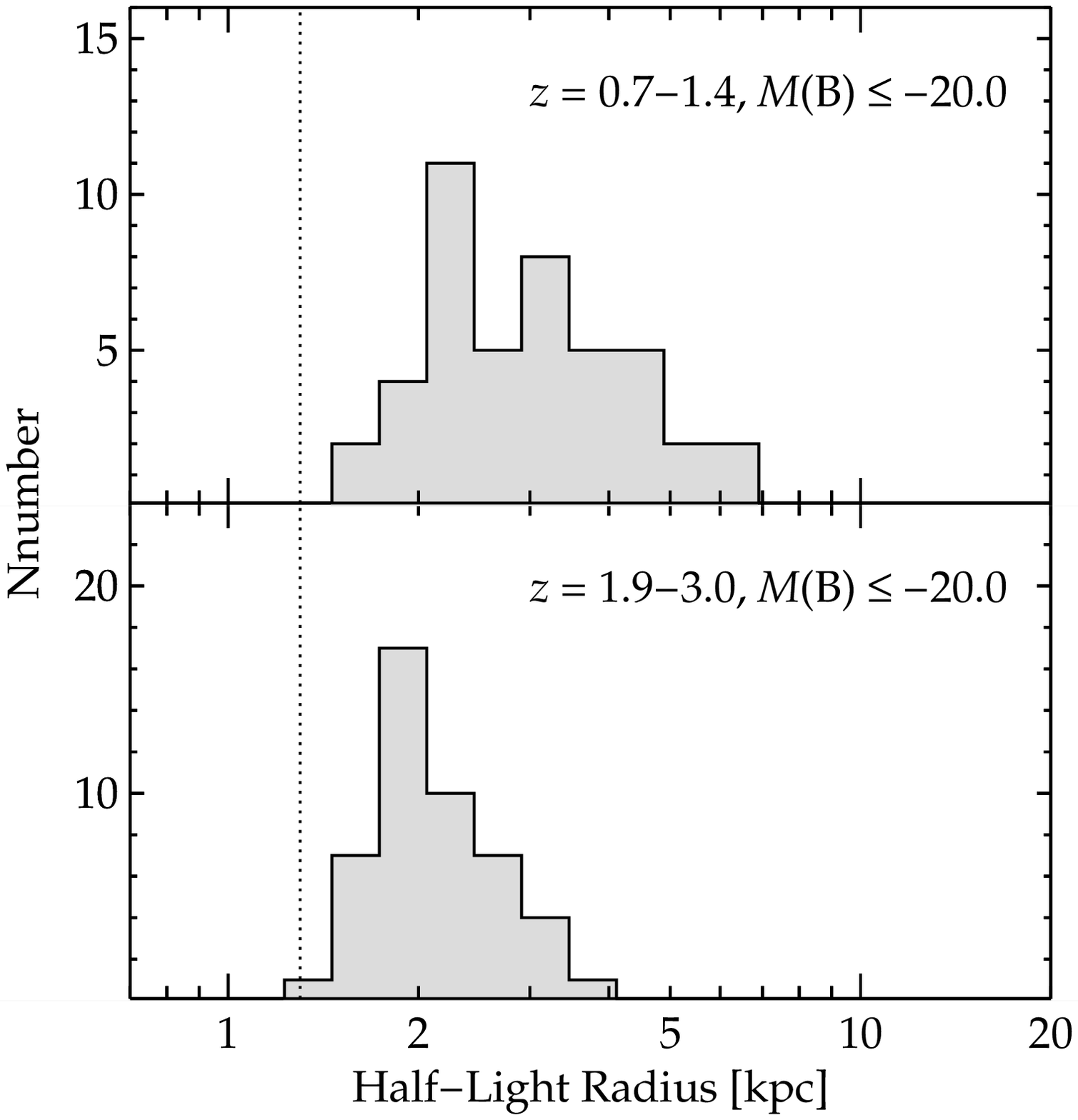}{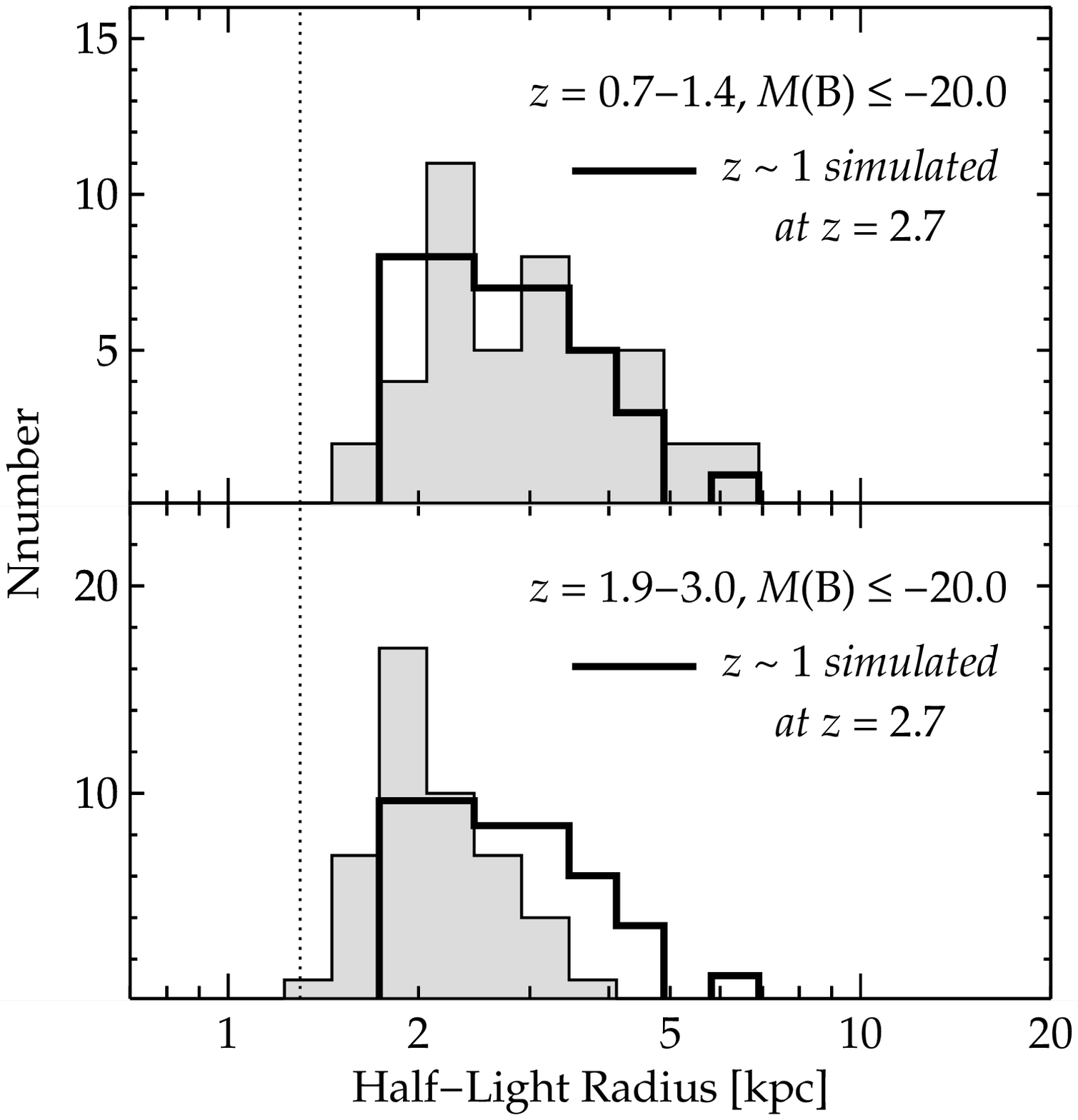}
\epsscale{1.0}
\caption{\figfourcap}
\end{figure*}
\fi

Figure~\ref{fig:radhists} shows the distribution of galaxy sizes with
$M(B) \leq -20.0$ for the galaxy samples, and for the the $z\sim 1$
sample simulated at $z = 2.7$.  The majority of the luminous $z\sim 1$
galaxies would be detected if they were present at $z\sim 2.7$.  The
mean half--light radius of the $z\sim 1$ galaxies simulated at $z=2.7$
is reduced slightly: $\langle r_{1/2,\mathrm{sim}}\rangle \simeq
3.1$~kpc (reduced by $<20$\% compared to the results above).   Using
the one--dimensional KS--test to the half--light distributions of the
$z\sim 2.3$ sample and the $z\sim 1$ sample simulated at $z=2.7$, we
find that these populations stem from different parent distributions
at the $3.9\sigma$ level.  Similarly, applying the two--dimensional
KS--test to the subset of bright galaxies with $M(B) \leq -20$ (as
done previously), the radius--luminosity distributions for the $z\sim
2.3$ galaxies and the $z\sim 1$ galaxies simulated at $z=2.7$ trace
different underlying distributions at the $\simeq 99.6$\% confidence
level ($\simeq 2.9\sigma$).  Thus the strong size evolution that
exists between galaxies at $z\sim 2.3$ and 1 is robust against
surface--brightness dimming effects.  
%Furthermore, the  co-moving
%volume spanned by $1.9 \leq z \leq 3.0$  is 2.3 times larger than that
%at $0.7 \leq z \leq 1.4$ by approximately a factor of 2.3.  
This fact that the $1.9 \leq z \leq 3.0$ redshift range contains more
co-moving volume than that at $0.7 \leq z \leq 1.4$ 
actually strengthens the significance of the result.   For example,
if we scale the number density of objects in each redshift interval to
the same co-moving volume, then the one--dimensional KS--test derives
a $>5\sigma$ likelihood that the half--light distributions do not stem
from the same parent sample.

%%%%%%%%%%%%%%%%%%%%%%%%%%%%%%%%%%%%%%%%%%%%%%%%%%%%%%%%%%%%%%%%%%%%%%

\subsection{The Internal Color Dispersion of Galaxies\label{section:icd}}

In \S~\ref{section:radlum} we describe the increase in the sizes of
galaxies from $z\sim 2.3$ to 1.  If galaxies grow monotonically in
size, then relatively young stars will form at larger radii at later
times.  An observable consequence of this growth may be that internal
color structure develops within galaxies at later epochs.  

To investigate possible internal color evolution in the \hdf\ galaxy
samples, we have quantified the galaxies' internal color dispersion
between (rest--frame) UV--optical wavelengths of the \hdf\ galaxies
using the statistic defined in P03. The internal color dispersion
quantifies the differences in galaxy morphology as a function of
wavelength.  In particular, the (rest--frame) UV--optical internal
color dispersion characterizes diversification in the spatial
configuration of young stellar populations (which dominate the
rest-frame UV emission) and older stars, which contribute more to the
optical light.  In summary (see P03 for details), the internal color
dispersion, $\xi$, is defined as the dispersion of the pixel
intensities between two bandpasses about a mean color, normalized to
the total flux of the galaxy, \ie,
\begin{equation}
\xi(I_1,I_2) = \frac{\sum(I_2 - \alpha I_1 - \beta)^2 - 
	\sum(B_2 - \alpha B_1)^2}
	{\sum(I_2 - \beta)^2 - \sum(B_2 - \alpha B_1)^2},
\end{equation}
where $I_1$ and $I_2$ are the pixel intensity values from each image
obtained in two bandpasses.  The scaling factor $\alpha$ is the flux
ratio, or \textit{color}, between the images, while the linear offset
$\beta$ adjusts (if necessary) for differences in the relative
background levels of the two images.  To account for internal color
dispersion resulting purely from fluctuations in the sky background,
we subtract the contribution from background regions (\ie, image
sections with no detected sources) in each image $B_1$ and $B_2$,
respectively.  The internal color dispersion, $\xi$, is
flux--independent, quantifies the morphological $K$--correction of a
galaxy between the bandpasses $I_1$ and $I_2$, and has an analytically
understood error distribution.

We have computed the internal UV--optical color dispersion for the
galaxies in the \hdf\ samples using the \wfv\ and \nicj\ images for
the $z\sim 1$ sample and \wfi\ and \nich\ images for the $z\sim 2.3$
sample.   The \hst\ images do not have a straightforward distribution
of background pixel values due to the correlated pixels from the
data--reduction process of ``drizzling'' the images, and due to the
fact the the WFPC2 and NICMOS F110W images were convolved to match the
PSF of the NICMOS F160W image.  Both properties of the data will tend
to suppress inter--pixel noise.  Thus, it is necessary to test that
the analytic uncertainties in the internal color dispersion given in
P03 accurately describe the statistical errors for galaxies in the
\hst\ data.   We repeatedly inserted artificial galaxies with known
internal color dispersion into blank regions of the \hdf\ images, and
then computed the distribution of measured internal color dispersion
values.  The width of this distribution provides an estimate of the
uncertainty on $\xi$. Indeed, our simulations indicated that the
analytically derived error $\delta(\xi)$ had underestimated the
intrinsic uncertainty.   We subsequently rescaled the derived
$\delta(\xi)$ values by factors that were empirically derived from the
simulations, \ie, $\delta^\prime(\xi) = C\, \delta(\xi)$, with $C =
5.98$ for $\xi(\wfv,\nicj)$ and $C=7.14$ for $\xi(\wfi,\nich)$.  It
should be noted that while the corrected errors provide a good
estimate for the statistical errors arising from the background
pixel--to--pixel noise, they likely still underestimate the real
uncertainties as they neglect systematic effects arising from, \eg,
inaccurate PSF matching, image registration, image calibration, etc.
In particular, errors in PSF--matching will have a stronger effect on
compact, and centrally peaked objects, and are probably the dominant
error source for the brightest objects.

\ifsubmode
\begin{figure}
\epsscale{1.0}
\plotone{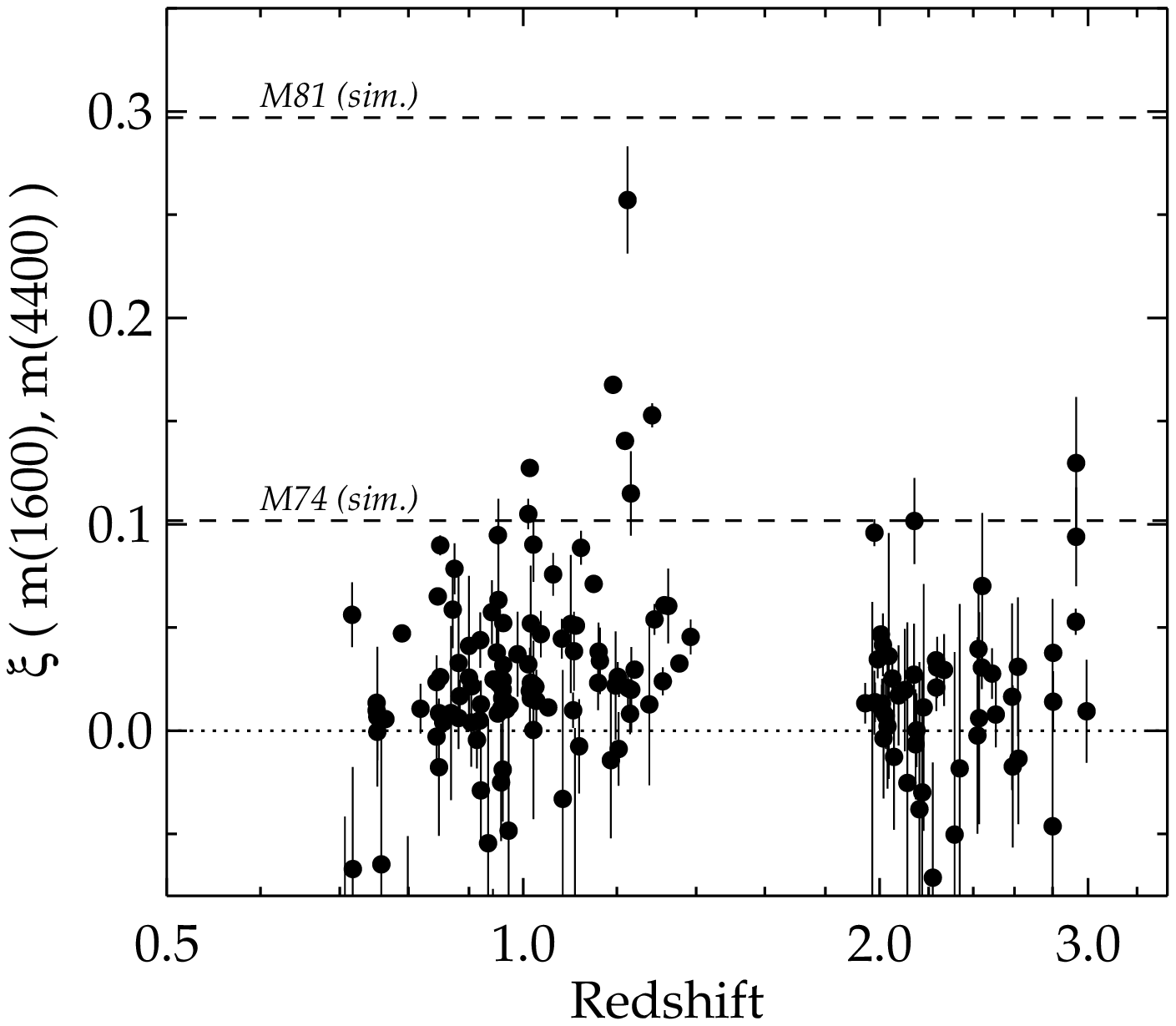}
\epsscale{1.0}
\caption{\figfivecap}
\end{figure}
\else
\begin{figure}
\epsscale{1.2}
\plotone{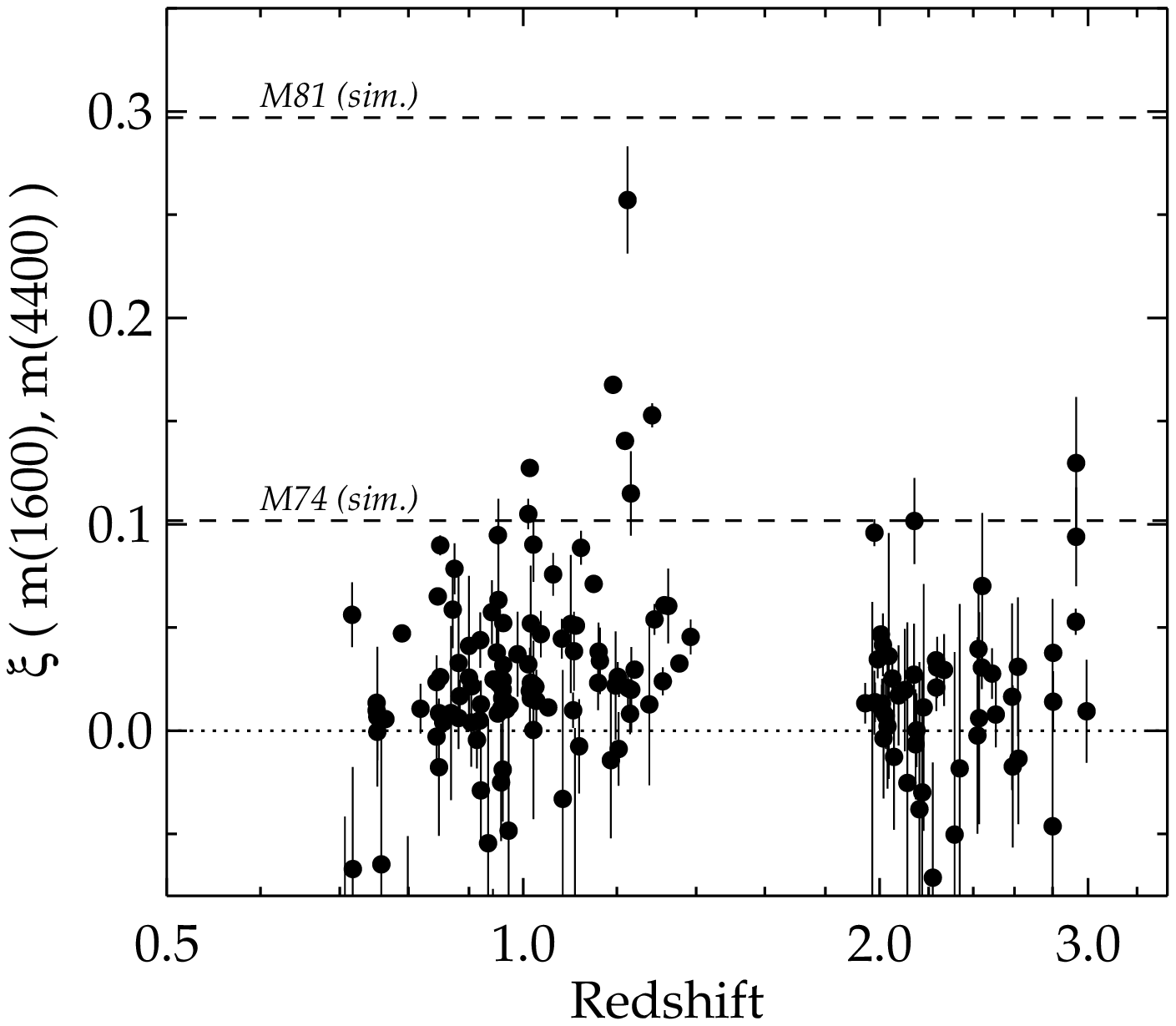}
\epsscale{1.0}
\caption{\figfivecap}
\end{figure}
\fi

\ifsubmode
\else
\begin{figure*}
\epsscale{1.0}
\plotone{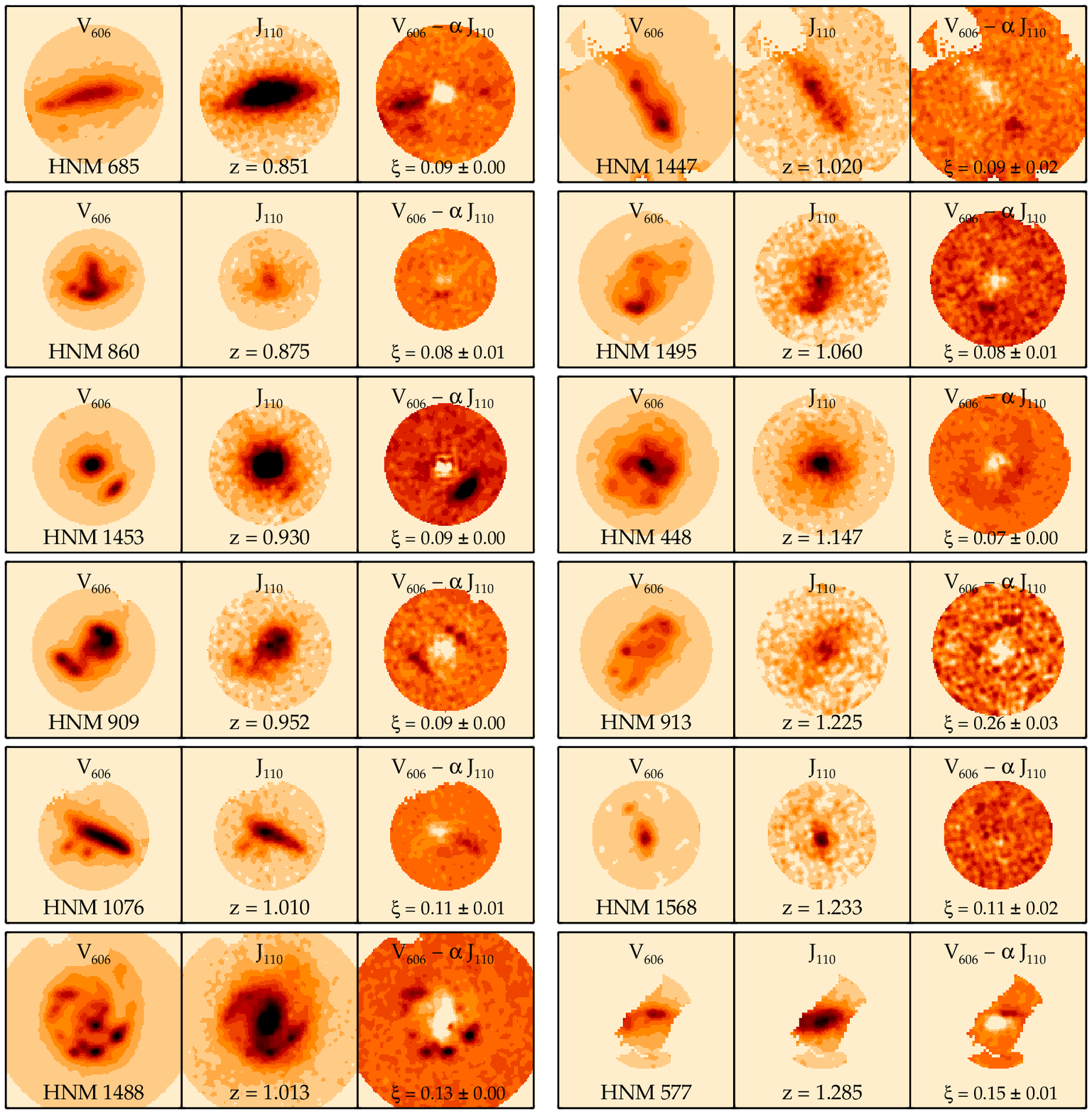}
\epsscale{1.0}
\caption{\figsixcap}
\end{figure*}
\begin{figure*}
\epsscale{1.0}
\plotone{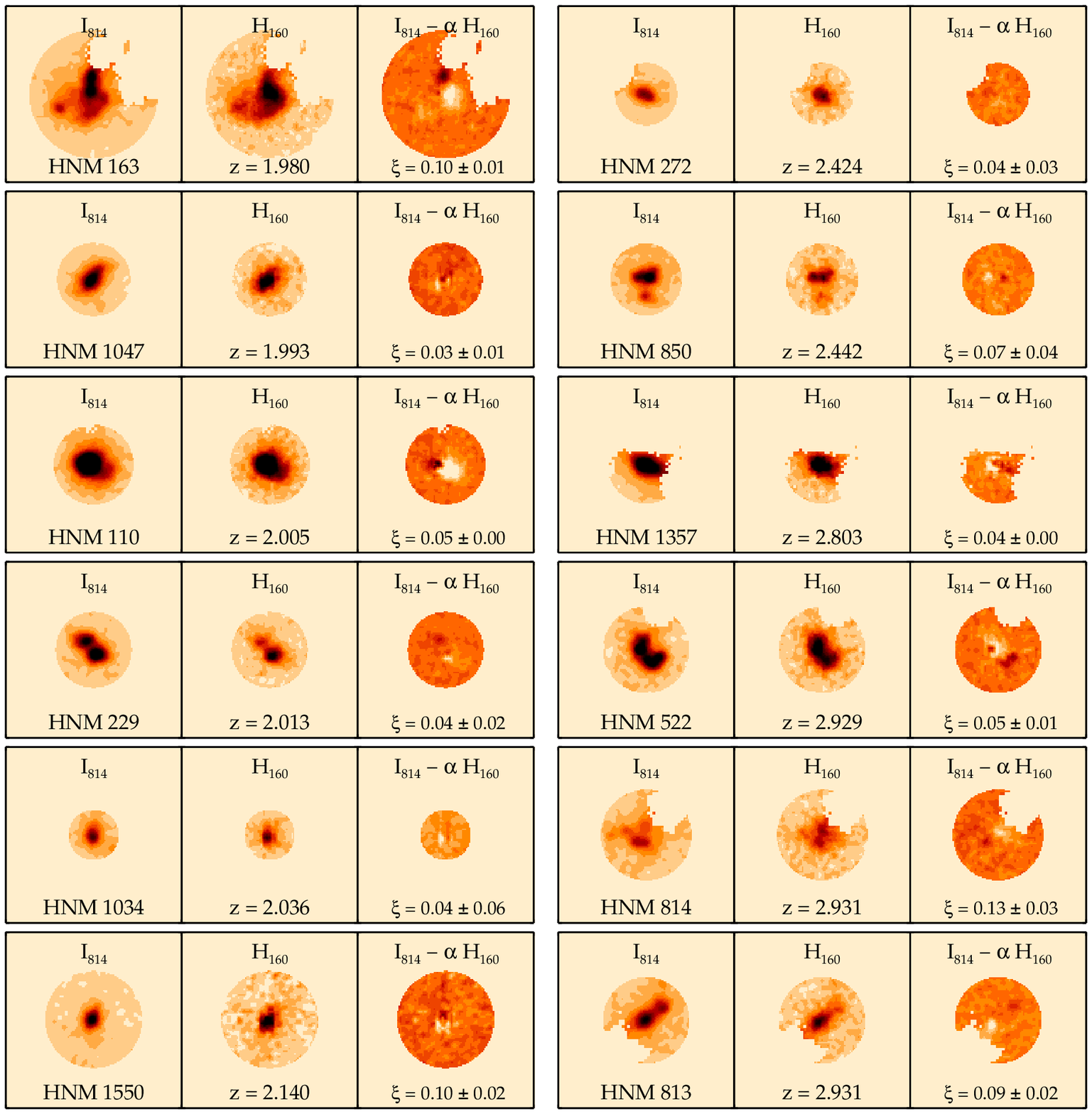}
\epsscale{1.0}
\caption{\figsevencap}
\end{figure*}
\fi

Figure~\ref{fig:hdfxi} shows the distribution of internal UV--optical
color dispersion as a function of redshift for the galaxies in the
\hdf\ samples.   Large uncertainties on $\xi$ are generally indicative
of objects with low signal to noise in a particular bandpass.  As a
fiducial reference, we use the internal color dispersion values for
two local galaxies observed in the same rest--frame bands, M81 (the
early--type spiral galaxy with the highest internal color dispersion
in the local P03 sample) and M74 (a ``typical'' mid--type spiral
galaxy in the P03 sample).  In Figure~\ref{fig:hdfxi}, we show the
derived internal color dispersion for these two galaxies after
convolving them with the NICMOS F160W PSF, and resampling them to the
same physical pixel scale as the high--redshift \hdf\ galaxies in the
\hst\ images.  This illustrates the internal color dispersion in these
fiducial local galaxies (with no evolution) relative to the
high--redshift objects.

There is an evident increase in the upper envelope of internal color
dispersion between  $z\sim 2.3$ and 1.  Formally, the mean and scatter
increase by roughly a factor two from $z\sim 2.3$ to 1: from  $\langle
\xi(\wfi,\nich) \rangle \simeq 0.02$ at $z\sim 2.3$ to $\langle
\xi(\wfv,\nicj) \rangle \simeq 0.04$ at $z\sim 1$; and from $\sigma
(\xi[\wfi,\nich]) \simeq 0.03$ at $z\sim 2.3$ to $\sigma (
\xi[\wfi,\nich]) \simeq 0.05$ at $z\sim 1$.  However, the change in
the number of galaxies with large internal color dispersion is more
dramatic.  Most of the galaxies at $z\sim 2.3$ have
$\xi(\wfi,\nich)\leq 0.05$ (46/53), with a small tail to $\xi \sim
0.1$ (7/53).  In contrast, the $z\sim 1$ sample contains many more
objects with high internal--color dispersion --- 28/113 (25\%) of
galaxies have $\xi(\wfv,\nicj) > 0.05$ --- including a sizable number
(8/113) with values comparable to, or higher than, the mid--type
spiral M74. 

Note that it is unlikely that we are missing some significant fraction
of galaxies at $z\sim 2.3$ with high internal color dispersion.  This
is based on the facts that the $1.9 \leq z \leq 3.0$ redshift interval
contains more than twice the co-moving volume as that of the $0.7
\leq z \leq 1.4$ (although cosmic variance uncertainties apply), the
galaxies have been selected with the same limiting surface--brightness
criteria, and the object resolution changes little for these redshifts
(see \S~2).  Furthermore, in \S~\ref{section:sb_icd} we argue that
surface--brightness dimming effects are not responsible for the change
in the internal color dispersion at these redshifts, which reinforces
our results from P03 that the  intrinsic rest--frame UV--optical
internal color dispersion of local galaxies is recoverable for $0 < z
\lsim 3$.  This evolution is not dominated by
surface--brightness--dimming effects. 

\ifsubmode
\begin{figure}
\epsscale{1.0}
\plotone{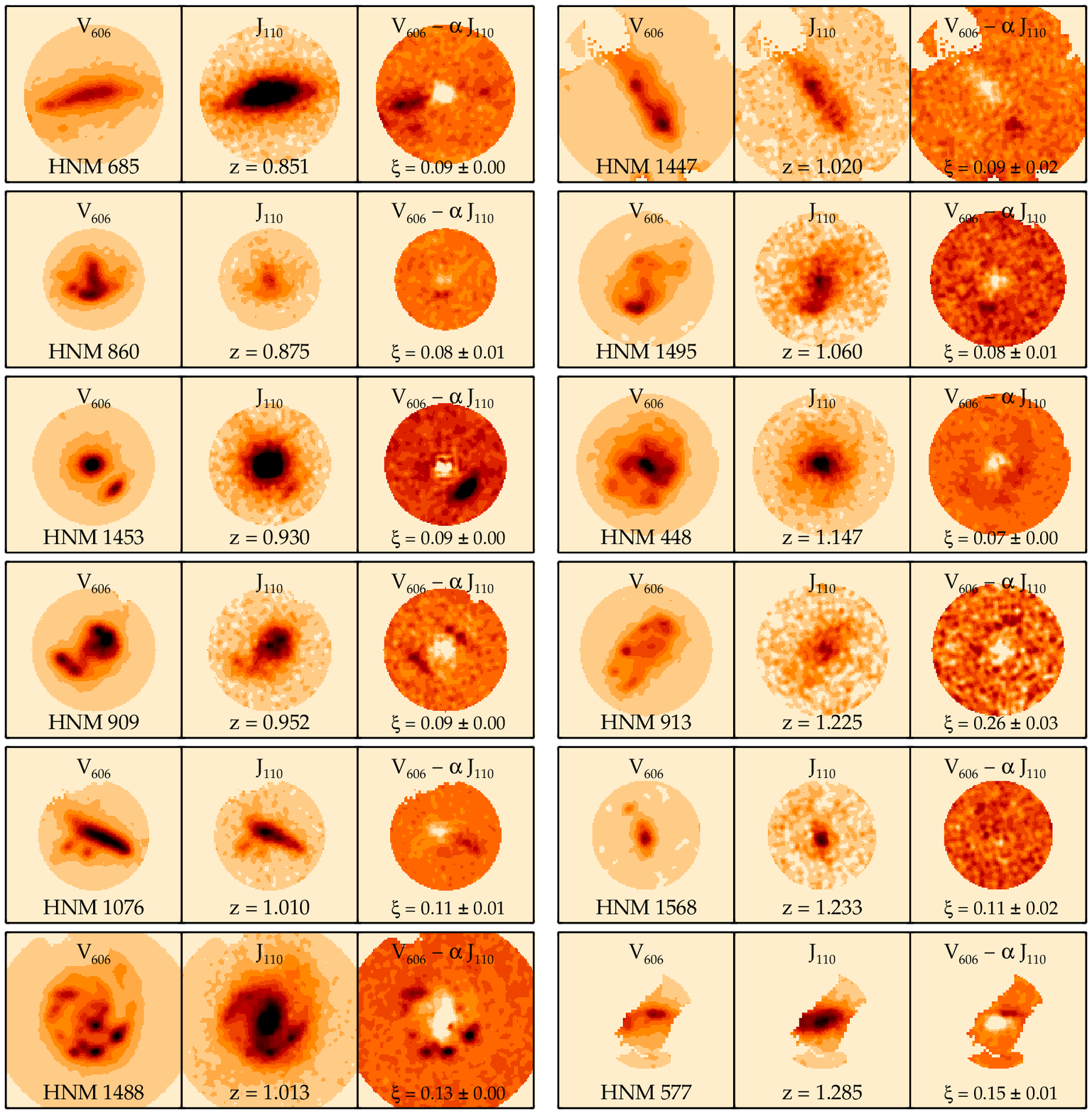}
\epsscale{1.0}
\caption{\figsixcap}
\end{figure}
\begin{figure}
\epsscale{1.0}
\plotone{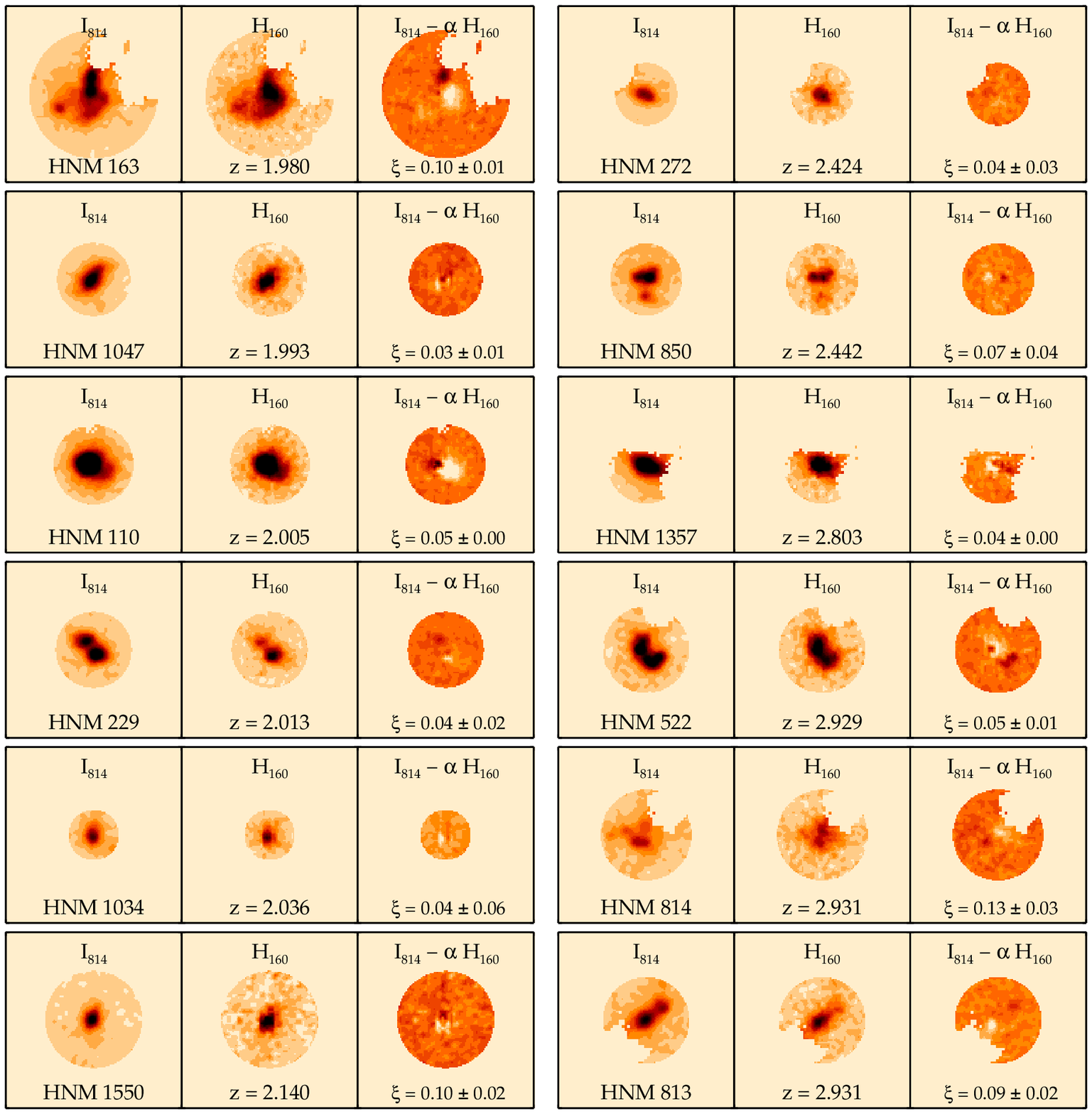}
\epsscale{1.0}
\caption{\figsevencap}
\end{figure}
\else
\fi

Figure~\ref{fig:montage_xi_lowz} shows the galaxies with the highest
$\xi(\wfv,\nicj)$ in the $z\sim 1$ sample.   Each object shows a
significant qualitative difference between the rest--frame UV and
optical images, which is quantified by the higher $\xi$ values.  Many
of these galaxies have features indicative of early--to--late--type
spiral galaxies (\eg, HNM~448, 860, 913, 1488, 1495), with a strong,
central core  in the \nicj\ images, and prominent star--forming spiral
arms in the \wfv\ images with relatively bluer colors.  These objects
have internal color dispersion and color residual images qualitatively
similar to that observed in local spirals, \eg, M81, M74, M100 (P03).
Several of the $z\sim 1$ galaxies in the figure show the appearance of
inclined disk systems (\eg, HNM~577, 685, 1076, 1447), and the
residual colors may suggest patchy dust and/or varied stellar content
similar to nearby analogs; \eg, M82 and UGC~06697 (P03).  There is
also an actively star--forming galaxy with a disturbed morphology at
all wavelengths and strong residual colors (HNM~909).  This galaxy has
a red core and blue outer regions, and it is possibly an spiral with
some interacting activity, and a diverse stellar content and variable
dust opacity.  HNM~1453 shows clear multiple components of very
different rest--frame UV--to--optical colors: a red elliptical galaxy
with an off--center, blue component.  This object has been identified
as a gravitational lens candidate of a background object \citep[see,
\eg,][]{zep97}.  Thus, it seems that this object consists of two
objects along the line of sight, which were not split in the HNM
catalog. We will exclude this object from further analysis.

Figure~\ref{fig:montage_xi_hiz} shows the galaxies in the  $z\sim 2.3$
sample with the highest $\xi(\wfi,\nich)$ values. Nearly all the
galaxies show evidence for disturbed morphologies, and  the internal
color dispersion probably arises from variations in the relative
star--formation and variable dust obscuration.  The exception is
possibly HNM~163, which shows evidence for a redder core and a bright,
off--center blue knot, and may result from variations in the stellar
content.  However, this galaxy has evidence for high dust obscuration
(Papovich et al.\ 2001), which may also dominate the internal color
dispersion.

Several pairs of the galaxies in $z\sim 2.3$ sample have been
``deblended'' by the SExtractor software into two components: HNM 162
+ 163; HNM 813 + 814; HNM 1357 + 1358 (see
Figure~\ref{fig:montage_hiz1}).  These galaxies all have spectroscopic
redshifts, but the multiple galaxy components are merged at typical
ground--based resolution ($\approx 1\arcsec$), and as a consequence
their spectra are blended as well.  However, the photometric redshifts
of the individual components \citep{bud00} are approximately identical
in each case ($\delta z/(1+z) \lsim 0.02$).  Although these individual
components have similar redshifts \citep[see][]{dic03}, it is unclear
whether these objects actually are multiple components of the same
parent galaxy or close galaxy pairs.  In the analysis above we have
kept the galaxies separated, but we have tested how merging them into
single galaxies would affect the derived sizes and internal color
dispersion.  In  three of the cases, the derived half--light radius
increases by a factor less than two, with the exception being the case
of HNM 162 + 163 where the half--light radius changes insignificantly.
This would shift the galaxies in the radius--luminosity diagram
(Figure~\ref{fig:radlum}), but would not alter any of the conclusions
from that plot.  In all four cases, the internal color dispersion of
the merged pairs equals essentially the luminosity--weighted average
of the internal color dispersion of the individual components.  This
has little effect on the distribution in Figure~\ref{fig:hdfxi}, and
reduces the internal color dispersion of HNM 163 and 814, which have
two of the largest values in the $z\sim 2.3$ sample.  Using the values
for the merged objects would actually augment the change in the
scatter of the internal color dispersion from $z\sim 2.3$ to 1.  We
therefore keep the objects separated in order to be conservative in
this evolution.

\ifsubmode
\subsection{Effects of Surface Brightness on the Internal Color
Dispersion}\label{section:sb_icd}
\else
\subsection{Effects of Surface Brightness on the \\ Internal Color
Dispersion}\label{section:sb_icd}
\fi

\ifsubmode
\else
\begin{figure}
\epsscale{1.0}
\plotone{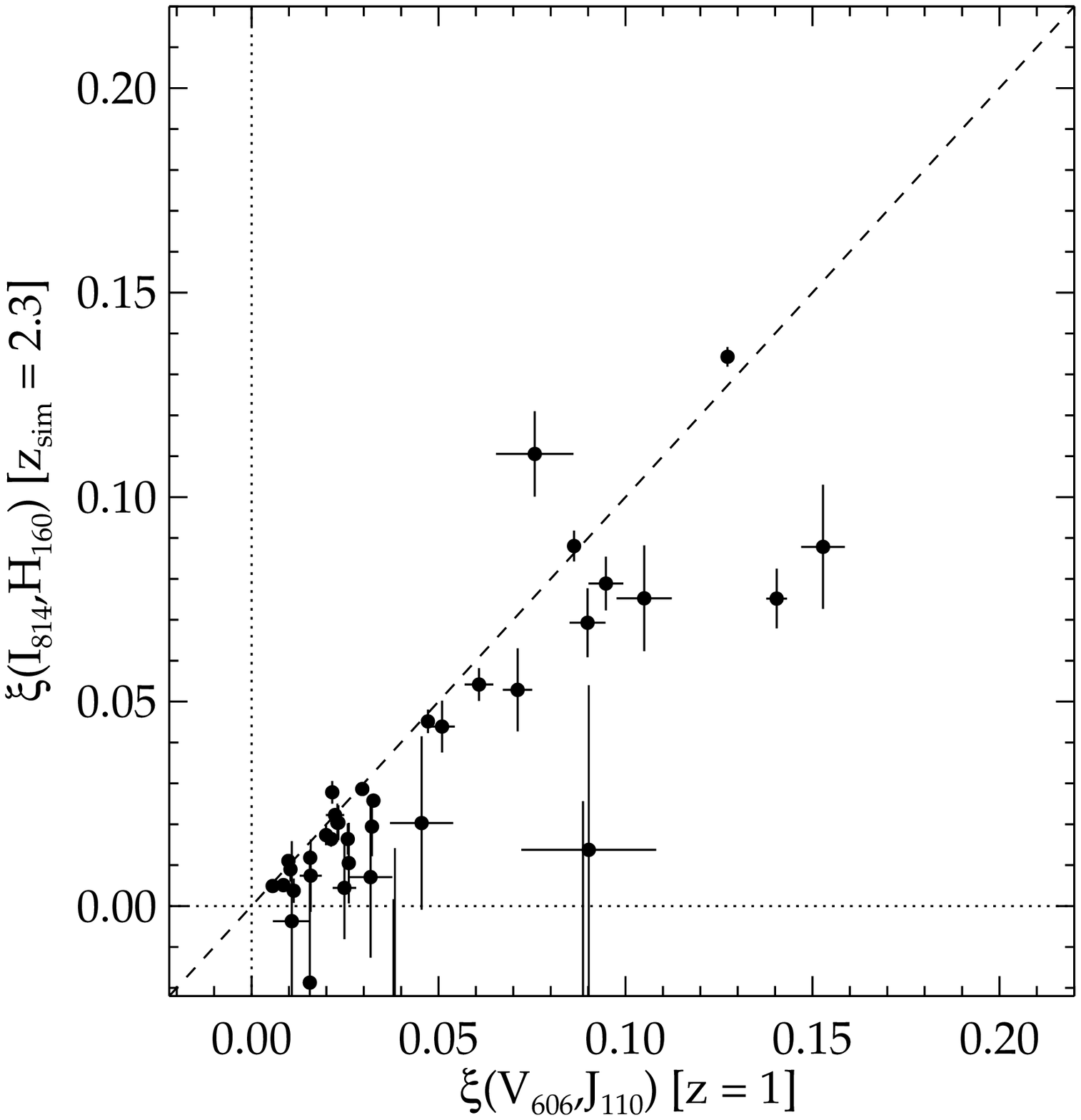}
\epsscale{1.0}
\caption{\figeightcap}
\end{figure}
\fi

In this section we consider the effects from surface--brightness
dimming on the internal color dispersion for increasing redshift.  In
P03, we tested this for local galaxies observed by \uit, and concluded
that the internal color dispersion was fairly robust to $z\lsim 3$ for
galaxies with sufficient S/N in images with \hst--like angular
resolution.  Here, we expand these tests by simulating the appearances
of bright ($M(B) \leq -20.0$) galaxies in the $z\sim 1$ \hdf\ sample.
We simulated the appearance of each $z\sim 1$ galaxy at $z = 2.3$ (the
median redshift of the high--redshift \hdf\ sample) by appropriately
resampling and dimming their images in the \wfv\ and \nicj\ bands, and
adding noise from blank regions of the \wfi\ and \nich\ images (see
\S~\ref{section:sb_sizes}, also P03).  We then compute the internal
color dispersion of these simulated galaxies using the procedures
above.

Figure~\ref{fig:xivxisim} compares the measured internal color
dispersion of the $z\sim 1$ galaxies with the measured values from
these galaxies simulated at $z = 2.3$ in \wfi\ and \nich\ images.  In
general, the internal color dispersion values of the simulated
high-redshift galaxies correlate with their measured values at $z\sim
1$.  There is a slight reduction in the internal color dispersion
values, which results from decreased signal--to--noise from 
cosmological surface brightness dimming.

\ifsubmode
\begin{figure}
\epsscale{1.0}
\plotone{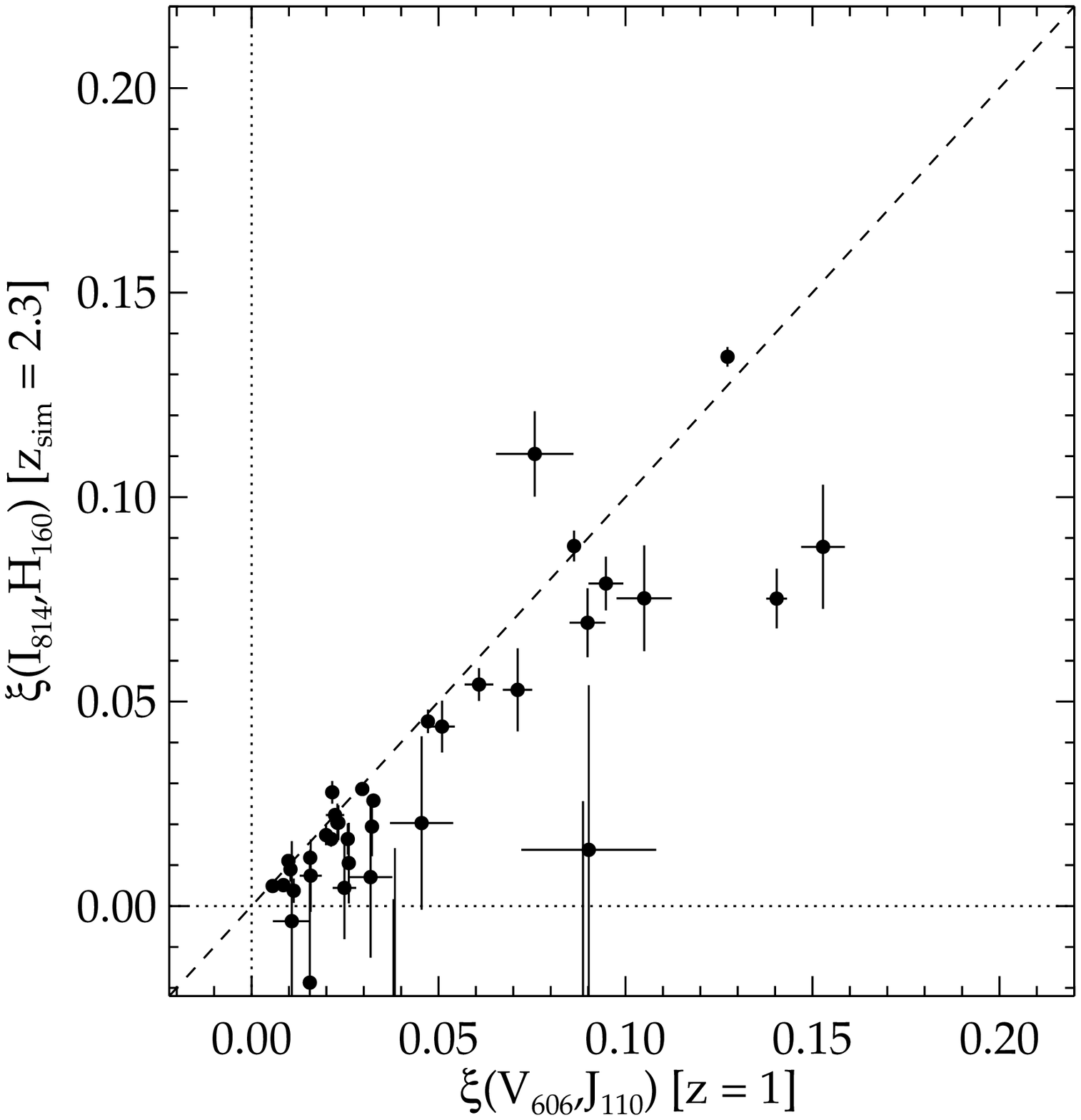}
\epsscale{1.0}
\caption{\figeightcap}
\end{figure}
\fi

Quantitatively, the number of galaxies with internal color dispersion
values above the fiducial value $\xi \geq 0.05$ drops from roughly 35\%
at $z\sim 1$, to 26\% for the simulated galaxies at $z=2.3$.  However,
this is still significantly greater than the 15\% of galaxies in the
$z\sim 2.3$ sample with $M(B) \leq -20$ and $\xi \geq 0.05$.
Furthermore, the reduction of the  internal color dispersion in the
simulated galaxies is possibly overestimated.  In
\S~\ref{section:discussion}, we argue that the \wfi--\nich\ internal
color dispersion at $z\sim 2.3$ is more sensitive to heterogeneities in
a galaxy's stellar populations compared to the \wfv--\nicj\ internal
color dispersion at $z\sim 1$.  This is a consequence of the fact that
at $z = 2.3$, the \wfi\ bandpass probes rest--frame $\approx
2400$~\AA, while at $z\sim 1$, the \wfv\ bandpass is sensitive to
slightly redder rest-frame wavelengths, $\sim 3000$~\AA.  This is not
taken into account in these simulations, and would increase the
$\xi(\wfi,\nich)$ values of the simulated $z = 2.3$ galaxies by as
much as $\sim 40$\% (see \S~\ref{section:discussion}).

%%%%%%%%%%%%%%%%%%%%%%%%%%%%%%%%%%%%%%%%%%%%%%%%%%%%%%%%%%%%%%%%%%%%%%

\section{Discussion}

\subsection{The Growth of Galaxy Sizes\label{section:sizes}}

The average size of luminous galaxies in the \hdf\ between $z\sim 2.3$
and $z\sim 1$ increases by roughly a factor of $\simeq 1.4$.
Characterizing this as a simple power--law with redshift, we find that
the galaxy sizes evolve as $r_{1/2} \sim (1+z)^{-1.2\pm 0.1}$ for
$z\sim 1 - 2.3$.  Hierarchical models predict that the characteristic
sizes of galaxy halos grow as, $r_\mathrm{vir}(z)
\propto M_\mathrm{vir}^{1/3}\, H(z)^{-2/3} \propto
V_\mathrm{vir}H(z)^{-1}$ \citep[\eg,][]{bou02,fer04}. For the default
cosmology, the Hubble constant evolves as $H(z) \sim (1+z)^{3/2}$,
which implies that galaxy sizes should evolve roughly as $r_{vir} \sim
(1+z)^{-(1 - 1.5)}$.  Assuming the half--light radius scales as the
virial radius, the observations here support this hierarchical picture
of matter assembly within halos for $z\sim 1-2.3$.  This result
broadly agrees with predictions from other semi--analytical
hierarchical models of galaxy formation
\citep[\eg,][]{bau98,mo99,som01}.

The radius--luminosity distribution of $z\sim 1$ galaxies is roughly
consistent with that of present--day galaxies and passive luminosity
evolution.  For example, for a monotonically evolving stellar
population model \citep[\eg, using the models of ][]{bru03}
formed in a burst of star--formation at $z_f \sim 4$, one expects
$\Delta M_B \simeq 1$~mag from $z = 1$ to $z = 0$.  Thus, the data
points for the $z\sim 1$ sample under this luminosity evolution would
shift to regions of the $M_B$ -- $r_{1/2}$ distribution that are
occupied by local galaxies.  However, it seems unavoidable that the
progenitors of the largest galaxies in the $z\sim 1$ sample and at
$z\sim 0$ are not consistent with simple luminosity evolution.  Pure
luminosity evolution would fade the $z\sim 2.3$ galaxies down the
luminosity axis of Figure~\ref{fig:radlum}, where they have sizes that
are too small compared with the observed distribution of galaxy sizes
at $z\sim 1$. This result does not suffer from a selection bias in the
sense that if the fully formed precursors to the $z\sim 1$ galaxies
were already in place at $z = 2.3$, they would be observed in the deep
NICMOS data (see \S~\ref{section:sb_sizes}).  Therefore,  the $z\sim
2.3$ population are not the fully--formed predecessors of the $z\sim
1$ galaxies.

\ifsubmode
\begin{figure}
\plotone{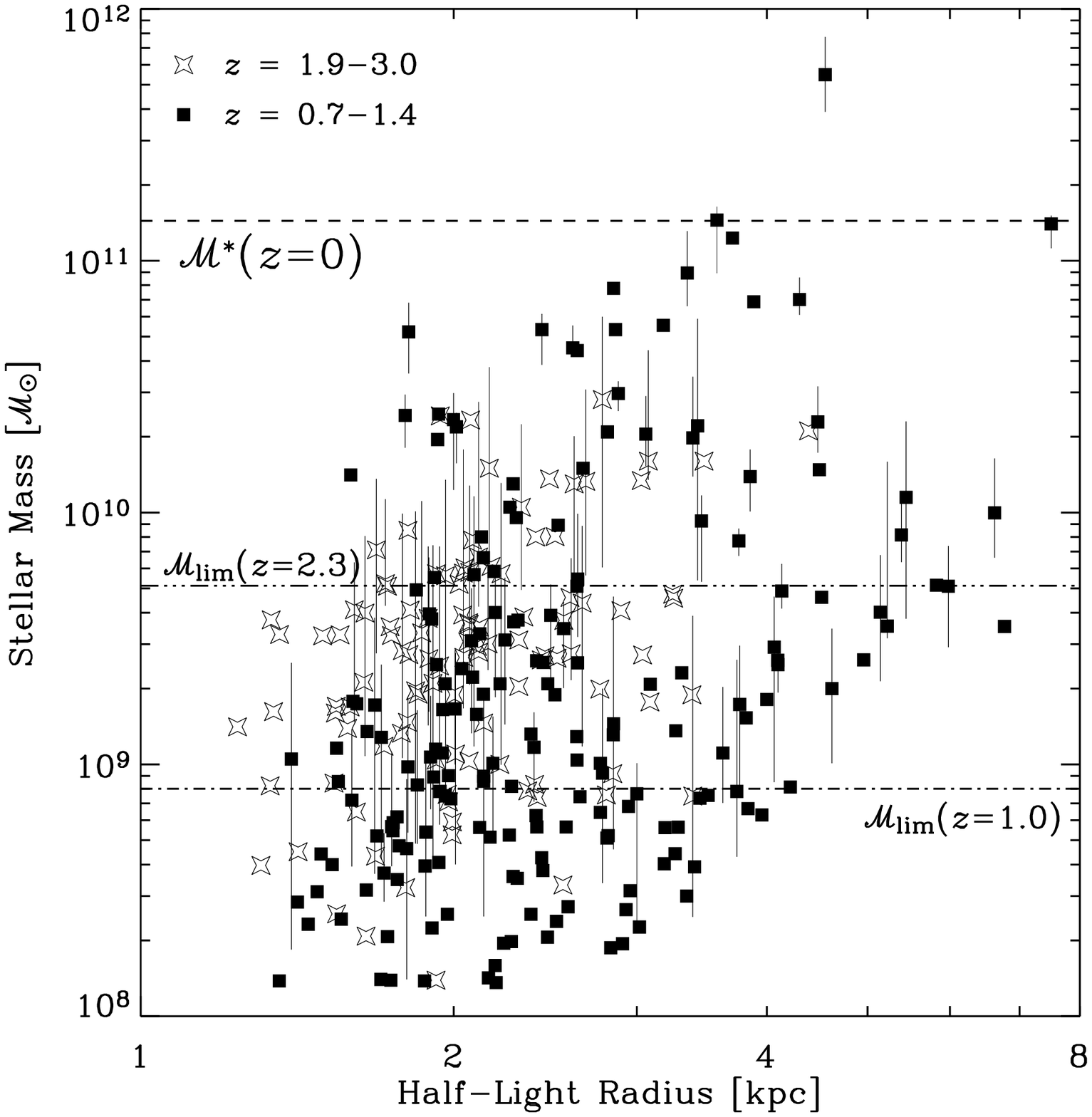}
\caption{\figninecap}
\end{figure}
\else
\fi

\ifsubmode
\else
\begin{figure}[bt]
\epsscale{1.2}
\plotone{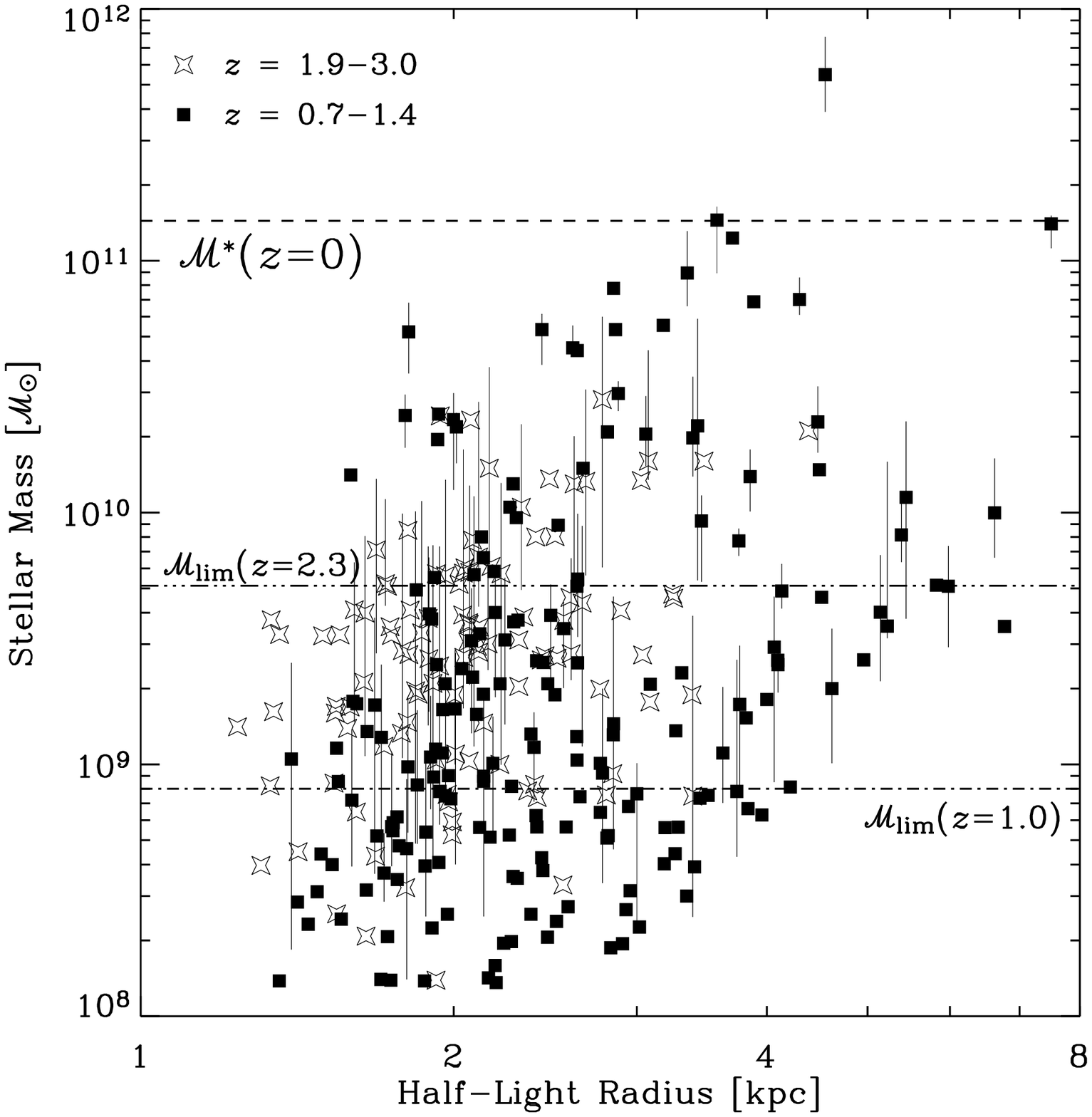}
\epsscale{1.0}
\caption{\figninecap}
\end{figure}
\fi

An alternative indication of growth for galaxies from $z\sim 2.3$ to 1
is evident from their radius--stellar-mass distribution, which is
shown in Figure~\ref{fig:r50vmass}.  Galaxy stellar masses have been
estimated by fitting a suite of stellar--population synthesis models
to the full spectrophotometry of the \hdf\ galaxies (see Papovich
\etal\ 2001; Dickinson \etal\ 2003).  The stellar mass of a galaxy
generally only increases with time (in the absence of interactions or
tidal stripping that could eject stellar material),\footnote{As
stars evolve off the main sequence they eject a   portion of their
stellar mass back into the galaxies' gas reservoir.   This amounts to
a decrease in the total stellar mass of $\approx  30$\% over the
lifetime of a passively evolving stellar population.} which implies
that evolution in a galaxies' stellar population manifests itself only
in the positive direction of the ordinate in
Figure~\ref{fig:r50vmass}.  Galaxy--size evolution  depends on the
assembly history.  Hierarchical models predict that the characteristic
galaxy sizes scale roughly proportional to cosmic time
\citep[\eg,][]{bou02} from the accretion of surrounding material.
This implies evolution in the positive direction along the abscissa in
Figure~\ref{fig:r50vmass} by a roughly factor of 1.5 from $z\sim 2.3$
to 1.  However,  the final half--\textit{mass} size depends on the
ratios of the sizes and masses of the progenitors.  For example, using
the relations presented in
\citet{col00} the radius of the merger product from two near--equal
mass progenitors with size ratios, $R_2/R_1 = (1/3,\, 1,\, 3)$, is $R
\simeq (0.7,\, 1.3,\, 2.2) \times R_1$, respectively.  Note that in
the case of $R_2/R_1 = 1/3$ that the final size is \textit{decreased}
relative to the size of the largest progenitor.  Therefore, in the
case of major mergers evolution in either direction along the abscissa
of the figure is possible.  As illustrated in
Figure~\ref{fig:r50vmass}, the distribution of $z\sim 1$ galaxies
extends to both larger stellar masses and sizes than galaxies at
$z\sim 2.3$.  The larger--stellar-mass galaxies ($\mathcal{M} \gsim
5\times 10^{10}$~\msol) at $z\sim 1$ are not present in the $z\sim
2.3$ sample.  At the same time, galaxies with larger half--light radii
($r_{1/2} \gsim 5$~kpc) appear at $z\sim 1$ in the \hdf\ but not at
higher redshifts.  These galaxies in general have $\mathrm{M} \lsim
10^{10}$~\msol, similar to that found for the most massive galaxies at
$z\sim 2.3$.  The lack of larger galaxies with higher stellar masses
at $z\sim 1$ is not unexpected simply because such objects are very
rare and unlikely to be present within the volume of the \hdf\ (see also
\S~\ref{section:sizes}). 

Previous studies of UV--bright $z\sim 3$ galaxies have concluded that
their space density is roughly that of present--day, massive,
early--type galaxies \citep{ste96,gia98}.  Through arguments of the
measured $z\sim 3$ clustering and angular separation, this population
may trace a one--to--one correspondence with massive dark matter halos
\citep{ade98,gia01,por01}.  Because the cores of such massive halos
will presumably evolve to become the locations of massive present--day
galaxies, it is natural to equate this high--redshift galaxy
population with that of present--day ``\lstar''--sized objects.
One may expect galaxies in massive haloes to have an accelerated
evolution, and thus they should assemble themselves at larger
look-back times, which may partly be responsible for some of the
observed evolution. This  scenario is plausible, with the additional
requirement that the galaxies necessarily undergo additional star
formation and growth at $z\lsim 2$ to match the $z\sim 1$ and
present--day galaxy radius--luminosity and radius--stellar-mass
distributions.

%%%%%%%%%%%%%%%%%%%%%%%%%%%%%%%%%%%%%%%%%%%%%%%%%%%%%%%%%%%%%%%%%%%%%%

\subsection{The Distribution of Galaxy Colors from $z\sim 1 -
  3$}\label{section:discussion}

The scatter in the distribution of UV--optical total colors and
internal color dispersion increases significantly in galaxies from
$z\sim 2.3$ to 1. Figure~\ref{fig:colorcolor} shows color--color plots
and the internal color dispersion as a function of total color for the
$z\sim 1$ and 2.3 galaxy samples.  The figure also shows the colors
from models of an evolving stellar population at $z=1$ and 2.3 with
three different monotonic star--formation histories (constant,
instantaneous burst, and exponentially decaying) as a function of age
from $10^7 - 10^{10}$~yr.  
				   
The $z\sim 2.3$ galaxies span a relatively small range of total
color--color space even though they come from a IR--selected, flux--limited
sample with complete photometric and spectroscopic redshift
information (see \S~\ref{section:hdfsample}).  The $z\sim 2.3$
galaxies have colors that are broadly consistent with those of a
young, constantly star--forming stellar population with moderate
amounts of dust reddening \citep[see also][]{saw98,pap01}.  Several
objects have highly reddened rest-frame UV colors ($\wfb-\wfi \sim 1$;
indicated by open symbols in the figure).  All of these sources have
redshifts $z\sim 2.9$ at which point the flux in the \wfb\--band is
affected by \lya\ absorption from the IGM.  This is not included in
the model colors.  \lya\ absorption does not affect the \wfi--\nich\
internal color dispersion.

\ifsubmode
\else
\begin{figure*}[t]
\epsscale{1.0}
\vbox{\plottwo{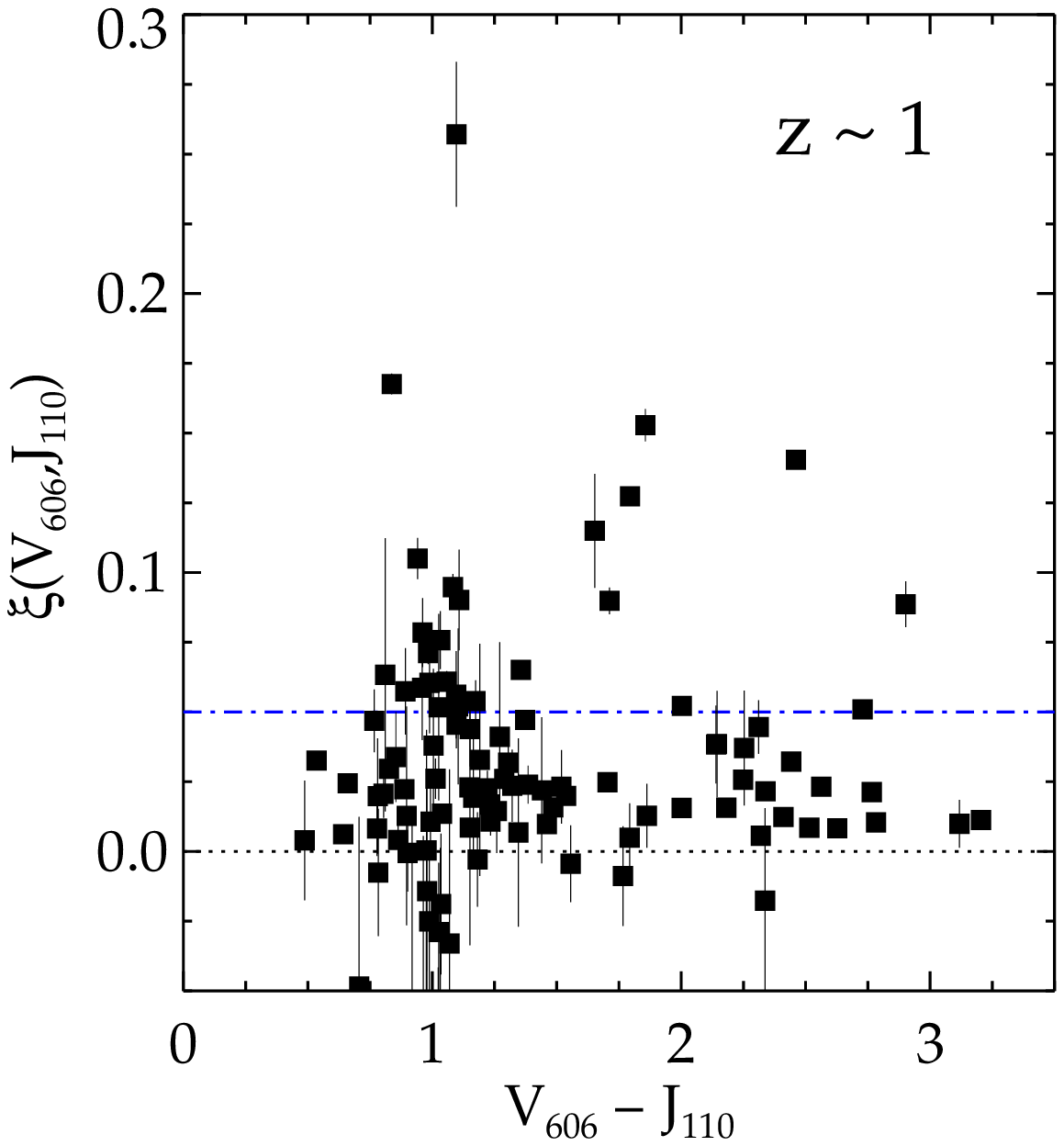}{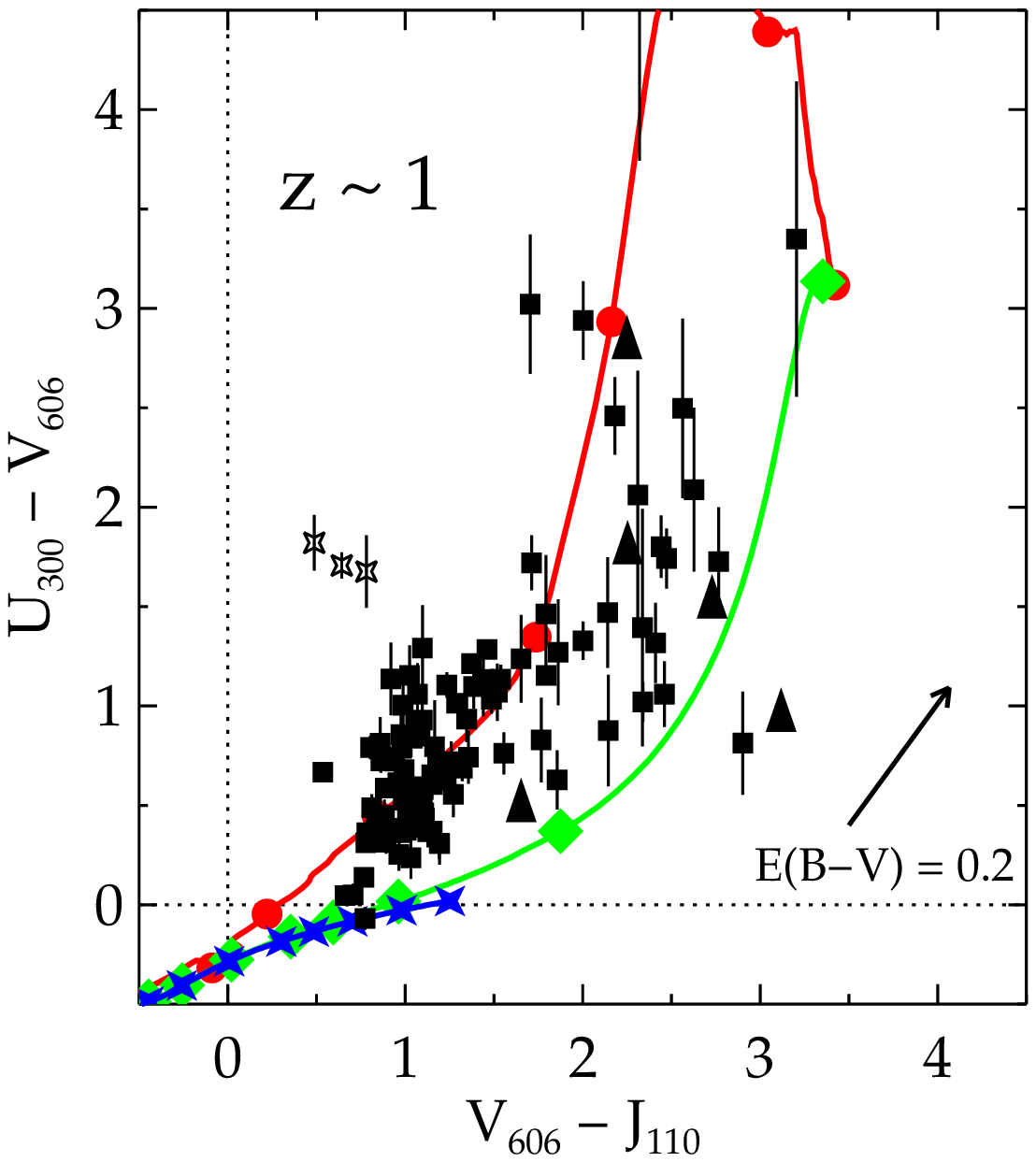}}
\vspace{0.1in}
\vbox{\plottwo{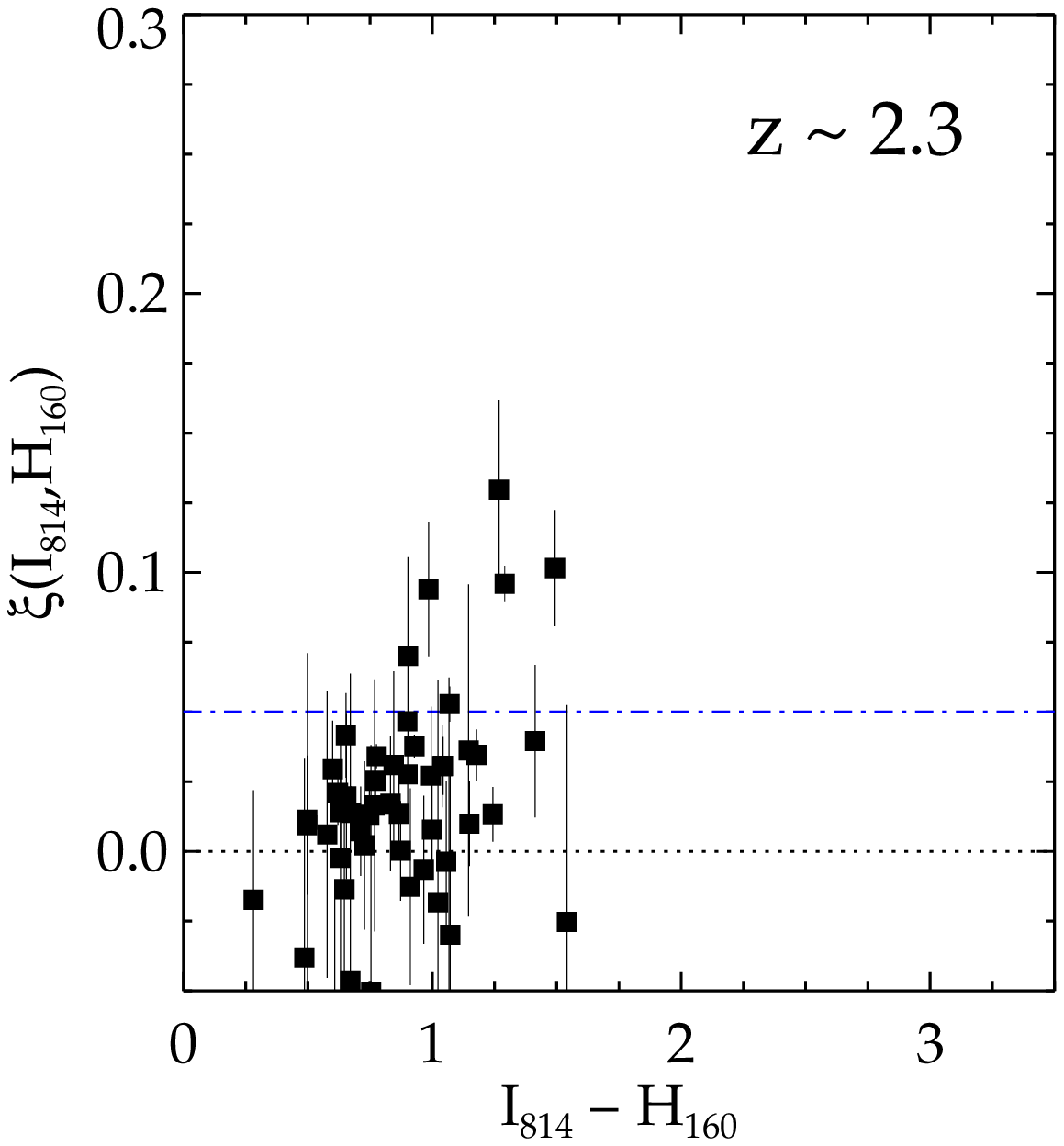}{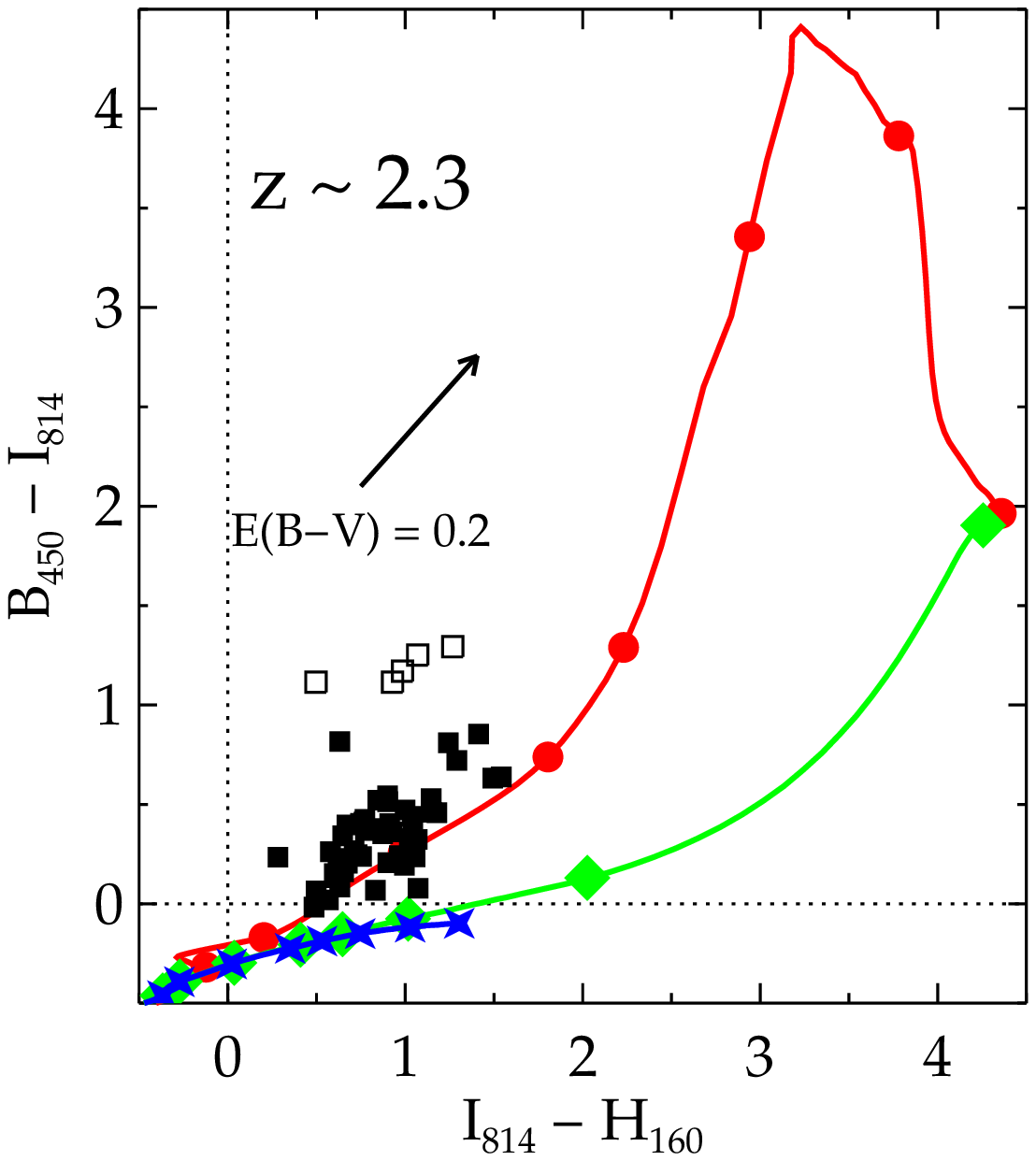}}
\epsscale{1.0}
\caption{\figtencap}
\end{figure*}
\fi

\ifsubmode
\begin{figure}[thp]
\epsscale{1.0}
\vbox{\plottwo{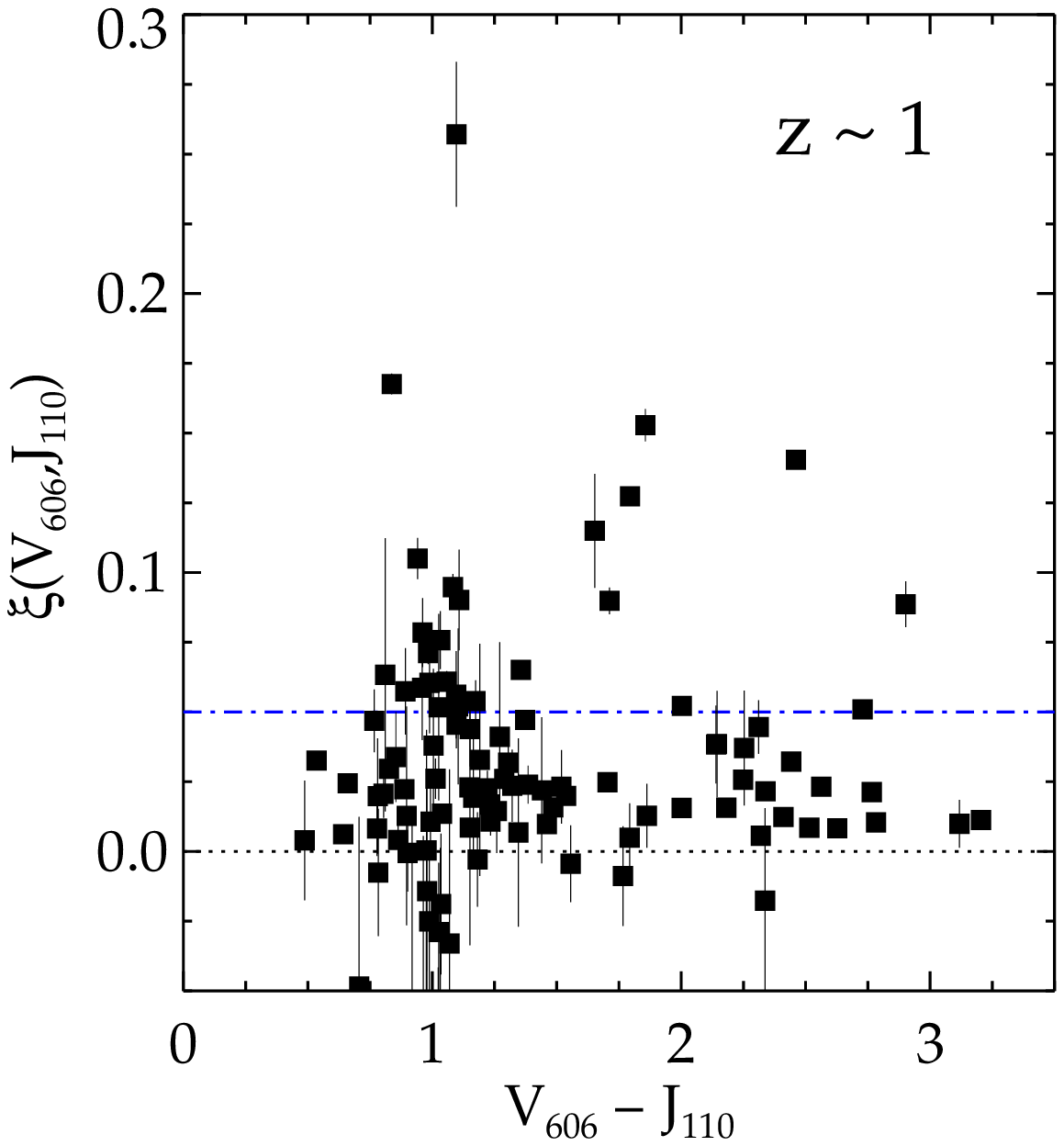}{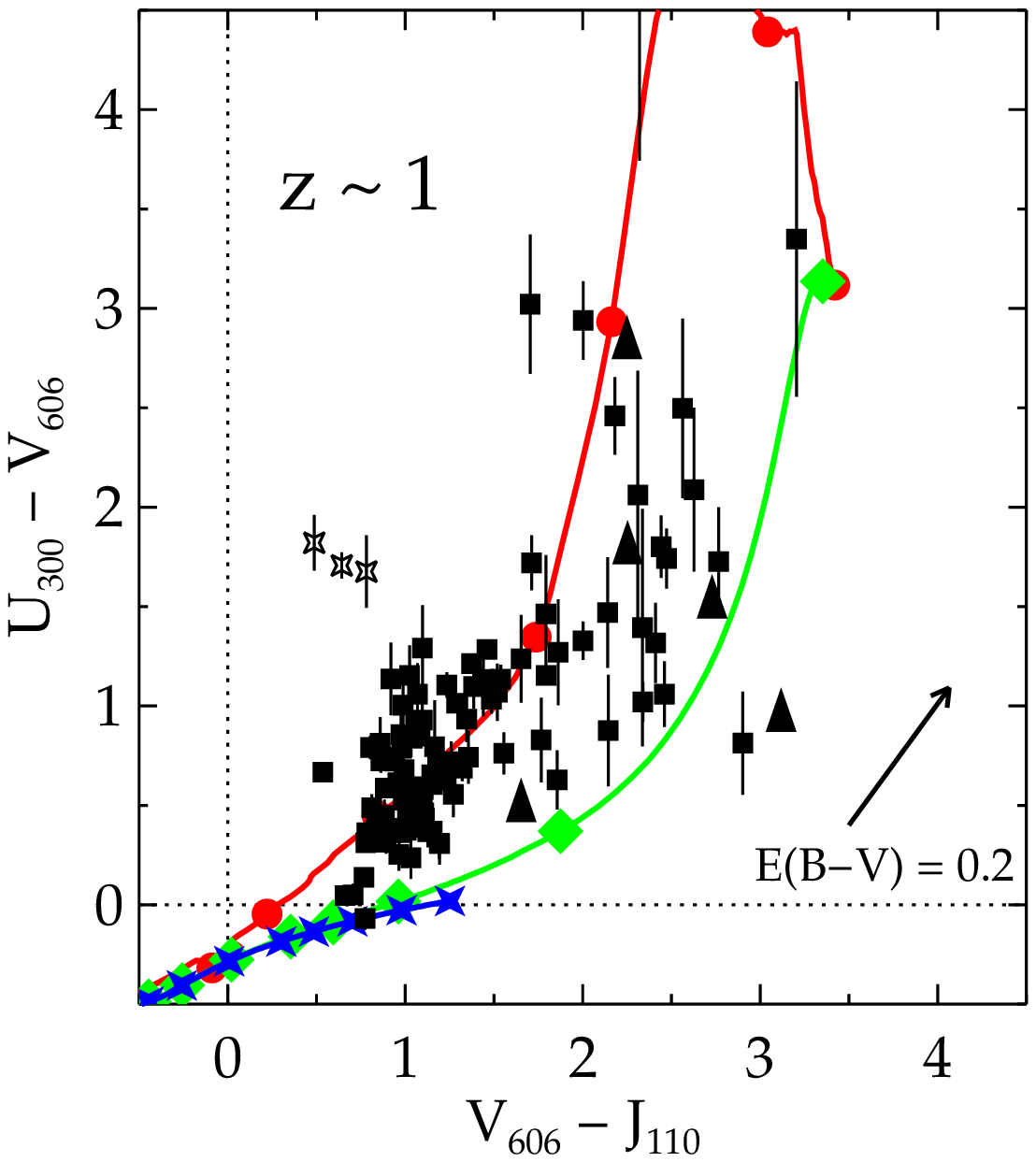}}
\vspace{0.1in}
\vbox{\plottwo{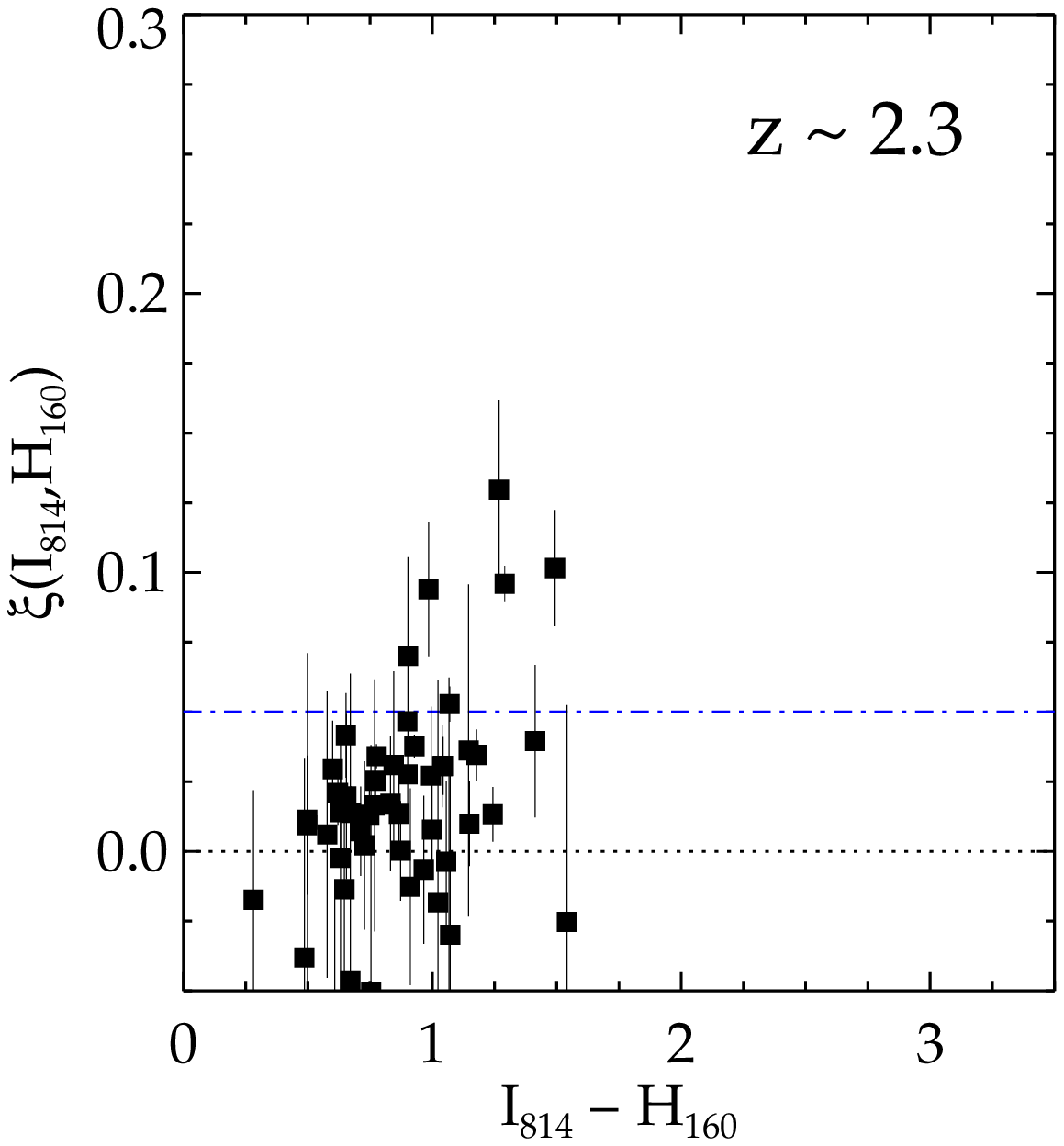}{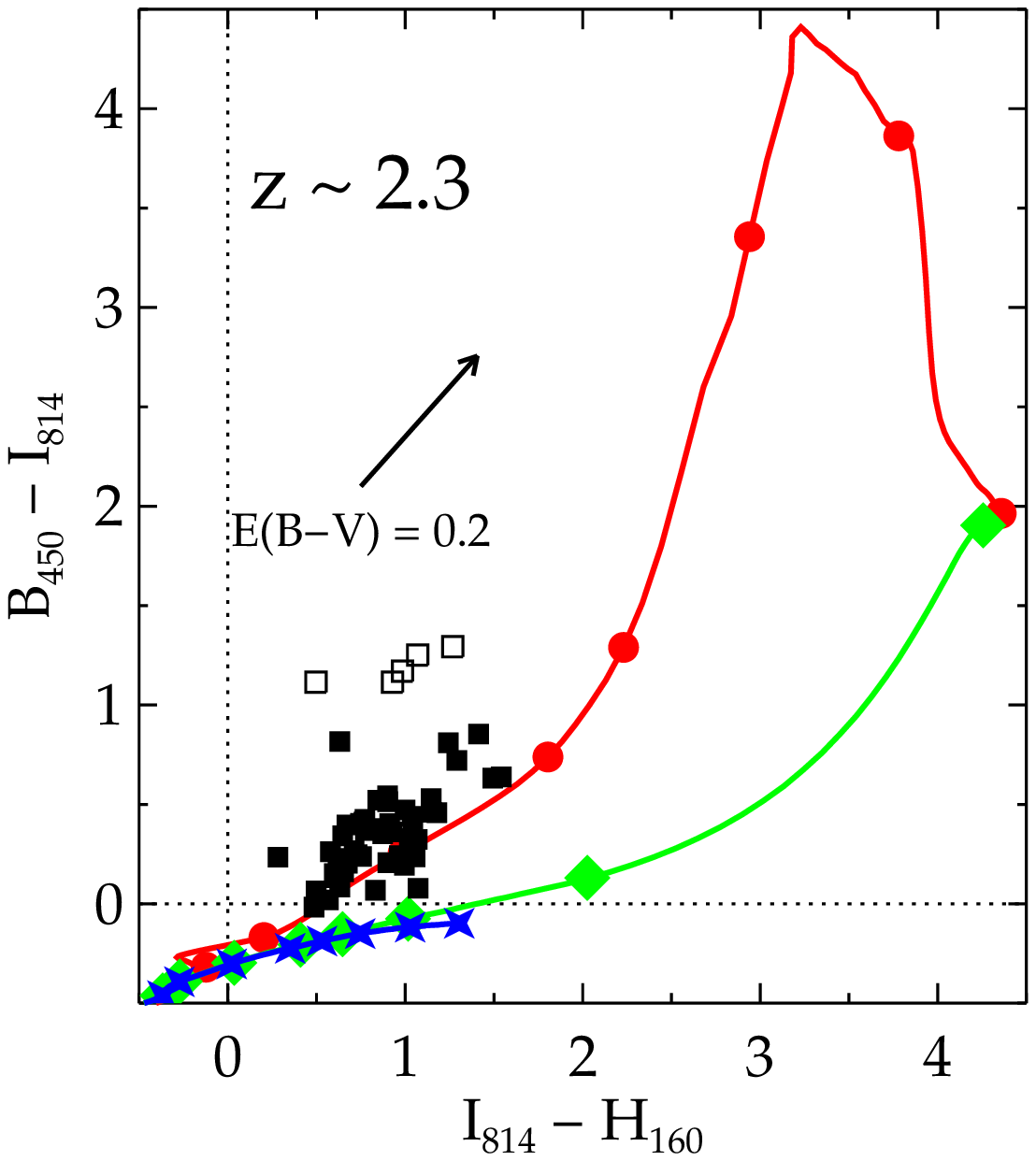}}
\epsscale{1.0}
\caption{\figtencap}
\end{figure}
\else
\fi

In contrast, the $z\sim 1$ galaxies show a greater scatter in their
total colors. A significant fraction of the population has $\wfu -
\wfv$ and $\wfv-\nicj$ colors that are well represented by stellar
populations with declining star--formation histories, and ages greater
than several gigayears.  The $z\sim 1$ galaxies have colors spanned by
the range of the models, including moderate dust extinction.  Many of
these objects have very red colors, $\wfu - \wfv \gsim 1.5$, $\wfv -
\nicj \gsim 2$, and $\mathrm{S/N}(\wfu) < 10$, which implies their
emission at UV--optical wavelengths is dominated by older stellar
populations with little contribution from recent star formation.
Several galaxies have rest-frame UV colors that are apparently
inconsistent with the models (\ie, galaxies with $\wfu-\wfi \gsim 1.5$
and $\wfv - \nicj \lsim 1$; indicated by open symbols in the plot).
Two of these objects, HNM 748 and 1078, have spectroscopic redshifts
$\simeq 0.9$, but photometric redshifts of 0.43 and 2.7, respectively.
Thus the spectroscopic values (which were used to define the galaxy
samples) are somewhat dubious.  The other object, HNM 1141, has a
photometric redshift with maximum likelihood at $z_\mathrm{ph} = 1.2$,
but with a poor goodness of fit.  There is a second maximum near
$z_\mathrm{ph}\sim 3$ that has a nearly--equal likelihood.  Given
their redshift ambiguities, we will exclude these objects in
subsequent analysis.

There is no clear correlation between total UV--optical color and
internal color dispersion. While the $\xi(\wfi,\nich)$ values of the
$z\sim 2.3$ galaxies are generally low, the galaxies with highest
internal color dispersion ($\xi \gsim 0.05$) have the reddest total
UV--optical colors ($\wfi - \nich \gsim 1$ and $\wfb - \wfi \gsim
0.5$).  This suggests that the red colors are due to either a mix of
the stellar populations and/or variations in dust attenuation in the
galaxies.  At $z\sim 1$, many of the objects with high internal color
dispersion have moderate rest--frame UV--optical colors, $\wfv -\nicj
\sim 1$.  This is similar to the trend between the internal color
dispersion and UV--optical color in local galaxies (see P03, their
figure~5), and suggests that high internal color dispersion results
from a heterogeneous mix of young and old stellar populations (which
produce the moderate total UV--optical colors) that are spatially
segregated within the galaxies.  The increase in the
internal color dispersion for the galaxies in the $z\sim 1$ \hdf\
sample --- as well as the increased scatter in total colors of the
galaxy population --- implies that the $z\sim 1$ galaxy population has
more diversity in its stellar populations compared to $z\sim 2.3$.  We
interpret this as evidence for old stellar populations that are not
prominent at $z \gsim 2$.

Previous authors have studied the spatially resolved colors and color
gradients of distant galaxies in order to interpret the
star--formation histories of the stellar populations in these objects
\citep[\eg,][]{abr99,cor01,men01,men04,mot02}.  These studies have generally
concluded that the spatial colors within galaxies reflect the
evolutionary histories and internal dynamics of the galaxies' stellar
populations.  \citet{abr99} and \citet{men01,men04} argued that the
internal UV--optical colors are more sensitive to the relative ages of
the galactic stellar populations than either dust or metallicity
effects, and homogeneity in an object's internal color distribution is
indicative of a broadly co-eval history for the galaxy's stellar
populations.   We reach a similar conclusion in our analysis of the
local galaxy sample (P03).  A more diverse star--formation history for
a galaxy's stellar populations produces heterogeneous internal colors.
For example, by investigating the internal colors of high--redshift ($z
\lsim 1$) galaxies in \hst\ images, both \citet{abr99} and
\citet{men01} find a large fraction ($\gsim 30-40$\%) of early--type
galaxies in the field that show evidence for recent star formation activity.

\ifsubmode
\else
\begin{figure*}[t]
\epsscale{1.0}
\plottwo{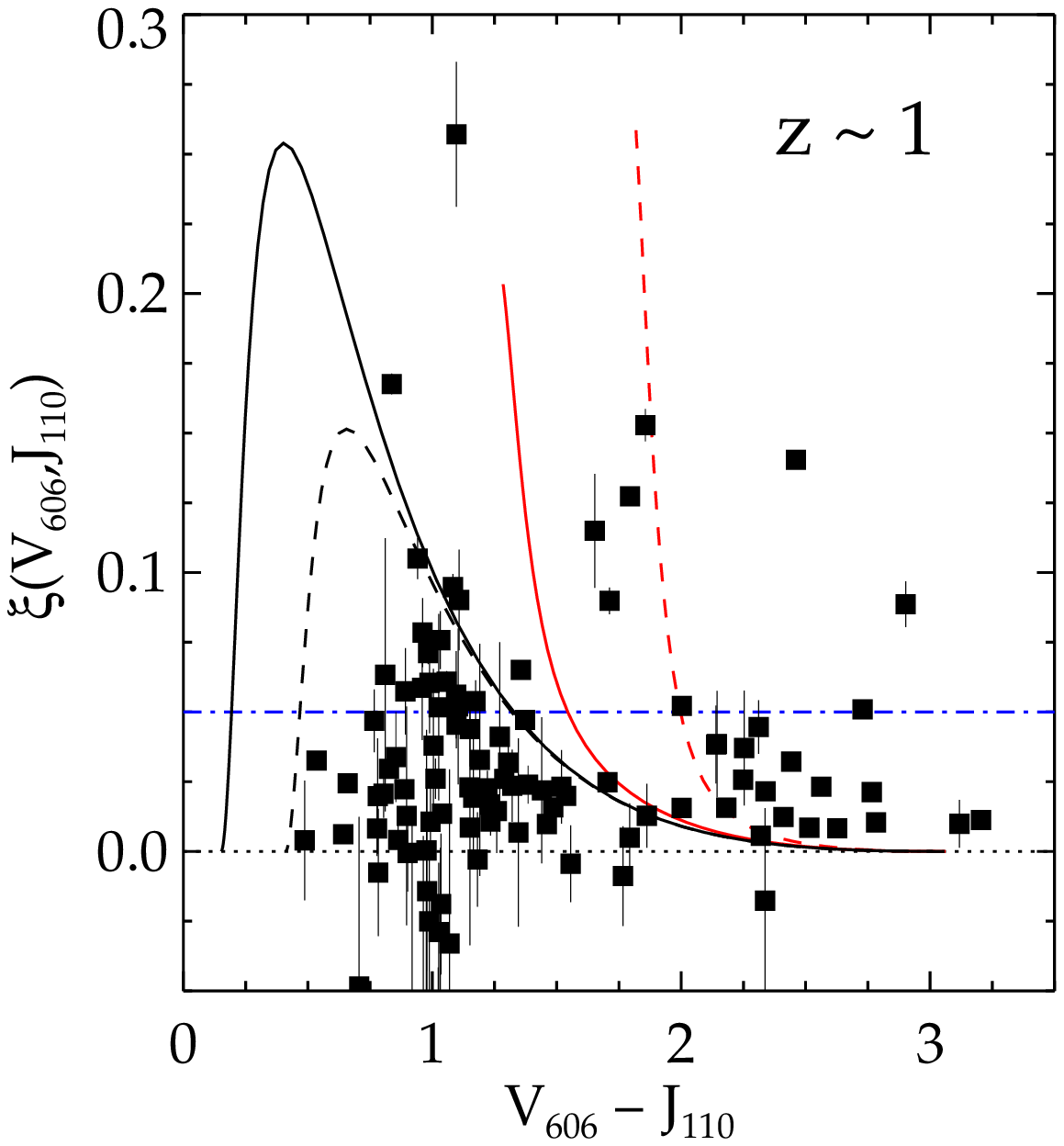}{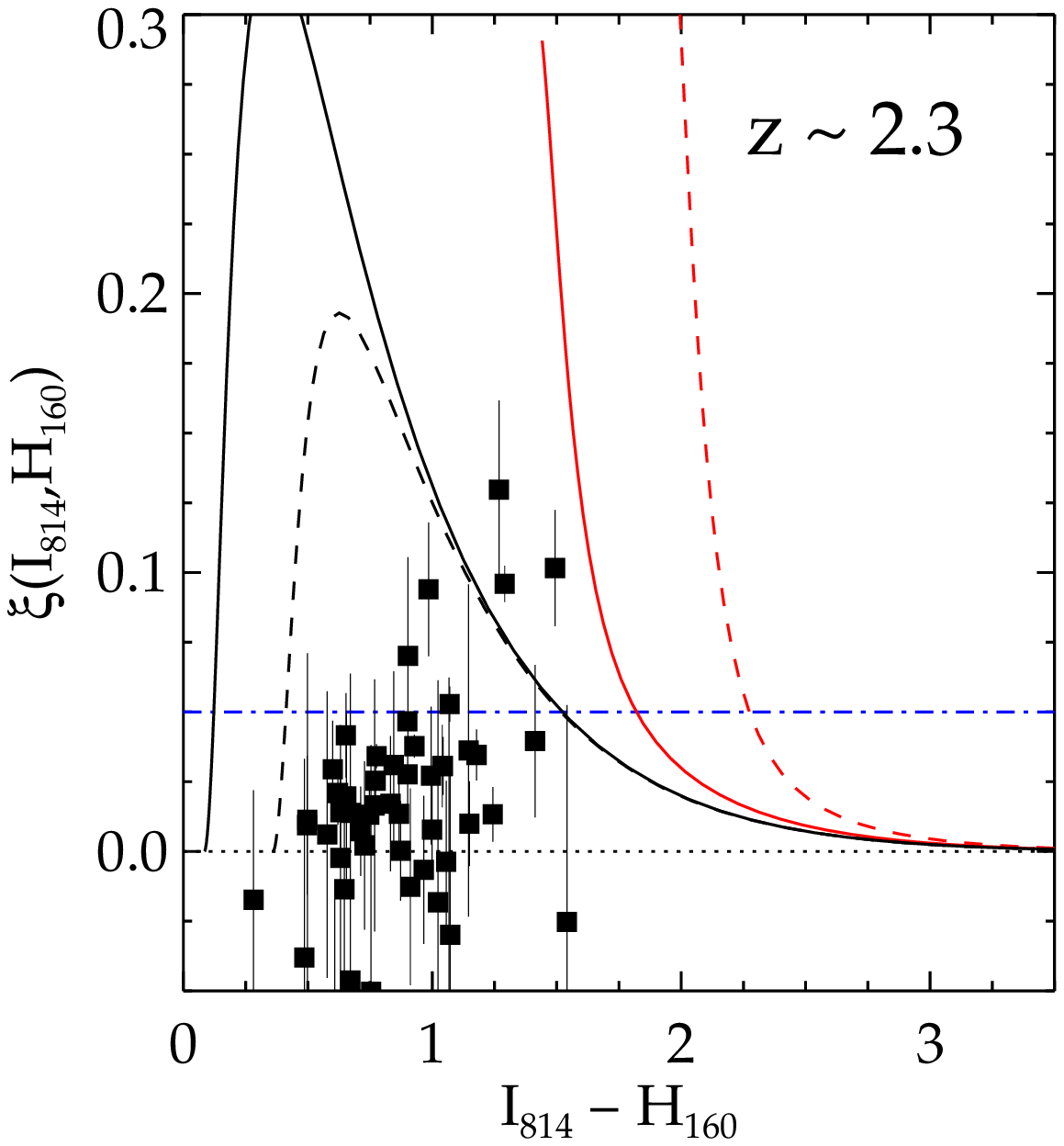}
\epsscale{1.0}
\caption{\figelevencap}
\end{figure*}
\fi

%These studies only constrain the presence of nascent star formation
%without placing it in the context of the galaxy assembly history.  

Clearly, galaxies with very strong internal color gradients and
heterogeneities as studied by these authors would probably also have
large internal color dispersion.  A small amount of ongoing star
formation in an already massive galaxy can dominate its UV flux
emission and give rise to strong variations in a galaxy's internal
colors, even though it may constitute a negligible total to the
galactic stellar mass (\eg, the ``frosting'' scenario of young stars
in elliptical and lenticular galaxies; see Trager \etal\ 2000). Our
measure of the internal color dispersion has an advantage that it can
quantify the amount of ongoing star forming relative to the previously
formed populations (see discussion below).  

%%%%%%%%%%%%%%%%%%%%%%%%%%%%%%%%%%%%%%%%%%%%%%%%%%%%%%%%%%%%%%%%%%%%%%

% edited to here

\subsection{Simulations of Galaxy Internal Colors\label{section:models}}

A fraction of the $z\sim 1$ galaxies in \hdf\ have morphologies of
early--to--mid-type spiral galaxies, and these objects exhibit the
highest internal color dispersion (see
figures~\ref{fig:montage_xi_lowz}).  This agrees with trends between
morphology and the internal color dispersion observed in local
galaxies (P03).

In most hierarchical merging models, the growth of rotationally
supported galaxy disks is expected as gas cools and accretes toward
the center of the underlying dark--matter halos \citep[\eg,][]{mo98},
and this is broadly supported by the evolution in galaxy sizes (see
\S~\ref{section:sizes}).  This process is expected to form
star--forming disks around spheroids produced in merger remnants, and create
spiral galaxies, which broadly reproduces the observed distribution of
colors, gradients, and masses in $z\sim 2-3$ galaxies
\citep{con98,ste02}.
% add something here about old stellar haloes around dwarfs/MW?

To investigate how the formation of new disks around older spheroids
affects the internal color dispersion, we considered a simple model to
represent the galaxies  at $z=1$ and 2.3.  We simulated galaxies with
two components, an old spheroidal component and a young star--forming
disk component, and assigned to these colors using the \citet{bru03}
stellar population synthesis models and assuming a Salpeter IMF and
solar metallicity (see Table~\ref{table:models}).  The spheroidal
component was given the colors of a stellar population formed in an
instantaneous burst viewed at an age of 2.8~Gyr and a $r^{1/4}$--law
surface--brightness profile.   The selected age corresponds to the
maximum possible at $z=2.3$, and approximately to the elapsed cosmic
time between $z=2.3$ and 1, and the half--light radius is consistent
with the observed properties of the $z\sim 2.3$ galaxy sample and
with the derived sizes of local galaxy bulges \citep{car99}.  The disk
component has the colors of relatively young stellar populations (see
Table~\ref{table:models}), and an exponential surface--brightness
profile.  The disk brightness is incrementally increased relative to
that of the spheroid to span the full range of bulge-to-disk ratios.

\ifsubmode
\begin{deluxetable}{lccccccc}
\tablethree
\end{deluxetable}
\else
\begin{deluxetable*}{lccccccc}
\tablethree
\end{deluxetable*}
\fi
\ifsubmode
\begin{figure}
\epsscale{1.0}
\plottwo{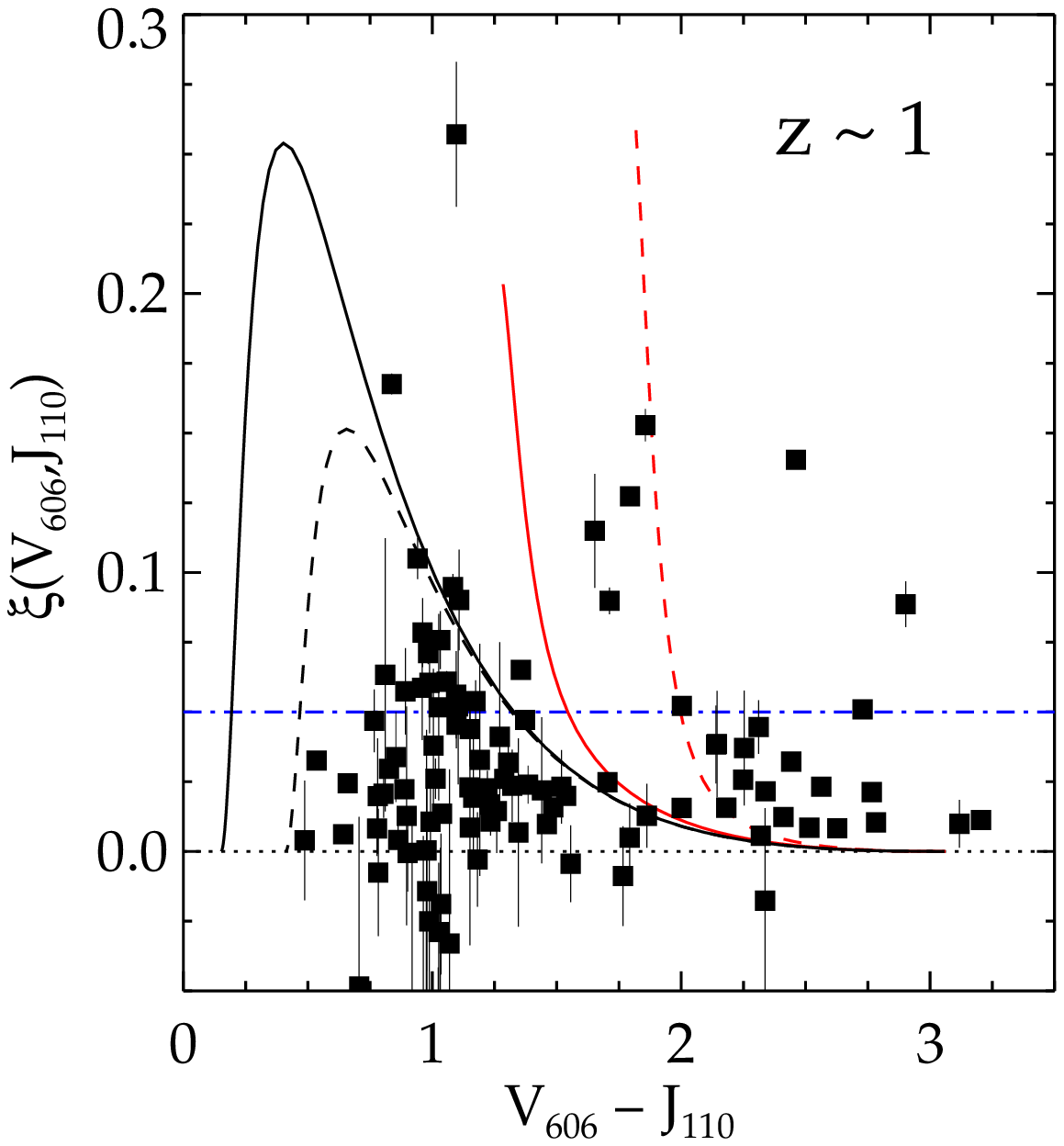}{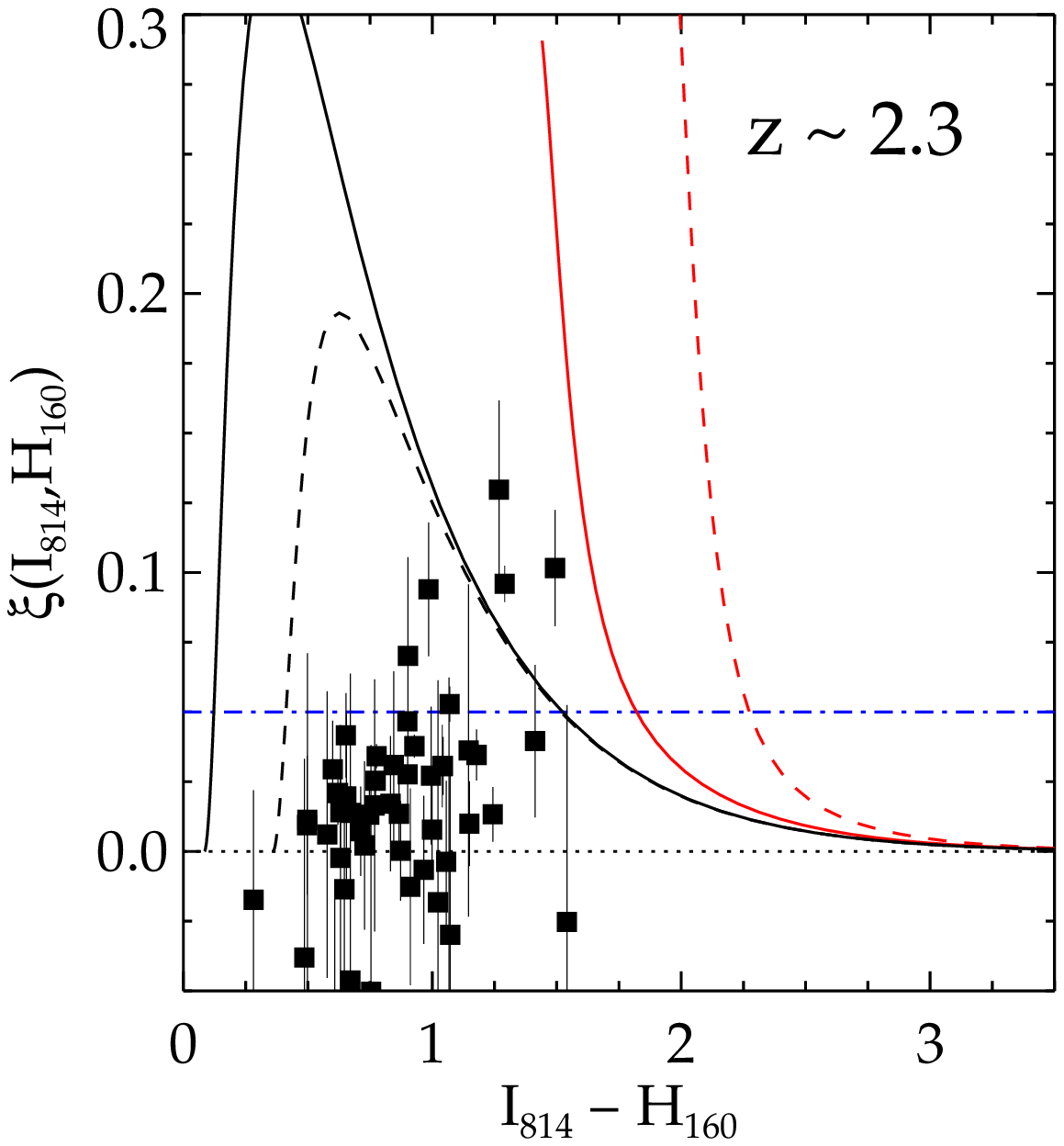}
\epsscale{1.0}
\caption{\figelevencap}
\end{figure}
\else
\fi

The total colors and internal color dispersion of these models are
plotted in Figure~\ref{fig:colorvxi_models} with the data.
Interestingly, the same model predicts larger $\xi(\wfi,\nich)$ values
at $z = 2.3$ relative to $\xi(\wfv,\nicj)$ at $z=1$ by $\approx 40$\%.
The $\wfi-\nich$ color at $z = 2.3$ is bluer in the rest frame and
covers a longer rest--frame wavelength baseline than the $\wfv-\nicj$
color at $z=1$.  Therefore, the $\wfi-\nich$ is slightly more
sensitive to the UV--optical color dispersion, and an object with the
same intrinsic color dispersion would have a greater measured
$\xi(\wfi,\nich)$ at $z=2.3$ than $\xi(\wfv,\nicj)$ at $z=1$.  This
actually \textit{strengthens} the conclusion that the range of the
UV--optical internal color dispersion increases from $z\sim
2.3$ to 1.

\ifsubmode
\else
\begin{figure}[tbh]
\plotone{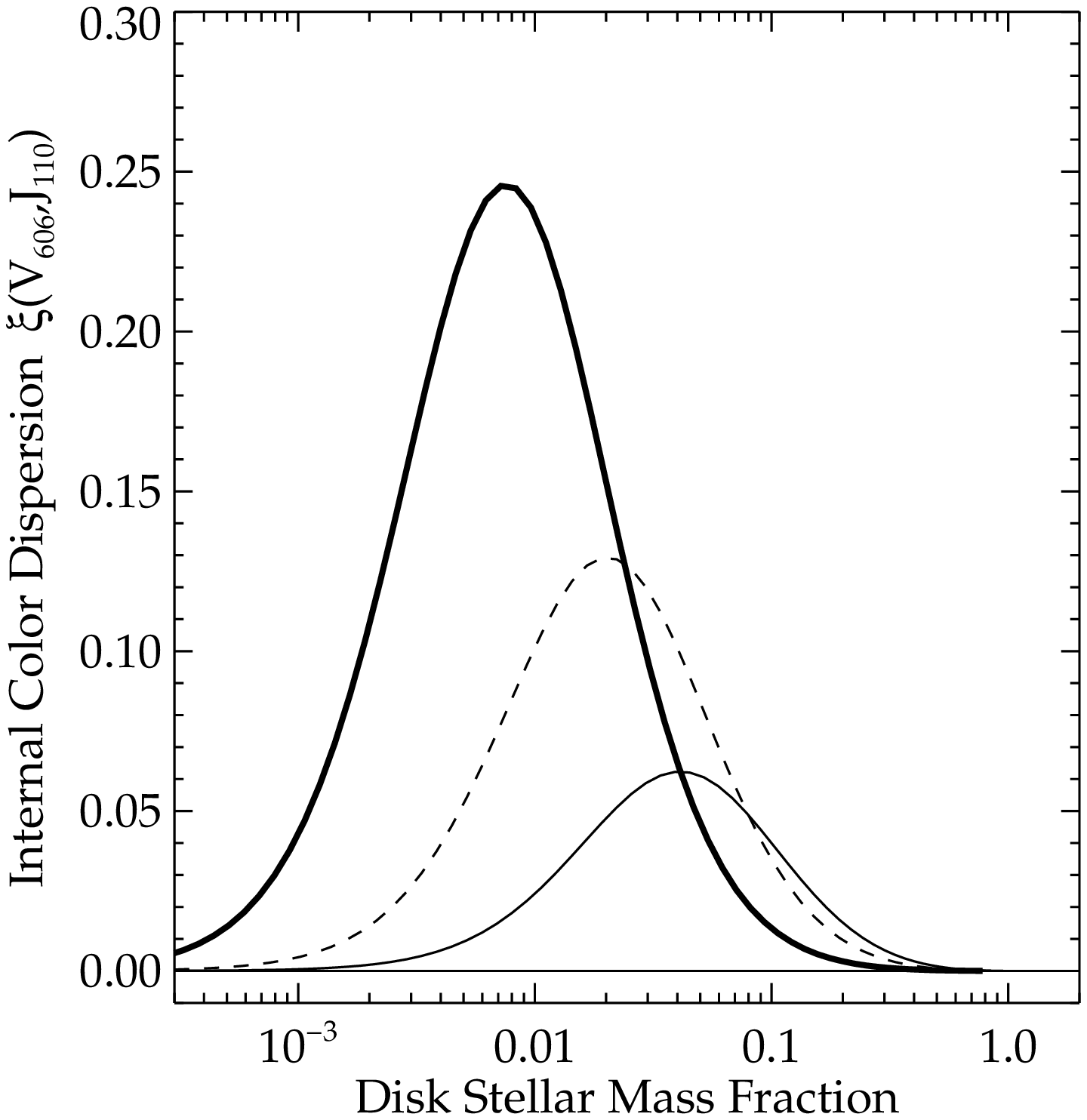}
\caption{\figtwelvecap}
\end{figure}
\fi

\ifsubmode
\else
\begin{figure*}[thb]
\plottwo{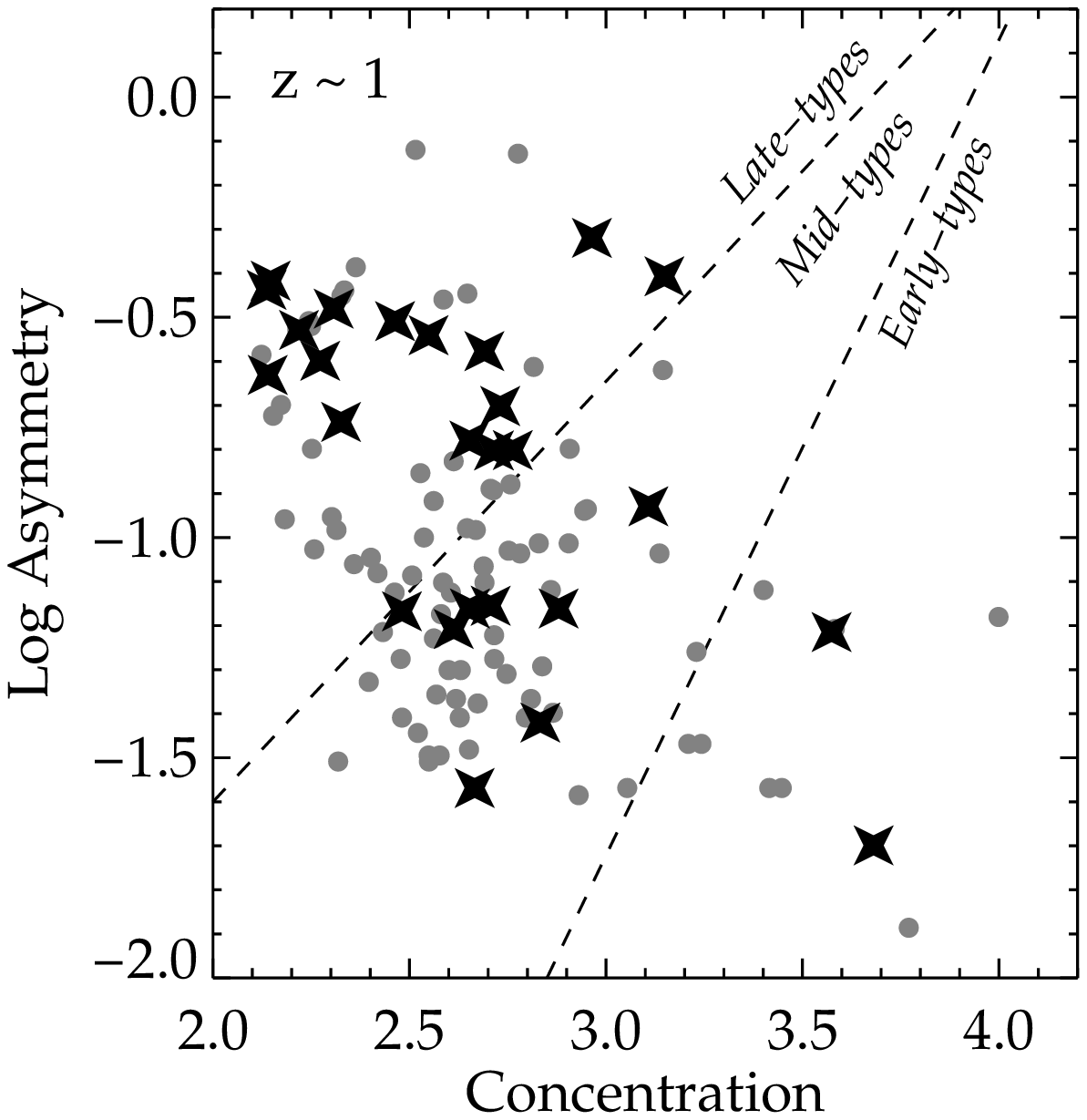}{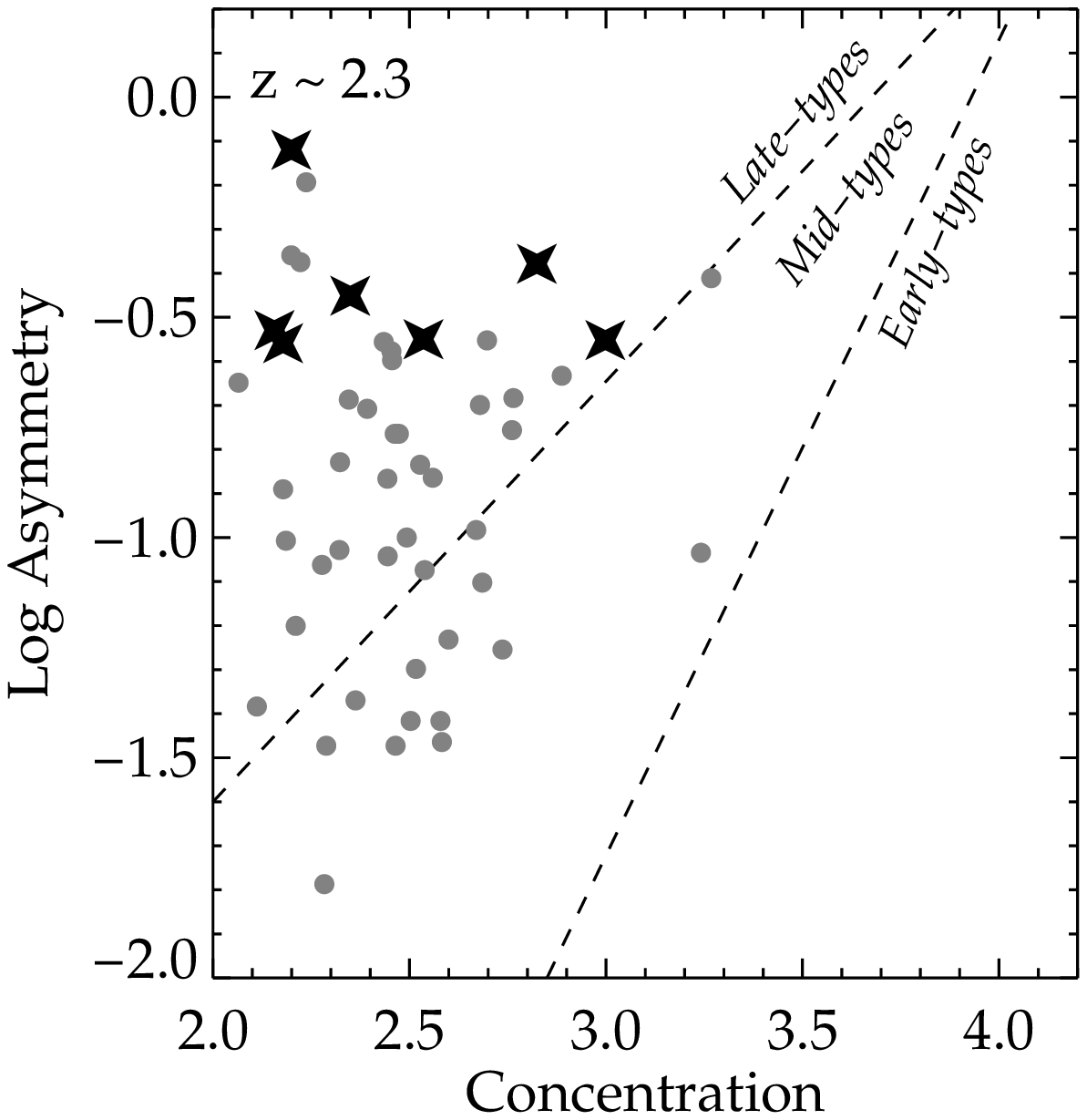}
%\epsscale{1.0}
\caption{\figthirteencap}
\end{figure*}
\fi

The simple models of a star--forming disk surrounding an older central
bulge span the range of total color and internal color dispersion
observed in most of the  galaxies at $z\sim 1$ that have moderate
UV--optical colors.   At low and high disk--to--bulge ratios, the
model colors are nearly uniform (i.e., dominated by the disk and bulge
respectively).  The internal color dispersion is largest for mixed
models.  The stellar populations in these galaxies are consistent with
this scenario.

However, there are many galaxies in the $z\sim 1$ sample with very red
colors ($\wfv - \nicj \gsim 1.5$) and high internal color dispersion,
and these are not accounted for by the simple models. The light from
these objects must be dominated by relatively evolved stellar
populations, possibly with small traces of ongoing star--formation to
induce large internal UV--optical color dispersion \citep{tra00}.  To
model this possibility, we considered the spheroid and disk models
with an additional source of very young star-forming knots
(``\ion{H}{2} regions'') typically located in galactic spiral arms and
starbursts  (described in Table~\ref{table:models}).  These models,
plotted in Figure~\ref{fig:colorvxi_models}, reproduce the high
internal color dispersion with redder total colors than the pure
disk/bulge models.  
%Therefore, the galaxies at $z\sim 1$ with red
%colors and high internal color dispersion are consistent with added
%amounts of embedded star formation within galaxies otherwise dominated
%by more--evolved stellar populations.  
The high internal color dispersion
is produced from spatially segregated and heterogeneous stellar
populations.  We therefore conclude that the increase in the internal
color dispersion between $z\sim 2.3$ to 1 corresponds to greater
diversity in the amounts of young and old stellar populations in these
galaxies.

The best example of this behavior in the $z\sim 1$ \hdf\ sample is the
giant, ``grand--design'' spiral galaxy, HNM 1488 (see
figures~\ref{fig:montage_lowz1} and \ref{fig:montage_xi_lowz}). Bulge
and possible bar features dominate the rest--frame $B$--band (\nicj\
image).  In the rest--frame UV (\wfv), these features fade relative to
the disk, spiral arms, and prominent star--forming knots embedded in
the disk.  The redder bulge dominates the rest-frame optical light,
giving a red total color, $\wfv - \nicj = 1.8$. The internal color
dispersion is very high, $\xi(\wfv,\nicj) = 0.13$, and arises
primarily from the color differences between the bulge, disk, and
\ion{H}{2} regions (Figure~\ref{fig:montage_xi_lowz}).  Based on the
models, these three features are required to produce both the red
total color and high internal color dispersion.

\ifsubmode
\begin{figure}
\plotone{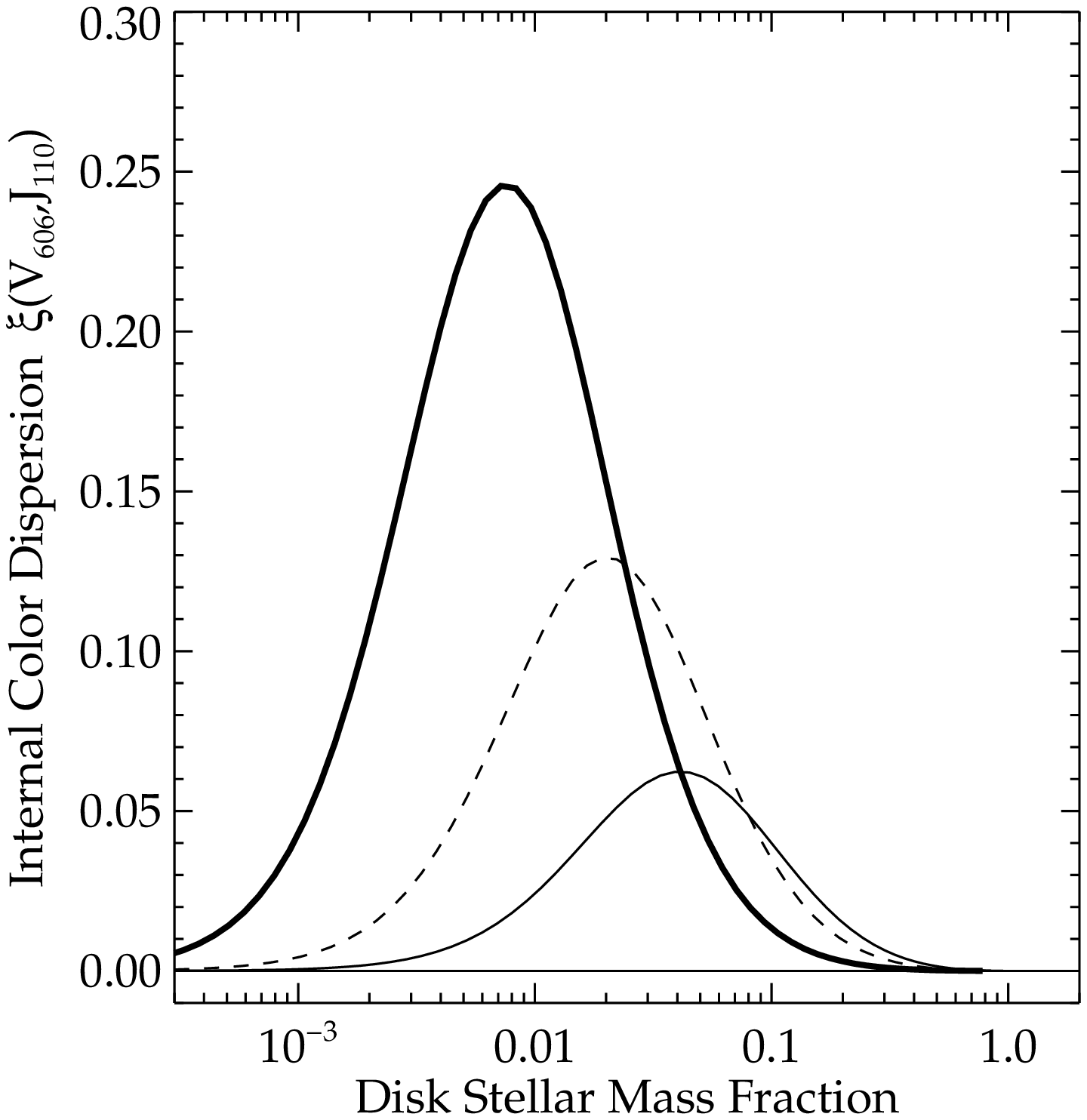}
\caption{\figtwelvecap}
\end{figure}
\else
\fi

The different galaxy components described in the simulations above
have very different stellar-mass--to--blue-light ratios, as indicated
in Table~\ref{table:models}.  Figure~\ref{fig:sim_massratio} shows the
internal color dispersion of the models as a function of the relative
amount of stellar mass in the different model components.    For the
disk models listed in Table~\ref{table:models}, the stellar mass of
the bulge exceeds that in the disk by up to a factor of $\sim 10-40$
when the luminosities of the bulge and disk are equal, and the internal
color dispersion reaches a maximum near this point.  Based on
Figure~\ref{fig:sim_massratio}, high values of the
internal color dispersion ($\xi \gsim 0.05$) occur when $\lsim 10$\%
of the stellar mass resides in recently formed stars (see also P03).

\ifsubmode
\begin{figure}[bt]
\plottwo{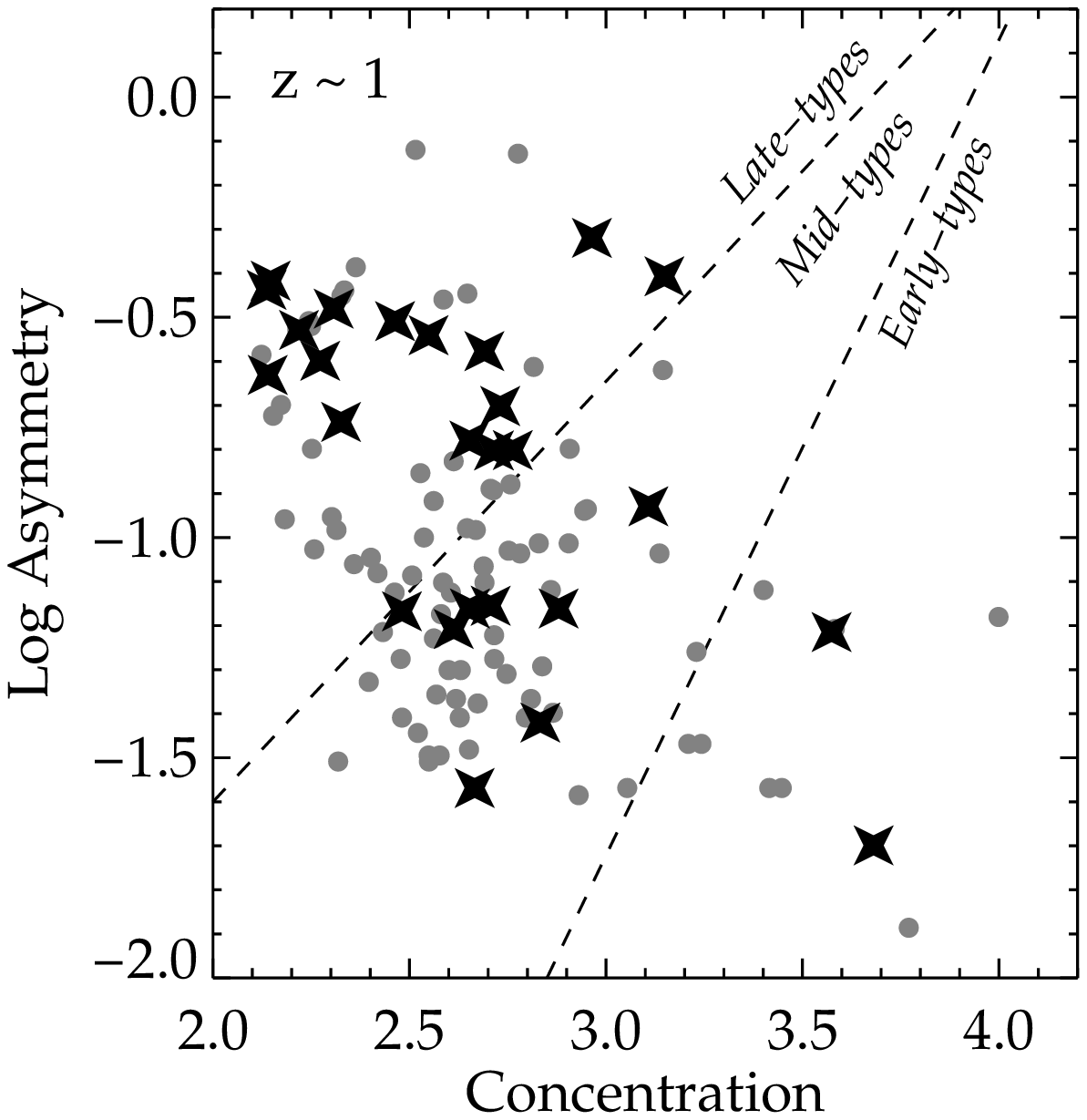}{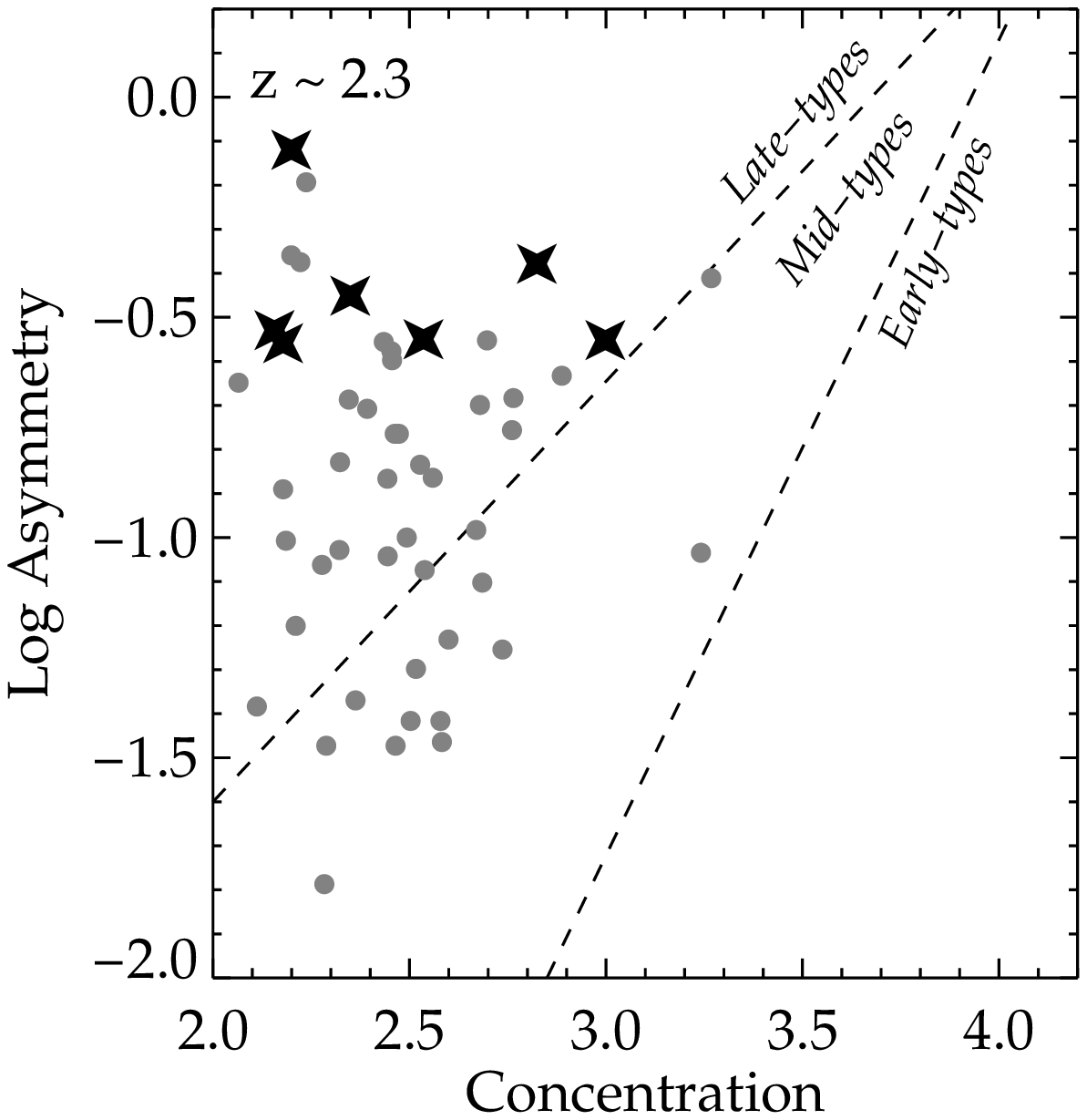}
%\epsscale{1.0}
\caption{\figthirteencap}
\end{figure}
\else
\fi

\subsection{Internal Color Dispersion as a Function of Morphology}

The increase in range of the UV--optical internal color dispersion
coincides with the emergence of large, early-- and mid--type
Hubble--sequence galaxies.  For example, \citet{abr99} and
\citet{con03} have shown that galaxy morphological type is generally
distinguished by their concentration and asymmetry indexes \citep[see
also][]{con03b}.  If the formation of Hubble--sequence galaxies
corresponds to an increase in the scatter of the UV--optical color
dispersion, then one expects there to be a relation between the
internal color dispersion and these morphological indicators.
Figure~\ref{fig:ca} shows the concentration and asymmetry values along
with internal color dispersion for the two \hdf\ samples
\citep{con03}. At $z\sim 2.3$ galaxies with high UV--optical internal color
dispersion occupy the upper range of asymmetry values, which is
indicative of irregular star--forming galaxies possibly due to
interactions and mergers \citep{con04b}. This suggests that the
internal color dispersion results from variations in the stellar
populations and dust opacity, \eg, HNM 813 + 814 (see
figures~\ref{fig:montage_hiz1} and \ref{fig:montage_xi_hiz}).  Because
galaxies with relatively high internal color dispersion correspond to
highly asymmetric galaxies, this supports the notion that they are
strongly interacting and/or merging.  At $z\sim 1$ galaxies with high
internal color dispersion span a range of morphological types as
traced by concentration and asymmetry.   There is still a group of
galaxies with high internal color dispersion and high asymmetry.
These objects are similar to the higher redshift star--forming
galaxies, and is consistent with a recent analysis of
\hst/ACS images, which suggests that star-forming galaxies at $z\sim
1.5$ and 4 have similar far--UV (1500~\AA) morphologies (Lotz \etal\
2005).   There are also a large number of galaxies with low asymmetry,
and higher concentration, which suggest that the high internal color
dispersion in these objects occurs in otherwise normal mid/early--type
galaxies.  It seems that the emergence of Hubble--sequence galaxies
coincides with larger internal color dispersion, and thus the assembly
of diverse, spatially segregated stellar populations.

Two of the galaxies, HNM 1214 and 1453, are classified as early--types
by high concentration and low asymmetry, and also have moderately high
internal color dispersion ($\xi \simeq 0.08$).  Upon visual inspection
in both cases very faint blue companions are visible at distances of
$\lsim 1\arcsec$, which were not split by SExtractor and increase
the internal color dispersion.  The companion near HNM 1214 is visible
roughly one arcsecond to the left of the galaxy in
Figure~\ref{fig:montage_lowz1}.  Companions only affect the internal
color dispersion in very few of objects (these two and HNM 1453, which
was eliminated from the sample for this reason, see
\S~\ref{section:icd}). The large internal color dispersion in the
majority of objects results from spatially segregated heterogeneous
stellar populations.

\subsection{Formation of Stellar Diversity in Galaxies from $z\sim
1-3$}\label{section:sf_mode} 

Many of the $z\sim 1$ galaxies have red total colors that are consistent with
the passively evolving stellar--population models with ages older than
$1-5$~Gyr.  Because these ages are greater than the
elapsed cosmic time from $z\sim 2.3$ to 1, some of the stellar
populations currently in $z\sim 1$ galaxies must have been present in
progenitors at the earlier epoch.  However, due to the large growth in
galaxy stellar mass and size over these redshifts, the $z\sim 1$
galaxies were generally not fully formed by $z\gsim 2$.  

%Conceptually, if the ancestors of the $z\sim 1$ population have simple 
%star--formation histories (\eg, instantaneous bursts or exponentially
%declining star--formation histories), and if these stellar populations
%are fully mixed, then they would exhibit no variation in their internal color
%dispersion.   

In addition to evolution in size and mass, the increase in the range
of galaxy internal color dispersion from $z\sim 2.3$ to 1 is evidence
that the diversity in the stellar populations in galaxies is
increasing.  Papovich \etal\ (2001) found that the relatively short
ages of galaxies at $2 \lsim z\lsim 3.5$ (less than a few hundred
megayears) coupled with the lack of older faded ``post-burst'' remnants
implied that these galaxies have recurring star--formation
episodes. The time scales of these episodes correspond to $\lsim
1$~Gyr such that the newly--formed stellar populations dominate the
rest--frame UV--optical wavelength regions of the galaxy SEDs.  One
cannot exclude the possibility that older (and possibly more
heterogeneous) stellar populations are hidden beneath the ``glare'' of
the young stars.  This is broadly consistent with small range of low
internal color dispersion observed in the $z\sim 2.3$ sample, but with
the additional constraint that the nascent stellar populations in
these galaxies dominate both the total galaxy colors and the internal
galaxy colors, at least over scales of $\simeq 1.5$~kpc (the NICMOS
resolution at these redshifts).

If either recurrent bursts such as those described above or
quasi--constant star formation is the norm for galaxies at $z\sim
2.3$, then eventually these galaxies will form sufficient stellar
mass, which will dominate the emitted optical luminosity.
Strong UV--optical internal color dispersion generally requires that
$\gsim 80$\% of the total stellar mass resides in older stellar
populations (see \S~\ref{section:models}).   At $z= 2.3$, the age
of the Universe is less than 3~Gyr, and thus there has been little
opportunity for enough star--formation events to have instilled strong
heterogeneity in the galaxies' stellar populations.  At $z\sim 1$ the
age of the Universe roughly doubles to $\sim 6$~Gyr, which permits a
greater number of star--formation episodes with this timescale and
allows for greater internal color dispersion.   
The increase in the internal color dispersion for $z\sim 2.3$ to 1 seems 
to require that enough time has elapsed for galaxies to have formed a
sufficient amount of their stellar mass relative to their current
star--formation rate such that the older stars dominate the optical
luminosity. 

The discussion above ignores the effects of dust extinction in
star--forming regions, which can contribute to the overall
``patchiness'' in galaxy internal colors and produce strong
morphological $K$--corrections \citep{win02}.  However, studies of the
variation in internal colors \citep{abr99,men04} and the internal
color dispersion of local galaxies (P03) broadly conclude that these
diagnostics are more sensitive to variations in the ages of the galaxy
stellar populations relative to variable dust opacity. Therefore,  we
attribute the strong variation in galaxy internal color dispersion
primarily to differences in the ages of the galaxy stellar populations.  

Because the half--light radius is growing from $z\sim 2.3$ to
1 (see \S~\ref{section:sizes}), we know that galaxies do not form
\textit{in situ} and evolve passively over this redshift range.
Furthermore, as the size increases, galaxies at later epochs will have
proportionally more young stars at larger radii.
% XXXX something about dwarfs, MW halo here?
So, size evolution will naturally lead to color gradients.   This is
supported by the fact that the $z\sim 1$ galaxies with the largest
internal color dispersion appear to have red centrally concentrated
cores with bluer regions at larger radii (see
Figure~\ref{fig:montage_xi_lowz}).

% edited to here

Although the large internal color dispersion values at $z\sim 1$ imply
a large stellar diversity at this epoch, the same is not the case at
$z\sim 2.3$.  Based on the simple models presented in
\S~\ref{section:models}, although there is sufficient time at $z = 2.3$ for
galaxies to have developed a heterogeneous stellar content with
significant internal color dispersion, none is observed (see, for
example, Figure~\ref{fig:colorvxi_models}). Therefore, we conclude
that galaxies at $z\gsim 2$ are \textit{not} forming a diverse stellar
content as rapidly as is allowed purely from timescale arguments.
This implies that galaxies at these epochs are not assembling their
stellar populations in the quasi--continuous, inside-out models
described in \S~\ref{section:models}.

One alternative to account for the fact that the $z\sim 2.3$ galaxies
do not exhibit strong internal color dispersion is that intense,
discrete starbursts such as those associated with actively merging
galaxies dominate the star--formation histories at this epoch.
Intense starbursts naturally dominate the UV--optical light at the
time the galaxy is observed, which accounts for the observed
properties of the galaxies at $z\sim 2-4$
\citep{low97,saw98,ste99,pap01,pap04}.  Furthermore, if these starbursts
arise from strong interactions or mergers, then during the merger one
expects some internal color dispersion to result from any previously
formed, heterogeneous stellar populations with some contribution from
varying dust opacity \citep[see, \eg, the large heterogeneity in both
the stellar populations and dust extinction in NGC
4038/4039;][]{whi99}.  The merger process will tend to erase any
diversity in the constituent stellar--populations by fully mixing them
on short timescales \citep[\eg,][]{mih96,ste02}, and thus the merger
product should have largely homogenized stellar populations. This is
consistent with the small derived values for the internal color
dispersion at $z\sim 2.3$.  The period while galaxies are strongly
interacting and/or actively merging may produce patchy star formation
with larger internal color dispersion.  Indeed, the objects at $z\sim
2.3$ with high internal color dispersion ($\xi \sim 0.05 -0.1$) have
large asymmetries (Figure~\ref{fig:ca}), supporting the hypothesis
that they are actively merging. We therefore interpret the
observations as evidence for major mergers associated with strong
starbursts.   Furthermore, these events probably occur frequently,
with the timescale between bursts $\lsim 1$~Gyr, in order to keep the
stellar content in these galaxies homogenized and to keep the total
UV--optical colors dominated by the young starburst.

By $z\sim 1$ the internal color dispersion in galaxies has increased
significantly, which we interpret as an increase in the diversity of
galaxies' stellar content.  This occurs partly because galaxies
have formed sufficient mass in older stellar populations that recent
star forming episodes contribute less to the total optical emission
(while still dominating the UV light; see the discussion above).
However, the heterogeneous stellar populations are spatially
segregated: the $z\sim 1$ galaxies with the highest internal color
dispersion typically have red central cores surrounded by bluer regions
(see Figure~\ref{fig:montage_xi_lowz}).  This signifies that major
mergers occur less frequently, because any merger activity would tend to
erase this diversity in the configuration of the stellar populations.
The growth of large disks at this epoch naturally increases the
internal color dispersion, and is consistent with the interpretation
of the ubiquitousness of thick disks in local galaxies \citep{dal02}.
Thus, the emergence of galaxies with large internal color dispersion
signifies great diversity in their stellar content, and we interpret
this as evidence for a decline in the major--merger rate in galaxies
at $z\lsim 1.5$ \citep[see also][]{vdb02,con04b}.  

The decline in the importance of bursts with decreasing redshift may
imply that more quiescent modes dominate the total star--formation
rate. This is consistent with the suggestion by van den Bergh (2002)
and reminiscent of the ``collisional starburst'' models of Somerville
et al.\ (2001), where bursts dominate at high redshifts, and quiescent
star formation dominates at lower redshifts ($z\lsim 0.8$).  The
larger range of internal color dispersion for the massive galaxies at
$z\sim 1$ implies that quiescent star--formation plays a more dominant
role at later times compared to major mergers.  The transition between
where these modes dominate occurs between $z\sim 1.4-1.9$.

%%%%%%%%%%%%%%%%%%%%%%%%%%%%%%%%%%%%%%%%%%%%%%%%%%%%%%%%%%%%%%%%%%%%%%

\section{Conclusions}

We have studied the evolution in the morphologies and colors of two
samples of high--redshift galaxies at $z\sim 2.3$ and $z\sim 1$ in
order to investigate how these galaxies assembled their stellar
populations.   From $z\sim 2.3$ to 1, galaxies  grow in size, stellar
mass, total color, and internal color dispersion, which we interpret
as evidence for an increase in the diversity of stellar populations as
galaxies assemble.

The majority of luminous galaxies ($M(B) \leq -20.5$ in the rest
frame) in the $z\sim 1$ sample have rest--frame optical morphologies
that are classifiable as early--to--mid Hubble types.  These galaxies
have regular, symmetric morphologies, and many show strong
transformations between their morphologies when viewed at rest--frame
UV and optical wavelengths.  In contrast, the rest--frame optical
morphologies of the luminous ($M(B) \leq 20.5$ in the rest--frame)
galaxies at $z\sim 2.3$ are irregular and/or compact, and there is
little difference in the galaxy morphologies from rest--frame far--UV
to optical wavelengths.  None of the $z\sim 2.3$ galaxies appear to be
normal Hubble--sequence galaxies.  Because the $z\sim 2.3$ galaxies
show little transformation in morphology from rest--frame UV to
optical wavelengths, the UV components are not generally the small,
star--forming pieces within larger host galaxies. 

The mean galaxy size increases from $z\sim 2.3$ to 1 by roughly 40\%,
in broad agreement with expectations from hierarchical models.
Furthermore, the number density of large galaxies, $r_{1/2} > 3$~kpc
and $M(B) \leq -20$, increases by a factor of $\approx 7$.   We have
tested that the size evolution is robust against surface--brightness
dimming effects.  The half--light radii of the $z\sim 1$ galaxies when
simulated at $z=2.7$ decrease by $< 20$\% compared to their
measured values at $z\sim 1$, but this decrease is small compared to
the observed evolution.  The radius--luminosity distribution of
galaxies at $z\sim 1$ is consistent with the local distribution for
pure luminosity evolution.  However, the size--luminosity, and
size--mass distributions of $z\sim 2.3$ galaxies indicate that they
are not the fully formed progenitors of large, present--day galaxies.
Galaxies at $z\gsim 2$ must continue to build, both in terms of size
and stellar mass.

We have analyzed the rest--frame UV--optical total colors, and the
internal color dispersion using the novel statistic developed in P03.
The internal color dispersion quantifies differences in galaxy
morphology as a function of wavelength, and between rest--frame UV and
optical wavelengths it constrains the amount of current
star--formation relative to the older stellar populations.   Both the
mean and scatter of total color and the internal color dispersion
increase from $z\sim 2.3$ to 1.   At $z\sim 1$, galaxies with high
internal color span a range of  morphological types.  Many are spiral
galaxies, with a few starbursting and interacting systems.  There is
no clear correlation between total color and the internal color
dispersion, although galaxies with the highest internal color
dispersion have moderate total UV--optical colors, which implies a mix
of heterogeneous stellar populations.  The $z\sim 1$ galaxies with the
highest internal color dispersion appear to be early--to--mid-type
spiral galaxies, and they exhibit strong transformations between their
UV and optical rest--frame morphologies. At $z\sim 2.3$ few galaxies
have high internal color dispersion, and those that do also have
interacting and disturbed morphologies, which implies that the
UV--optical light from these galaxies is dominated by young, largely 
homogeneous stellar populations.  We interpret the evolution in galaxy
color and internal color dispersion as evidence that the diversity in
the stellar populations of galaxies is increasing from $z\sim 2.3$ to
1 as the older stellar components build up over time.

To interpret the evolution in the properties of galaxies, we have
modeled the stellar populations of galaxies in spheroidal and disk
components, and star--forming \ion{H}{2} regions.   These simple
models broadly support the conclusion that the range of internal color
dispersion corresponds to increased diversity in the galaxies' stellar
populations.  Galaxies with significant bulge components with
star--forming disks and/or compact, \ion{H}{2} regions produce high
internal color dispersion, and broadly reproduce the colors, sizes,
and internal color dispersion of the high--redshift galaxies.  We find
that large values of the internal color dispersion require galaxy to
have formed  spatially segregated, diverse stellar populations where
old stars dominate the optical light and stellar mass (\ie, $\gsim
80$\% of the stellar mass resides in stars $> 1$~Gyr).  For smaller
fractions of old stars, young stellar populations dominate the
UV--optical internal colors. 

The scatter in the internal color dispersion of the $z\sim 2.3$
galaxies is smaller than what is allowed under these models and basic
timescale arguments.  We interpret this as evidence that brief,
discrete and recurrent starburst episodes dominate the star--formation
history of galaxies at this epoch.  If these arise from strong interactions
or mergers of gas--rich constituents, then they will erase any
heterogeneity in the stellar content that otherwise develops, and this
is supported by the low internal color dispersion observed in galaxies
at this epoch.   The greater range of internal color dispersion that forms
at $z\sim 1$ suggests that major mergers are less frequent and more
quiescent star--formation mechanisms are the norm.   

The greater diversity in the stellar populations of these
high--redshift galaxies coincides with the emergence with large Hubble
sequence galaxies.   Large values of the UV--optical internal color
dispersion require a diverse and spatially unmixed stellar
population, and this occurs when galaxies have formed most of their stellar
mass.  In order to maintain the spatial heterogeneity between the
young and old stellar populations also requires that major mergers are
less common at this epoch.   These conditions all occur at $z\lsim
1.4$, which naturally allows galaxies with Hubble--sequence
morphologies to develop. 

\acknowledgements

We wish to thank our colleagues for stimulating conversations, the
other members of the HDF--N NICMOS GO team who contributed to many
aspects of this program, and the STScI staff for their optimal
planning of the observations and efficient processing of the data.  We
are grateful to Tam\'as Budav\'ari for helpful assistance and for
providing  the photometric redshifts, and to Eric Bell and George
Rieke for their comments on this manuscript and many interesting
discussions.  We thank the anonymous referee for insightful and prompt
comments, which improved the quality and clarity of the conclusions in
this paper.  We also wish to acknowledge the very generous hospitality
and science--conducive environment of the Aspen Center for Physics,
where much of this work was finished.  CP also thanks STScI for
hospitality on several visits while this research was completed.
Partial support for this work was provided by NASA through grant
GO-07817.01-96A from the Space Telescope Science Institute, which is
operated by the Association of University for Research in Astronomy,
Inc., under NASA contract NAS5-26555.  CP acknowledges partial support
by NASA through Contract Number 960785 issued by JPL/Caltech.

%%%%%%%%%%%%%%%%%%%%%%%%%%%%%%%%%%%%%%%%%%%%%%%%%%%%%%%%%%%%%%%%%%%%%%

%%%%%%%%%%%%%%%%%%%%%%%%%%%%%%%%%%%%%%%%%%%%%%%%%%%%%%%%%%%%%%%%%%%%%%

\end{document}